\documentclass[twoside]{bour}
\usepackage{model_h}
\usepackage{amsfonts,amsmath,stmaryrd, amssymb,amsthm,bbm, cite}
\usepackage[german, french, english,]{babel}
\usepackage{mathrsfs,lscape,rotating, rotate, times, upgreek}
\numberwithin{equation}{section}

\begin{document}
\title
{SPIN\\ {OR, ACTUALLY: SPIN AND QUANTUM STATISTICS}\thanks{Notes prepared with efficient help by K. Schnelli and E. Szabo}}
\author{J\"urg Fr\"ohlich\\ Theoretical Physics \\ ETH Z\"urich and IH\'ES\thanks{Louis-Michel visiting professor at IH\'ES / email: juerg@itp.phys.ethz.ch}}

\maketitle
\begin{abstract}{ The history of the discovery of electron spin and the Pauli principle and the mathematics of spin and quantum statistics are reviewed. Pauli's theory of the spinning electron and some of its many applications in mathematics and physics are considered in more detail. The role of the fact that the tree-level gyromagnetic factor of the electron has the value $g_{\mathrm{e}}=2$ in an analysis of \emph{stability} (and instability) \emph{of matter} in arbitrary external magnetic fields is highlighted. Radiative corrections and precision measurements of $g_{\mathrm{e}}$ are reviewed. The general connection between spin and statistics, the CPT theorem and the theory of braid statistics, relevant in the theory of the quantum Hall effect, are described.
}
\end{abstract}
\vspace{1cm}
\noindent ``He who is deficient in the art of selection may, by showing nothing but the truth, produce all the effects of the grossest falsehoods. It perpetually happens that one writer tells less truth than another, merely because he tells more `truth'.''\\
(T. Macauley, `History', in \emph{Essays}, Vol. 1, p 387, Sheldon, NY 1860)\\

\vspace{1cm}
\noindent Dedicated to the memory of \emph{M. Fierz}, \emph{R. Jost}, \emph{L. Michel} and \emph{V. Telegdi}, teachers, colleagues, friends.

\setlength{\unitlength}{1cm}

\newcommand{\op}[1]{\Hat{#1}} 
\newcommand{\adj}{^*} 
\newcommand{\grad}{\overset{{\scriptscriptstyle\boldsymbol{\rightarrow}}}{\phantom{-}}\mspace{-14mu}\nabla}
\newcommand{\divA} {\grad\cdot\vecA}
\newcommand{\rotA} {\grad\wedge\vecA}
\newcommand{\rotV} {\grad\wedge\vecV}
\newcommand{\deq}{\mathrel{\mathop:}=}
\newcommand{\eqd}{=\mathrel{\mathop:}}
\newcommand{\jadj}{ j \mspace{-7mu}\overline{\phantom{j}}\mspace{-1.5mu}}
\newcommand{\iadj}{ i \mspace{-7mu}\overline{\phantom{j}}\mspace{-1.5mu}}
\newcommand{\adjop}[1]{#1\mspace{-13mu}\overline{\phantom{d}}\mspace{2mu}}
\newcommand{\adjOp}[1]{#1\mspace{-13mu}\overline{\phantom{J}}\mspace{2mu}}
\newcommand{\vecA}{\mspace{1.5mu}{\overset{{\scriptscriptstyle\boldsymbol{\rightarrow}}}{\phantom{-}}}\mspace{-15.5mu}A\mspace{1.5mu}}
\newcommand{\vecB}{\mspace{1.5mu}\overset{{\scriptscriptstyle\boldsymbol{\rightarrow}}}{\phantom{-}}\mspace{-16.5mu}B\mspace{1.5mu}}
\newcommand{\vecV}{\mspace{1.5mu}{\overset{{\scriptscriptstyle\boldsymbol{\rightarrow}}}{\phantom{-}}}\mspace{-14.5mu}V\mspace{1.5mu}}
\newcommand{\vecOmega}{\mspace{1.5mu}\overset{{\scriptscriptstyle\boldsymbol{\rightarrow}}}{\phantom{-}}\mspace{-12.5mu}\Omega\mspace{1.5mu}}
\newcommand{\vecH}{\mspace{2.5mu}\overset{{\scriptscriptstyle\boldsymbol{\rightarrow}}}{\phantom{-}}\mspace{-16.5mu}H\mspace{0.5mu}}
\newcommand{\vecM}{\mspace{4.5mu}\overset{{\scriptscriptstyle\boldsymbol{\rightarrow}}}{\phantom{-}}\mspace{-18.5mu}M\mspace{1.5mu}}
\newcommand{\vecW}{\mspace{4.5mu}\overset{{\scriptscriptstyle\boldsymbol{\rightarrow}}}{\phantom{-}}\mspace{-17.5mu}W\mspace{1.5mu}}
\newcommand{\vecD}{\mspace{1.5mu}\overset{{\scriptscriptstyle\boldsymbol{\rightarrow}}}{\phantom{-}}\mspace{-16mu}D\mspace{1.5mu}}
\newcommand{\vecX}{\mspace{1.5mu}\overset{{\scriptscriptstyle\boldsymbol{\rightarrow}}}{\phantom{-}}\mspace{-17.5mu}X\mspace{1.5mu}}
\newcommand{\vecL}{\mspace{1.5mu}\overset{{\scriptscriptstyle\boldsymbol{\rightarrow}}}{\phantom{-}}\mspace{-14.5mu}L\mspace{1.5mu}}
\newcommand{\vecS}{\mspace{1.5mu}\overset{{\scriptscriptstyle\boldsymbol{\rightarrow}}}{\phantom{-}}\mspace{-15mu}S\mspace{1.5mu}}
\newcommand{\vecE}{\mspace{1.5mu}\overset{{\scriptscriptstyle\boldsymbol{\rightarrow}}}{\phantom{-}}\mspace{-16mu}E\mspace{1.5mu}}
\newcommand{\niceunderline}[1]{#1\mspace{-15mu}\underline{\phantom{J}}\mspace{2.5mu}}
\newcommand{\barfunction}[1]{#1\mspace{-7mu}\overline{\phantom{j}}}
\newcommand{\adjpartial}{\partial\mspace{-8.5mu}\overline{\phantom{j}}}
\newcommand{\widetildeSO} {\mathrm{SO}(2) \mspace{-52mu}\widetilde{\phantom{iiiiiiiii}}\mspace{-1.5mu}}

\newcommand{\umat}{\mathbbmss{1}} 

\newcommand{\exalg}{{\bigwedge}^{\textbf{\large{.}}}}
\newcommand{\exalgOmega}{{\Omega}^{{}^{\textbf{\large{.}}}}}
\newcommand{\be} {\begin{equation}}
\newcommand{\bel}[1]{\begin{equation}\label{#1}}
\newcommand{\ee} {\end{equation}}
\renewcommand{\epsilon}{\varepsilon}

\newcommand{\R} {\mathbb{R}}
\newcommand{\C} {\mathbb{C}}
\newcommand{\N} {\mathbb{N}}
\newcommand{\Z} {\mathbb{Z}}
\newcommand{\Q} {\mathbb{Q}}
\newcommand{\E} {\mathbb{E}}
\newcommand{\M} {\mathcal{M}} 
\newcommand{\Spinn} {\mathrm{Spin}(n)}
\newcommand{\Hilbert} {\mathscr{H}}
\newcommand{\HO}[1] {H^{(#1)}}
\newcommand{\xv} {\vec{x}}
\newcommand{\yv} {\vec{y}}
\newcommand{\vv} {\vecV}
\newcommand{\Vv} {\vecV}
\newcommand{\Av} {\vecA}
\newcommand{\gmf} {g_{\mathrm{e}}}
\newcommand{\scalarr}[2]{(#1 \mspace{2mu}, #2)}

\newcommand{\scalar}[2]{\langle{#1} \mspace{2mu}, {#2}\rangle}
\newcommand{\scalarb}[2]{\big\langle{#1} \mspace{2mu}, {#2}\big\rangle}
\newcommand{\scalarB}[2]{\Big\langle{#1} \,\mspace{2mu},\, {#2}\Big\rangle}
\newcommand{\bra}[1]{\langle #1 |}
\newcommand{\brab}[1]{\big\langle #1 \big|}
\newcommand{\braB}[1]{\Big\langle #1 \Big|}
\newcommand{\brabb}[1]{\bigg\langle #1 \bigg|}
\newcommand{\braBB}[1]{\Bigg\langle #1 \Bigg|}

\newcommand{\ket}[1]{| #1 \rangle}
\newcommand{\ketb}[1]{\big| #1 \big\rangle}
\newcommand{\ketB}[1]{\Big| #1 \Big\rangle}
\newcommand{\ketbb}[1]{\bigg| #1 \bigg\rangle}
\newcommand{\ketBB}[1]{\Bigg| #1 \Bigg\rangle}
   \newtheoremstyle{anttiit}
     {7pt}
     {7pt}
     {\itshape}
     {}
     {\scshape}
     {.}
     {.5em}
     {}
   \newtheoremstyle{antti}
     {7pt}
     {7pt}
     {\normalfont}
     {}
     {\scshape}
     {.}
     {.5em}
     {}

\theoremstyle{antti} 
\newtheorem{definition}{Definition}[section]
\newtheorem*{definition*}{Definition}
\newtheorem{example}{Example}[section]
\newtheorem*{example*}{Example}
\newtheorem{remark}{Remark}[section]
\newtheorem*{remarks}{Remarks}
\newtheorem*{remark*}{Remark}
\theoremstyle{anttiit} 
\newtheorem{theorem}{Theorem}[section]
\newtheorem*{theorem*}{Theorem}
\newtheorem{lemma}[theorem]{Lemma}
\newtheorem*{lemma*}{Lemma}
\newtheorem{corollary}[theorem]{Corollary}
\newtheorem*{corollary*}{Corollary}

\newenvironment{Proof}[1][Proof]{\begin{proof}[\sc{#1}]}{\end{proof}}

\newpage
\tableofcontents
\newpage

\section[Introduction to `Spin']{Introduction to `Spin'\footnote{I have to refrain from quoting literature in this introductory section -- apologies!}}

The 21$^{\textrm{st}}$ Century appears to witness a fairly strong decline in Society's -- the public's, the politicians', the media's and the younger generations' -- interest in the hard sciences, including Physics, and, in particular, in fundamental theoretical science based on precise mathematical reasoning. It is hard to imagine that reports on a discovery like the deflection of light in the gravitational field of the sun and on the underlying theory, \emph{general relativity}, along with a photograph of its creator, \emph{Albert Einstein}, would make it onto the front pages of major daily newspapers, as it did in 1919.\\
\indent This development is, of course, not entirely accidental, and I could easily present a list of reasons for it. But let's not!\\
\indent While the amount of economic wealth  and added value that have been and are still being created on the basis of Physics-driven discoveries of the 19$^{\textrm{th}}$ and 20$^{\textrm{th}}$ Century is truly \emph{gigantic}, and while one may expect that this will continue to be the case for many more years to come, fundamental physical science is confronted with a certain decline in public funding, e.g., in comparison with the life sciences. Physics is perceived to have entered a baroque state, with all the beauty that goes with it.\\
\indent In this situation, it is laudable that our French colleagues are doing something to document the continuing importance and the lasting beauty of Physics: the `\emph{S\'eminaire Poincar\'e}' (or ``Bourbaphy'')! I hope that the organizers of the `S\'eminaire Poincar\'e' will find the right format and the right selection of topics for their series, and that their seminar will be accompanied by complementary activities aimed at a broader public.

\indent This time, the topic of the `S\'eminaire Poincar\'e' is `\emph{Spin (and Quantum Statistics)}'. This choice of topic is not unreasonable, because, on one hand, it involves some interesting and quite fundamental experiments and theory and, on the other hand, it is connected to breathtakingly interesting and important practical applications. The scientific community sees me in the corner of mathematical physics and, thus, I have been asked to present an introductory survey of, primarily, the mathematical aspects of `Spin and Quantum Statistics'. I am only moderately enthusiastic about my assignment, because, as I have grown older, my interests and activities have shifted more towards general theoretical physics, and, moreover, I have contributed a variety of results to, e.g., the theory of magnetism and of phase transitions accompanied by various forms of magnetic order that I cannot review, for lack of space and time.\\
\indent In this short introduction, I attempt to highlight the importance of `Spin and Quantum Statistics' for many phenomena in physics, including numerous ones that have found important technological applications, and I wish to draw attention to some of the many unsolved theoretical problems.\\

\indent Our point of departure is found in the facts that electrons, positrons, neutrinos, protons and neutrons are particles with \emph{spin $\frac{1}{2}$} obeying \emph{Pauli's exclusion principle}. With the exception of neutrinos, they have a \emph{non-vanishing magnetic dipole moment}. Moreover, those particles that carry electric charge experience Coulomb- and Lorentz forces. In a magnetic field their magnetic moments and spins precess (like tops in the gravitational field of the Earth). All fundamental forces appear to be mediated by exchange of bosons of spin 1 (gauge bosons) or  helicity 2 (gravitons). These facts, when exploited within the framework of quantum theory, are at the core of our theoretical description of a \emph{vast number of phenomena} some of which we will now allude to. They are, in their majority, not very well understood, mathematically.\\

	\indent(1) \emph{Chemistry}. That electrons have spin $\frac{1}{2}$ and obey the Pauli principle, i.e., are \emph{fermions}, is one of the most crucial facts underlying all of chemistry. For example, it is the basis of our understanding of covalent bonding. If electrons were \emph{spinless} fermions not even the simplest atoms and molecules would be the way they are in Nature: Only ortho-helium would exist, and the hydrogen molecule would not exist.\\	
	\indent If electrons were not fermions, but bosons, there would exist ions of large negative electric charge, matter would form extremely dense clumps, and bulk matter would not be thermodynamically stable; (see section 4).\\
	\indent Incidentally, the hydrogen molecule is the only molecule whose stability has been deduced directly from the Schr\"odinger-Pauli equation with full mathematical rigour\footnote{by \emph{G.M. Graf, J.M. Richard, M. Seifert} and myself.}. \emph{Hund's 1$^{\textrm{st}}$ Rule} in atomic physics, which says that the total spin of the electrons in an only partially filled $p$-, $d$-, $\ldots$ shell of an atom tends to be as large as possible, is poorly understood, \emph{mathematically}, on the basis of the Schr\"odinger-Pauli equation.\\
	\indent We do not understand how \emph{crystalline} or \emph{quasi-crystalline order} can be derived as a consequence of \emph{equilibrium quantum statistical mechanics}.\\
	\indent All this shows how little we understand about `emergent behavior' of many-particle systems on the basis of fundamental theory. We are \emph{not} trying to make an argument against reductionism, but one in favour of a \emph{pragmatic attitude:} We should be reductionists whenever this attitude is adequate and productive to solve a given problem and `emergentists' whenever this attitude promises more success!\\
	
	\indent(2) \emph{`Nuclear and hadronic chemistry'}. At the level of fundamental theory, our understanding of binding energies, spins, magnetic moments and other properties of nuclei or of the life times of radioactive nuclei remains quite rudimentary. Presently more topical are questions concerning the `chemistry of hadrons', such as: How far are we in understanding, on the basis of QCD, that a color-singlet bound state of three quarks (fermions with spin $\frac{1}{2}$), held together by gluons, which forms a proton or a neutron, has spin $\frac{1}{2}$? How, in the world, can we reliably calculate the magnetic dipole moments (the gyromagnetic factors) of hadrons? How far are we in truly understanding low-energy QCD? These are questions about strongly coupled, strongly correlated physical systems. They are notoriously hard to answer. \\

	\indent(3) \emph{Magnetic spin-resonance}. The fact that electrons and nuclei have spin and magnetic dipole moments which can precess is at the basis of \emph{Bloch's spin-resonance phenomenon}, which has enormously important applications in the science and technology of \emph{imaging}; (Nobel Prizes for \emph{Felix Bloch, Edward Purcell, Richard Ernst, Kurt W\"uthrich}...). Of course, in this case, the basic theory is simple and well understood.\\

	\indent(4) \emph{Stern-Gerlach experiment}: a direct experimental observation of the spin and magnetic moment of atoms. Theory quite easy and well understood.\\

	\indent(5) \emph{Spin-polarized electron emission from magnetic materials}. This is the phenomenon that when massaged with light certain magnetic materials emit  \emph{spin-polarized electrons}. It has been discovered and exploited by \emph{Hans-Christoph Siegmann} and collaborators and has important applications in, e.g., particle physics.\\

	\indent(6) \emph{Electron-spin precession in a Weiss exchange field}. When a spin-polarized electron beam is shot through a spontaneously magnetized iron-, cobalt or nickel film the spins of the electrons exhibit a \emph{huge} precession. This effect has been discovered by H.-C. Siegmann and his collaborators and might have important applications to ultrafast \emph{magnetic switching}. Theoretically, it can be described with the help of the Zeeman coupling of  the electrons' spin to the \emph{Weiss exchange field} (\emph{much larger} than the magnetic field) inside the magnetized film. This effect can be interpreted as a manifestation of the $\mathrm{SU}(2)_{\textrm{spin}}$-\emph{gauge-invariance} of Pauli's electron equation; (see also section 3.3).\\
	\indent Related effects can presumably be exploited for the production of spin-polarized electrons and for a Stern-Gerlach type experiment for electrons.\\

	\indent(7) \emph{Magnetism}. There are many materials in Nature which exhibit magnetic ordering at low temperatures or in an external magnetic field, often in combination with metallic behavior. One distinguishes between \emph{paramagnetism, diamagnetism, ferromagnetism, ferrimagnetism, anti-ferromagnetism}, etc. In the context of the quantum Hall effect, the occurrence of \emph{chiral spin liquids} and of chiral edge spin currents has been envisaged; $\ldots$. \\
	\indent The theory of paramagnetism is due to Pauli; it is easy. The theoretical basis of diamagnetism is clear. The theory of \emph{anti-ferromagnetism} and \emph{N\'eel order} at low temperatures in insulators is relatively far advanced. But the theory of ferromagnetism and the appearance of spontaneous magnetization is disastrously \emph{poorly understood}, \emph{mathematically}. Generally speaking, it is understood that spontaneous (ferro- or anti-ferro-) magnetic order arises, at low enough temperature, by a conspiracy of electron spin, the Pauli principle and Coulomb repulsion among electrons. The earliest phenomenological description of phase transitions accompanied by the appearance of magnetic order goes back to \emph{Curie} and \emph{Weiss}. \emph{Heisenberg} proposed a quantum-mechanical model inspired by the idea of direct electron exchange interactions between neighboring magnetic ions (e.g. $\textrm{Fe}$) in a crystalline back ground. While it has been shown, mathematically, that the classical Heisenberg model (large-spin limit) and the \emph{Heisenberg anti-ferromagnet} exhibit the expected phase transitions\footnote{in work by \emph{Simon, Spencer} and myself, and by \emph{Dyson, Lieb} and Simon; and followers.}, \emph{no} precise understanding of the phase transition in the \emph{Heisenberg ferromagnet} (finite spin) has been achieved, yet.\\
	\indent Most of the time, the microscopic origin of \emph{exchange interactions} between spins in magnetic materials remains poorly understood, mathematically. No mathematically precise understanding of ferromagnetic order in models of \emph{itinerant electrons}, such as the weakly filled one-band \emph{Hubbard model}, has been reached, yet. However, there is some understanding of N\'eel order in the half-filled one-band Hubbard model (`\emph{Anderson mechanism}') and of ferromagnetic order in Kondo lattice models with a weakly filled conduction band (\emph{Zener's mechanism} of \emph{indirect exchange}), which is mathematically rather precise at \emph{zero} temperature.\\
	\indent Realistic \emph{spin glasses} are extremely poorly understood, theory-wise. \\
	\indent Alltogether, a general theory of magnetism founded on basic equilibrium quantum statistical mechanics still remains to be constructed!\\
	\indent Of course, magnetism has numerous applications of great importance in magnetic data storage, used in computer memories, magnetic tapes and disks, etc.\\

	\indent(8) \emph{Giant and colossal magneto-resistance}. The discoverers of giant magneto-resistance, \emph{Albert Fert} and \emph{Peter Gr\"unberg}, have just been awarded the 2007 Nobel Prize in Physics. Their discovery has had phantastic applications in the area of data storage and -retrieval. It will be described at this seminar by Fert and collaborators. Suffice it to say that electron spin and the electron's magnetic moment are among the main characters in this story, and that heuristic, but quite compelling theoretical understanding of these phenomena is quite advanced.\\

	\indent(9) \emph{Spintronics}. This is about the use of electron spin and multi-spin entanglement for the purposes of quantum information processing and quantum computing. Presently, it is a hot topic in mesoscopic physics. Among its aims might be the construction of scalable arrays of interacting quantum dots (filled with only few electrons) for the purposes of quantum computations; (the spins of the electrons would store the Qbits).\\

	\indent(10) \emph{The r\^ole of electron spin and the Weiss exchange field in electron -- or hole -- pairing mechanisms at work in layered high-temperature superconductors}. This is the idea that the Weiss exchange field in a magnetic material can produce a strong attractive force between two holes or electrons (introduced by doping) in a spin-singlet state, leading to the formation of Schafroth pairs, which, after condensation, render such materials superconducting.\\

	\indent(11) \emph{The r\^ole played by spin and by particle-pairing in the miraculous phase diagram of} ${}^{3}\textrm{He}$ \emph{and in its theoretical understanding}. The r\^ole played by spin in the physics of `heavy fermions'.\\

	\indent(12) \emph{The r\^ole of the Pauli principle (and spin, in particular neutron spin) in the physics of stars}. The theory of the Chandrasekhar limit for white dwarfs and neutron stars is based on exploiting the Pauli principle for electrons or neutrons in an important way. The superfluidity expected to be present in the shell of a neutron star is a phenomenon intimately related to the spin of the neutron, neutron pairing and pair condensation.\\

\indent Many of these topics have been close to my heart, over the years, and I have written hundreds of pages of scientific articles that have been read by only few people. One could easily offer a one-year course on these matters. But, in the following sections, I really have to focus on just a few \emph{basic} aspects of `Spin and Quantum Statistics'.\\

\indent\emph{Acknowledgments.} I thank C. Bachas, B. Duplantier and V. Rivasseau for inviting me to present a lecture at the `S\'eminaire Poincar\'e' and my teachers and numerous collaborators for all they have taught me about `Spin and Quantum Statistics', over many years. I am very grateful to K. Schnelli for his help.\\

\indent\emph{Remark.} These notes have been written at a `superluminal' speed and are therefore likely to contain errors and weaknesses, which I wish to offer my apologies for.


\section{The Discovery of Spin and of Pauli's Exclusion Principle, Historically Speaking}
My main sources for this section are~\cite{R1,R2,R3,R4,R5,R6}. Let us dive into a little history of science, right away.

\subsection{Zeeman, Thomson and others, and the discovery of the electron}
Fairly shortly before his death, in 1867, \emph{Michael Faraday} made experiments on the influence of `strong' magnetic fields on the frequency of light emitted by excited atoms or molecules. He did this work in 1862 and did not find any positive evidence for such an influence. In the 1880's, the American physicist \emph{Henry Augustus Rowland} invented the famous `Rowland gratings', which brought forward much higher precision in measuring wave lengths of spectral lines.

In 1896, \emph{Pieter Zeeman}, a student of \emph{Kamerlingh Onnes} and \emph{Hendrik Antoon Lorentz}, took up Faraday's last experiments again, using Rowland gratings. He found that the two sodium D-lines are broadened when the magnetic field of an electromagnet\footnote{Concerning electromagnets, one could embark on a report of the important contributions and inventions of \emph{Pierre Weiss}, once upon a time a professor at ETH Zurich.} is turned on. He proposed to interpret the effect in terms of Lorentz' theory of charges and currents carried by fundamental, point-like particles. In 1895, Lorentz had introduced the famous \emph{Lorentz force} acting on charged particles moving through an electromagnetic field. When Zeeman had discovered the effect named after him Lorentz proposed a model of harmonically bound charged particles of charge $e$. When a magnetic field $\vecH$ is turned on in a direction perpendicular to the plane of motion of such a particle the angular frequency of its motion changes by the amount

\begin{equation*}
\varDelta\omega = \frac{e}{mc}|\vecH|\,,
\end{equation*}

\noindent where $m$ is its mass and $c$ is the speed of light. Using Lorentz' formula, Zeeman inferred from the broadening of the sodium lines that

\begin{equation*}
\frac{e}{m}\simeq 10^7 \mathrm{emu}/\mathrm{g}\qquad(1.76\times10^7\mathrm{emu}/\mathrm{g})\,.
\end{equation*}

In 1897, Zeeman discovered a \emph{splitting} of the blue line of cadmium, in rough agreement with Lorentz' theoretical expectations. From polarization effects he inferred that $e$ is \emph{negative}. \emph{George Stoney} had earlier provided an estimate for the elementary electric charge $e$. Thus, Zeeman could have predicted the mass of the charged particle that emits electromagnetic radiation from the `interior' of an atom or molecule, the \emph{electron}.

In the same year, the quotient $\frac{e}{m}$ was measured in experiments with \emph{cathode rays}, first by \emph{Emil Wiechert}\footnote{Of fame also in connection with the Li\'enard-Wiechert potentials.}, who conjectured that such rays consist of charged particles with a very small mass $m$ (=mass of an electron); then - with very high accuracy - by \emph{Walter Kaufman} and, more or less simultaneously, by \emph{J.J. Thomson}, who also proposed Wiechert's charged-particle picture. In 1899, Thomson measured the value of $e$ by cloud chamber experiments, and, in 1894, he had obtained some bounds on the speed of propagation of cathode rays, showing that this speed is considerably smaller than the speed of light. This combination of accomplishments led to the common view that J.J. Thomson is the discoverer of the \emph{electron}.

After the discovery of relativistic kinematics in 1905, by \emph{Einstein}, experiments with electrons became the leading tool to verify the kinematical predictions of the \emph{special theory of relativity}.
\hyphenation{Abraham Pais}
\subsection{Atomic spectra}
\begin{quote}``Spectra are unambiguous visiting cards for the gases which emit them.'' (Abraham Pais \cite{R1})
\end{quote}
Spectroscopy started in Heidelberg with the work of \emph{Gustav Kirchhoff} (1859) and \emph{Robert Bunsen}. Against the philosophical prejudices of \emph{Auguste Comte}, Kirchhoff concluded with the help of \emph{absorption spectroscopy} that the solar atmosphere must contain sodium\footnote{``It's not philosophy we are after, but the behaviour of real things.'' (R.P. Feynman)}. Kirchhoff and Bunsen are the fathers of modern optical spectroscopy and its application as an exploratory tool.

The first three lines of the \emph{hydrogen spectrum} were first observed by \emph{Julius Pl\"ucker} in 1859, then, more precisely, by \emph{Anders \AA ngstr\"om} in 1868. Searches for patterns in spectral lines started in the late 1860's. The first success came with \emph{Stoney} in 1871. The break-through was a famous formula,

\begin{equation*}
\lambda_n=\frac{Cn^2}{n^2-4}\,,
\end{equation*}

\noindent where the $\lambda_n$ are wave lengths of light emitted by hydrogen, C is some constant, and $n=3,4,\ldots$, discovered by \emph{Johann Jakob Balmer} in 1885. In 1892, \emph{Carl Runge} and \emph{Heinrich Kayser} made precise measurements of spectral lines of 22 elements. \emph{Runge} and \emph{Friedrich Paschen} discovered the spectra of ortho- and parahelium. A precursor of the \emph{Rydberg-Ritz combination principle} was discovered in 1889 by \emph{Johannes Rydberg}, its general form was found by \emph{Walther Ritz} in 1908.

Precursors of \emph{Rutherford's planetary model} of the atom (1911) can be found in remarks by \emph{Heinrich Hertz} (lectures about the constitution of matter in Kiel), \emph{Hermann von Helmholtz}, \emph{Jean Perrin} (1901), \emph{Hantaro Nagaoka} (1903), and \emph{J.J. Thomson} (1906).

In 1913, \emph{Niels Bohr} came up with his quantum theory of the hydrogen atom\footnote{His theory has a more incomplete precursor in the work of \emph{Arthur Erich Haas} (1910).}, with the idea that atomic spectra arise by photon emission during transitions of an electron from one `stationary state' (a term introduced by Bohr) to another, and with the \emph{Bohr frequency condition}, which has a precursor in \emph{Einstein's} work of 1906 on Planck's law for black-body radiation. Bohr's results provided a quantum-theoretical `explanation' of Balmer's formula and of a special case of the Rydberg-Ritz combination principle. 

Subsequent to Bohr's discoveries, in attempts to interpret the so-called `fine structure' of atomic spectra discovered by \emph{Albert Michelson} (1892) and Paschen (1915), Bohr's quantum theory was to be married with the special theory of relativity. The pioneer was \emph{Arnold Sommerfeld} (1916). He introduced the \emph{fine structure constant}

\begin{equation*}
\alpha=\frac{e^2}{\hbar c}\,.
\end{equation*}

\noindent Sommerfeld's formula for the relativistic hydrogen energy spectrum is 

\begin{equation}
E_{n,l}=-\mathrm{Ry}\left[\frac{1}{n^2}+\frac{\alpha^2}{n^3}\left(\frac{1}{l+1}-\frac{3}{4n}\right)\right]+\mathcal{O}(\alpha^4)\,,
\label{relativisticenergy}
\end{equation}

\noindent where $n=1,2,3,\ldots, l= 0,1,\ldots,n-1$ and $\mathrm{Ry}$ is the Rydberg constant. Of course $\vecL$, with $|\vecL|\simeq \hbar (l+1)$, is the (quantized) angular momentum of the electron orbiting the nucleus.

In trying to explain experimental results of Paschen, \emph{Bohr} and, independently, \emph{Wojciech Rubinowicz} (a collaborator of Sommerfeld) found the \emph{selection rule}

\begin{equation}
\varDelta l = \pm 1
\label{selectionrule}
\end{equation}
\noindent for transitions between stationary states.\newline
\indent This rule did not work perfectly. In 1925, in their first publication and after ground-breaking work of \emph{Wolfgang Pauli}, \emph{George Uhlenbeck} and \emph{Samuel Goudsmit} proposed a modification of the Bohr-Rubinowicz selection rule: In (\ref{relativisticenergy}), write
\begin{equation}
l+1=j+\frac{1}{2}\,,
\end{equation}
\noindent with $j$ half-integer, and replace (\ref{selectionrule}) by 
\begin{equation}
\varDelta j = 0,\pm 1\,.
\end{equation}

\noindent This reproduced data for the fine structure of the ${\textrm{He}}^{+}$ spectrum perfectly. Here, the half-integer quantum number $j$ appears. Similar ideas were proposed independently by \emph{John Slater}.

Of course, the half-integer nature of $j$ (for atoms or ions with an odd number of bound electrons) is related to \emph{electron spin}; as everybody knows nowadays. Actually, half-integer quantum numbers were first introduced systematically by \emph{Alfred Land\'e} in an analysis of the Zeeman effect and correctly interpreted, by Pauli, as ``due to a peculiar classically not describable two-valuedness of the quantum theoretical properties of the valence electron'', in 1924. 

We have now reached the period when \emph{electron spin} enters the scene of physics. I shall briefly sketch how it was discovered by Pauli towards the end of 1924.
\subsection{Pauli's discovery of electron spin and of the exclusion principle}
Pauli's papers on electron spin and the exclusion principle are [7,8,10]. In~\cite{R7}, he analyzes what is known as the \emph{`anomalous Zeeman effect'}, namely the Zeeman effect in \emph{weak} magnetic fields (when relativistic spin-orbit terms dominate over the Zeeman term in the atomic Hamiltonian). This theme is taken up again in [8,10] and leads him to discover electron spin and the exclusion principle. Let us see how this happened!\newline

\indent In [7], Pauli started from the following facts and/or assumptions; (I follow modern notation and conventions).
\begin{itemize}
	\item[(1)] Spectral terms (energies corresponding to stationary states) can be labeled by \emph{`quantum numbers':}
		\begin{itemize}
			\item[(i)] A principal quantum number, $n$, (labeling shells).
			\item[(ii)] $L=0,1,2,3,\ldots(S,P,D,F,\ldots)$ with $L<n$ -- our orbital angular momentum quantum number -- and $M_L=-L,\,-L+1,\ldots,\,L$ --  the magnetic quantum number.
			\item[(iii)] $S=0, 1/2, 1,\ldots$, and $M_S=-S,\,-S+1,\ldots,\,S$.
			\item[(iv)] The terms of a multiplet with given $L$ and $S$ are labeled by a quantum number $J$ (our total angular momentum quantum number), whose possible values are $J= L+S, L+S-1,\ldots,|L-S|$, and a magnetic quantum number $M=-J,\,-J+1,\ldots,\,J$.
		\end{itemize}
	\item[(2)]There are \emph{selection rules} for the allowed transitions between stationary states:
	\begin{itemize} \item[] $\varDelta L = \pm 1,\,\varDelta S = 0,\,\varDelta J = 0,\pm 1$ (with $J=0\rightarrow J=0$ forbidden).
	\end{itemize}
	\item[(3)] Denoting by $Z$ the atomic number of a \emph{neutral} atom, one has the \emph{correspondence}
	\begin{eqnarray*}
	Z&\textrm{even}&\longleftrightarrow\;\,S,\,J\:\textrm{integer}\,,\\ \nonumber
	Z&\textrm{odd}&\longleftrightarrow\;\,S,\,J\:\textrm{half-integer}\,.
	\end{eqnarray*}
	\item[(4)] \emph{Bohr's frequency condition} (i.e., the formula for the frequency of light emitted in a transition from one stationary state to a lower-lying one.)
	\item[(5)] \emph{Line splittings in a magnetic field} $\vecH$.
		If Zeeman splitting dominates fine structure splitting (\emph{Paschen-Back effect}) then the energy splitting is given by
			\begin{equation}\label{energyshift}
			\varDelta E\simeq (M_L+2M_S)\mu_0|\vecH|\,,
			\end{equation}			
 		where $\mu_0=\frac{e\hbar}{2mc}$ is \emph{Bohr's magneton} (actually introduced by Pauli in 1920). \newline\newline
		If fine structure (spin-orbit interactions)  dominates over Zeeman splitting (\emph{anomalous Zeeman effect}) a term with quantum number $J$ splits into $2J+1$ equidistant levels labeled by a `magnetic quantum number' $M=-J,\,-J+1,\ldots,\,J$, and the energy splitting for a term with given $L,\,S,\,J$ and $M$ is given by
			\begin{equation*}
			\varDelta E= Mg\mu_0|\vecH|\,,
			\end{equation*}
		where $g$ is the \emph{Land\'e factor},
			\begin{equation}
			g=\frac{3}{2}+\frac{S(S+1)-L(L+1)}{2J(J+1)}\,.
			\label{gfactor}
			\end{equation}
		The selection rules for transitions are given by
			\begin{equation*}
			\varDelta M = 0, \pm 1\,.
			\end{equation*}
	
\end{itemize}

Starting from the Paschen-Back effect, Pauli postulates that the sum of energy levels in a multiplet with given $L$ and $M$ is a \emph{linear} function of $|\vecH|$ when one passes from strong to weak magnetic fields. He then determines Land\'e's $g$-factors uniquely from the energy splittings in large fields and the `sum rule' just stated. Nowadays, these calculations are an elementary exercise in the algebra of quantum-mechanical angular momenta (see, e.g.,~\cite{R6}), which I will not reproduce. Pauli concludes his paper~\cite{R7} with prophetic remarks that a derivation of the `laws' he analyzed within the principles of the (old) quantum theory then known, does not appear to be possible; that the connection between angular momentum and magnetic moment predicted by Larmor's theorem does not generally hold $(\gmf=2\,!)$; and that the appearance of half-integer values of $M$ and $J$ goes beyond the quantum theory of quasi-periodic mechanical systems.

Soon afterwards, Pauli started to think about the problem of completion of electron shells in atoms and the doublet structure of alkali spectra. This led him to his important paper~\cite{R8}.  Before I sketch the contents of~\cite{R8}, I recall a standard calculation of the gyromagnetic ratio between magnetic moment $\vecM$, and angular momentum $\vecL$. We consider a distribution of rotating, charged, massive matter. If we assume that the charge and mass densities are proportional to each other then

\begin{equation}
\frac{|\vecM|}{|\vecL|}=\frac{|q|}{2mc}\,,
\end{equation}

\noindent where $q$ is the total charge and $m$ the total mass. Apparently, the Land\'e factor is $g=1$. If the same calculation is done using relativistic kinematics (as Pauli did in~\cite{R8}) one finds that

\begin{equation}
\frac{|\vecM|}{|\vecL|}=\frac{|q|}{2mc}\cdot(\overline{\gamma})^{-1}\,,
\end{equation}

\noindent where $\gamma = (1-\frac{v^2}{c^2})^{-1/2}$, $v$ is the speed of a mass element, and $\overline{(\cdot)}$ denotes a suitable average. Note that $(\overline\gamma)^{-1}<1$!

When Pauli worked on paper~\cite{R8} the prejudice was that, for alkaline metals, the quantum number $S$ was related to the angular momentum of the \emph{core} (filled shells) of an atom. It was then to be expected that it would correspond to a magnetic moment $\vecM$ with

\begin{equation*}
|\vecM|=\frac{e}{2mc}(\overline\gamma)^{-1}S\,.
\end{equation*}

\noindent Thus, the Land\'e factor of the core should have come out to be

\begin{equation}
g_{\textrm{core}}=(\overline\gamma)^{-1}<1\,.
\end{equation}

\noindent Since the electrons in the core of large-$Z$ elements are relativistic, the prediction of the `core model' would have been that $g_{\textrm{core}}$ is \emph{measurably smaller} than 1.

However, formula (\ref{energyshift}), well confirmed, for large $|\vecH|$, in experiments by Runge, Paschen and Back for large-$Z$ elements, and Land\'e's formula (\ref{gfactor}) were only compatible with 
\begin{equation*}
g_{\textrm{core}}=2\,.
\end{equation*}
\noindent Pauli concluded that $S$ could not have anything to do with the angular momentum of the core (filled shells) of an atom. He goes on to propose that filled shells have angular momentum 0 and do \emph{not} contribute to the magnetic moment of the atom. By studying experimental data for the Zeeman effect in alkali atoms, he arrives at the following \emph{key conclusion}:

\begin{quote}
``The closed electron configurations shall not contribute to the magnetic moment and angular momentum of the atom. In particular, for the alkalis, the angular momenta of, and energy changes suffered by, the atom in an external magnetic field shall be regarded exclusively as an effect of the valence electron (`Leuchtelektron'), which is also the source of the magneto-mechanical anomaly\footnote{$\gmf=2\,!$}. The doublet structure of the alkali spectra, as well as the violation of the Larmor theorem are, according to this point of view, a result of a classically not describable two-valuedness of the quantum-theoretical properties of the valence electron.''
\end{quote}

\noindent Thus, Pauli had discovered the \emph{spin of the electron} and the \emph{`anomaly'} in its $g$-factor, $\gmf=2$. (See~\cite{R8bis} for a recent study why $g=2$ is the natural value of the tree-level gyromagnetic ratio of charged elementary particles.)\newline

\indent Soon, \emph{Ralph Kronig} and, independently, \emph{Uhlenbeck} and \emph{Goudsmit} interpreted the quantum number $S$ as due to an intrinsic rotation of electrons, picturing them as little charged balls. Kronig explained his idea to Pauli, who thought it was nonsense\footnote{One might say: \emph{correctly}, (since $s=\frac{1}{2}$ is far away from the classical limit $s=\infty$).}, and Kronig did not publish it. Uhlenbeck and Goudsmit were confronted with objections by \emph{Lorentz} against their idea related to the fact that $\gmf=2$, and wanted to withdraw their paper from publication, but \emph{Ehrenfest} convinced them to go ahead and publish it.\newline

\indent Now comes the problem of the \emph{Thomas precession}: As had been discovered by Einstein and explained by him to his colleagues working on the quantum theory, an electron traveling through an electric field $\vecE$ with a velocity $\vec{v}$ feels a magnetic field

\begin{equation}
\vecB'=-\frac{\vec{v}}{c}\wedge\vecE+\mathcal{O}\left(\frac{v^2}{c^2}|\vecE|\right)
\label{bfield}
\end{equation}

\noindent in its rest frame. If its magnetic moment in the rest frame is denoted by $\vecM$ one expects that its spin $\vecS$, will exhibit a precession described, in its rest frame, by 

\begin{equation}
\frac{d\vecS}{dt}=\vecM\wedge\vecB'\,,
\label{rotation}
\end{equation}

\noindent corresponding to a magnetic energy 

\begin{equation}
U'=-\vecM\cdot\vecB'\,.
\label{magneticenergy}
\end{equation}

\noindent For an electron in the Coulomb field of a nucleus

\begin{equation}
e\vecE=-\frac{\vec{x}}{r}\frac{\mathrm{d}V(r)}{\mathrm{d}r}\,,
\label{coulombenergy}
\end{equation}

\noindent where $r$ is the distance to the nucleus, and $V$ is the Coulomb potential. Plugging (\ref{coulombenergy}) into (\ref{bfield}) and (\ref{bfield}) into (\ref{magneticenergy}), we find that

\begin{equation*}
U'=\frac{\gmf}{2(mc)^2}\big(\vecS\cdot \vecL\big)\frac{1}{r}\frac{\mathrm{d}V(r)}{\mathrm{d}r}\,,
\end{equation*}

\noindent where $\vecL$ is the orbital angular momentum, the well-known spin-orbit interaction term. If this formula is taken literally and compared with Sommerfeld's calculation of the fine structure (see eq. (\ref{relativisticenergy})) one finds that $\gmf$ must be 1. This is a contradiction to the value $\gmf=2$ found in the analysis of the Zeeman effect for alkali atoms.

This contradiction vexed many people, foremost Pauli, and \emph{Heisenberg} communicated it to Uhlenbeck and Goudsmit when he saw their paper, (``ihre mutige Note''). It was resolved by \emph{Llewellyn Thomas}, in February 1926. Thomas pointed out that the rest frame of an electron moving in the Coulomb field of a nucleus is actually \emph{rotating} relative to the laboratory frame. The angular velocity of that rotation is denoted by $\vec{\omega}_T$. Then the equation for the precession of the electron's spin in a \emph{non-rotating} frame moving with the electron is given by

\bel{eq2.14}
\left(\frac{\mathrm{d}\vecS}{\mathrm{d}t}\right)_{\textrm{non-rotating}}=\left(\frac{\mathrm{d}\vecS}{\mathrm{d}t}\right)_{\textrm{rest frame}}+\vec{\omega}_T\wedge\vecS\,,
\ee

\noindent with $\left(\frac{\mathrm{d}\vecS}{\mathrm{d}t}\right)_{\textrm{rest frame}}$ given by (\ref{rotation}). The `magnetic energy' in the non-rotating frame is then given by

\begin{equation}\label{eq2.15}
U=U'+\vecS\cdot\vec{\omega}_T\,.
\end{equation}

\noindent The problem now boils down to calculating $\vec{\omega}_T$. This is an exercise in composing Lorentz boosts whose solution can be looked up, e.g., in~\cite{R9}. The formula for $\vec{\omega}_T$ is

\begin{equation}
\omega_T=\frac{1}{2}\frac{\vec{a}\wedge\vec{v}}{c^2}\left(1+\mathcal O\left(\frac{v^2}{c^2}\right)\right)\,,
\label{omegaT}
\end{equation}

\noindent where $\vec{a}$ is the acceleration of the electron, which, in an electric field, is given by $-\frac{e}{m}\vecE$, up to corrections $\mathcal O\left(\frac{v}{c}\right)$. Then $U$ is given by

\bel{eq2.17}
U\simeq\frac{(\gmf-1)e}{2mc}\,\vecS\cdot\left(\frac{\vec{v}}{c}\wedge\vecE\right)\,,
\ee

\noindent which, in the Coulomb field of a nucleus, becomes

\begin{equation}
U\simeq\frac{(\gmf-1)e}{2(mc)^2}\vecS\cdot\vecL\frac{1}{r}\frac{\mathrm{d}V}{\mathrm{d}r}\,.
\end{equation}

\noindent This expression reproduces the correct fine structure. Expression (\ref{omegaT}) for the Thomas precession frequency and the second term on the R.S. of (\ref{eq2.15}) have been verified, experimentally, in data for spectra of nuclei (where the Land\'e $g$-factor does \emph{not} take the value $g=2$).

Thomas' observations convinced people, including Einstein and Pauli, and boosted the acceptance of the naive interpretation of electron spin proposed by Uhlenbeck and Goudsmit in the physics community. 

I conclude my excursion into the history of the discovery of spin with comments on precursors.\newline

\indent In 1900, \emph{George Francis FitzGerald} had raised the question whether magnetism might be due to a rotation of electrons. In 1921, \emph{Arthur Compton} proposed that ``it is the electron rotating about its axis which is responsible for ferromagnetism''; (see~\cite{R1}, page 279). The same idea was proposed by \emph{Kennard}, who also argued (independently of \emph{Abraham}), that $\gmf$ could have the value 2. In 1924 (before he wrote the papers~\cite{R8} and~\cite{R10}), Pauli proposed that the atomic nucleus must, in general, have a non-vanishing angular momentum, which was relevant for an explanation of hyperfine splitting. (Whether his idea influenced Uhlenbeck and Goudsmit, or not, is unclear but rather unlikely.) Independently of (and priorly to) Uhlenbeck and Goudsmit, \emph{Kronig} and \emph{Urey} anticipated their idea, and \emph{Bose} had the idea that photons carry an intrinsic  `spin' (or helicity, as we would call it nowadays). 

Almost all these ideas were somewhat flawed or incomplete. For example, we understand -- since Heisenberg's proposal of a model of ferromagnetism -- that the \emph{Pauli principle} plays as important a r\^ole in explaining ferromagnetism as electron spin.\newline

\indent Thus, let me briefly recall the history of the discovery of \emph{Pauli's exclusion principle}\footnote{a name introduced by \emph{Dirac} in 1925.}. This discovery was made on the basis of Bohr's work on the periodic table of elements, in particular his \emph{`permanence principle'} (electrons in the shell of an ion keep their quantum numbers when further electrons are added), and of an important paper by \emph{Edmund Stoner}~\cite{R11}. Stoner classified electron  configurations corresponding to given values of the quantum numbers $L$ and $J$ and found, for alkali atoms, that the \emph{total} number of electrons in such a configuration is identical to the number of terms in the Zeeman spectrum of these atoms, namely $2(2L+1)$, for every $L<n$ (=principal quantum number). Pauli accidentally came across Stoner's paper. Considering alkali spectra, Pauli notices that ``the number of states in a magnetic field for given values of $L$ and $J$ is $2J+1$, the number of states for both doublets together, with $L$ fixed, is $2(2L+1)$''. Using Bohr's permanence principle, he extends his counting of states to more complicated atoms and to \emph{all} electrons in the hull of an atom. He concludes that ``every electron in an atom can be characterized by a principal quantum number $n$ and three additional quantum numbers $(L, J,m_J)$'', (with $J=L\pm \frac{1}{2}$). He notices that, for $L=0$, there are four possible states for two electrons with different principal quantum numbers, but only one when their principal quantum numbers agree. He then goes on to explain Stoner's and his observations by postulating that each state characterized by quantum numbers $(n,L,J,m_J)$ can be occupied by at most \emph{one} electron. (Pauli had actually defined $L,J=L\pm\frac{1}{2}$, and $m_J=J,J-1,\ldots,-J$ for \emph{single} electrons.) This is the \emph{exclusion principle}. Pauli concludes his paper with the sentence:
\begin{quote}
``The problem of a more coherent justification of the general rules concerning equivalent electrons in an atom here proposed can probably only be attacked successfully after a further deepening of the fundamental principles of quantum theory.''
\end{quote}

Further deepening of the fundamental principles of quantum theory was to come forward, just a few months later, starting with the work of \emph{Heisenberg}~\cite{R12}, followed by a paper by \emph{Max Born} and \emph{Pascual Jordan}~\cite{R13}, the ``Drei-M\"anner-Arbeit''~\cite{R14}, Dirac's first contributions to the new matrix mechanics~\cite{R15} (published before he earned his PhD degree under \emph{Fowler} in 1926), and, finally, by \emph{Schr\"odinger's} work on wave mechanics, in 1926; see~\cite{R16}. When Heisenberg started to do his fundamental work resulting in the paper~\cite{R12}, his friend Pauli was momentarily fed up with quantum theory and worked on Kaluza-Klein theory.\newline

\indent The quantum mechanics of angular momentum, including half-integer angular momentum, was fully developed in~\cite{R14}. Pauli's exclusion principle was reformulated, quantum mechanically, as saying that many-electron states (wave functions) must be \emph{totally anti-symmetric} under permutations of the positions and spins of individual electrons. An early contribution in this direction was in a paper by Heisenberg, the general formulation is due to Dirac (1926) and, in its definitive version, to \emph{Eugene Wigner} (1928), who profitted from his friend's, \emph{John von Neumann}, knowledge of the permutation groups and their representations. The first applications to statistical mechanics were made by \emph{Jordan}\footnote{Jordan was apparently first in discovering Fermi-Dirac statistics. But the editor of `Zeitschrift f\"ur Physik', Max Born, forgot to send Jordan's paper to the publisher during his stay in America. I thank N. Straumann for communicating this to me.}, \emph{Fermi} and Dirac, in 1926, (Fermi-Dirac statistics).

\emph{Bose-Einstein statistics} (for particles with integer spin) was introduced by \emph{Bose} (for photons) and \emph{Einstein} (for ideal monatomic quantum gases) in 1924. Its quantum-mechanical reformulation says that wave functions of many \emph{identical} bosons must be totally symmetric under permutations of these particles. Einstein predicted \emph{Bose-Einstein condensation} for non-relativistic Bose gases (and used a \emph{wave picture} for the atoms in the gas) in 1924.

It should be added that the spin and the value $\gmf=2$ of the gyromagnetic factor of the electron, as well as the fine structure of the hydrogen spectrum that led to the discovery of the Thomas precession, all found a natural explanation when \emph{Dirac} discovered his relativistic electron equation named after him, in 1927; see~\cite{R17}. We will briefly return to this equation, later.

I will now leave the history of the discoveries of spin and quantum statistics and proceed to sketching  some highlights, mathematical and physical ones, that emerged from these discoveries, \emph{not} attempting to provide a historical perspective and jumping over many important developments. I try to provide a glimpse at the usefulness of Mathematics in formulating and understanding the laws of Physics.


\section[Some of the Mathematics of Spin and a Theorem of Weyl]{Some of the Mathematics of Spin and a Theorem of Weyl\footnote{Sources for the material in this section are~\cite{R6,R18,R19,R20,R21,R22,R23}.}}

The model of space and time underlying non-relativistic quantum mechanics is inherited from \emph{Newtonian mechanics:} Physical space is homogeneous and isotropic, and an appropriate model is three-dimensional Euclidian space $\mathbb{E}^3$. Time is modelled by the real line, with the standard order relation and metric. Space-time $\mathcal{N}$ is given by $\mathbb{E}^3\times\mathbb{R}$. Events are identified with points in $\mathcal{N}$. The time difference between two events and the spatial distance between them are invariants. Dynamical symmetries of autonomous physical systems are described by the group of Euclidian motions, boosts and time translations, the so-called \emph{Galilei group}.

The model of space-time underlying special-relativistic quantum theory (gravity neglected) is usually taken to be the one proposed by \emph{Poincar\'e} and \emph{Minkowski}. Space-time is denoted by $\mathcal{N}\simeq\mathbb{R}^4$, events are labeled by points in $\mathcal{N}$, and the only invariant for a pair of events labeled by the points $(t,\vec{x})$ and $(t',\vec{x}')$ is given by
\begin{equation*}
c^2(t-t')^2-|\vec{x}-\vec{x}'|^2\,,
\end{equation*}
\noindent where $c$ is the speed of light.
If this quantity is positive then $\mathrm{sign}(t-t')$ is an invariant, too. Symmetries of autonomous physical systems are described by the Poincar\'e transformations of $\mathcal{N}$, which form the \emph{Poincar\'e group}.

The Galilei group is recovered from the Poincar\'e group by `group contraction', as the `deformation parameter' $1/c$ tends to 0.
As long as recoil on the gravitational field is neglected and this field is treated as an \emph{external field}, there are many good models of Lorentzian space-times that can serve as receptacles for a quantum theory. But a good model of space-time underlying a quantum theory of matter and gravitation is not known, yet!\newline

What has all this got to do with spin? Both the Galilei and the Poincar\'e group in $d=n+1$ dimensions (with $n=3$, in nature) contain the group $\mathrm{SO}(n)$ of spatial rotations as a subgroup: Generally speaking, if physical space is isotropic spatial rotations are dynamical symmetries of autonomous non-relativistic and special relativistic quantum-mechanical systems, and we must ask how these symmetries are represented on the space of states of such a system, and what this has got to do with spin.\\

\indent Let $G$ be any group of symmetries of a quantum-mechanical system with a Hilbert space $\mathscr{H}$ of pure state vectors. \emph{Eugene Wigner} has shown that symmetry transformations labeled by elements of $G$ are represented as unitary or anti-unitary operators acting on $\mathscr{H}$, and that these operators must define a \emph{projective representation} of $G$ on $\mathscr{H}$, (because the phase of a vector in $\mathscr{H}$ is not observable; the space of \emph{pure states} being given by projective space over $\mathscr{H}$). \emph{Valentin Bargmann} has shown that if $G$ is a \emph{connected}, \emph{compact} Lie group then all projective representations of $G$ are given by unitary representations of the \emph{universal covering group} $\tilde{G}$ associated with $G$.\\

\indent If $G=\mathrm{SO}(n), n=2,3,4,\ldots,$ (the rotation group in $n$ dimensions), then
\begin{equation*}
\widetilde{G} = 
\begin{cases}
\,\R\,,\qquad&n=2\\
\,\mathrm{SU}(2)\,,&n=3\\
\,\mathrm{Spin}(n)\,,&n\;\mathrm{general}\,.\\
\end{cases}
\end{equation*}
\indent The \emph{spin} of a quantum-mechanical particle is viewed as its intrinsic angular momentum and is thus described in terms of the generators of rotations in an irreducible, unitary representation of the quantum-mechanical rotation group $\Spinn$, where $n$ is the dimension of physical space. For $n=2$, these representations are given by the characters of the group $\R$, i.e., labeled by a \emph{real number} $s$, called the `spin of the representation'. For $n=3$, the representation theory of (the Lie algebra of) $\mathrm{Spin}(3)=\mathrm{SU}(2)$ has been worked out in~\cite{R14} and is taught in every course on introductory quantum mechanics. Irreducible representations are labeled by their `spin' $s=0,\,\frac{1}{2},\,1,\,\frac{3}{2},\ldots$. For general $n$, we refer, e.g., to~\cite{R23}. We do not have to go into this theory in any detail. We just note that, for $n\geq3$, $\Spinn$ is a two-fold cover of $\mathrm{SO}(n)$ and that, as a consequence, there are irreducible representations of $\Spinn$ that are \emph{single-valued representations} of $\mathrm{SO}(n)$ (rotations through an angle $2\pi=$identity) labeled by `$\sigma=1$', and representations of $\Spinn$ that are `\emph{double-valued representations}' of $\mathrm{SO}(n)$ (rotations through an angle $2\pi=-$identity) labeled by `$\sigma=-1$'.\\

For an understanding of differential-geometric aspects of `spin' it is useful to consider the quantum mechanics of a single non-relativistic particle with spin moving in a physical space described by a rather general $n$-dimensional manifold. Of course we are mainly interested in the examples $n=2$ (planar physics) and $n=3$; but, for purposes of applications in mathematics, it pays to be a little general, here. We are interested in formulating non-relativistic quantum mechanics on a space-time $\mathcal{N}$ of the form
\begin{equation*}
\mathcal{N}=\mathcal{M}\times\mathbb{R}\,,
\end{equation*}
where physical space, $\mathcal{M}$, is a general smooth, orientable $\mathrm{spin}^{\mathbb{C}}$ manifold, equipped with a Riemannian metric $g$, and $\mathbb{R}$ denotes time. Our goal is to derive Pauli's wave equation for a non-relativistic electron with spin moving in $\mathcal{M}$ under the influence of an external electromagnetic field and to also consider the quantum mechanics of positronium (a bound electron-positron pair). For the standard choice $\mathcal{M}=\mathbb{E}^3$ of direct interest in physics, Pauli's wave equation was discovered in~\cite{R18}.

\subsection{Clifford algebras and spin groups}\label{sect3.1}

Let $\mathcal{F}_k$ be the unital ${}^*$algebra generated by elements $b^{1},\ldots,b^{k}$ and their adjoints $b^{1\ast},\ldots,b^{k\ast}$ satisfying the canonical anti-commutation relations (CAR)
\begin{equation}
\big\lbrace b^i,b^j\big\rbrace=\big\lbrace b^{i\ast},b^{j\ast}\big\rbrace=0\,,\hspace{1cm}\big\lbrace b^{i},b^{j\ast}\big\rbrace=\delta^{ij}\,,
\label{eq3.1}
\end{equation}
where $\big\lbrace A,B \big\rbrace\deq AB+BA$. The algebra $\mathcal{F}_k$ has a unique (up to unitary equivalence) irreducible unitary representation on the Hilbert space $S\deq\mathbb{C}^{2^k}$ given by
\begin{align}
b^j &=\tau_3\otimes\cdots\otimes\tau_3\otimes\tau_{-}\otimes\umat_{2}\otimes\cdots\otimes\umat_2\,,\nonumber\\
\label{eq3.2}\\
b^{j\ast} &=\tau_3\otimes\cdots\otimes\tau_3\otimes\tau_{+}\otimes\umat_{2}\otimes\cdots\otimes\umat_2\,,\nonumber
\end{align}
with $\tau_{\pm}\deq\frac{1}{2}(\tau_1\pm \mathrm{i}\tau_2)$ in the $j^{\textrm{th}}$ factor; $\tau_1\,,\tau_2\,\mbox{and }\tau_3$ are the usual $2\times 2$ Pauli matrices. The representation~\eqref{eq3.2} is faithful, and hence $\mathcal{F}_k\simeq\ M\left(2^k,\mathbb{C}\right)$, the algebra of $2^k\times2^k$ matrices over the complex numbers.\newline

Let $V$ be a real, oriented, $n$-dimensional vector space with scalar product $\scalar{\cdot}{\cdot}\,$. The complexified Clifford algebra $\mathrm{Cl}(V)$ is the algebra generated by vectors $c(v),c(w)$, linear in $v,\,w$, with $v$ and $w$ in $V\otimes\mathbb{C}$, subject to the relations
\begin{equation}\label{eq3.3}
\big\lbrace c(v),c(w)\big\rbrace=-2\scalar{v}{w}\,.
\end{equation}
If $e^1,\ldots,e^n$ is an orthonormal basis of $V$, $n=\mathrm{dim}\,V$, then~\eqref{eq3.3} implies that
\begin{equation*}
\big\lbrace c(e^i),c(e^j)\big\rbrace=-2\delta^{ij}\,.
\end{equation*}
A ${}\adj$operation is defined by 
\begin{equation}
c(v)\adj=-c(\bar{v})\,,
\end{equation}
$v\in V\otimes\C$. Let $n=2k+p$, where $p=0\mbox{ or }1$ is the parity of $n$. Setting
\begin{align}
c(e^{2j-1})&\deq b^j-b^{j\ast}\,,\nonumber\\
&\label{eq3.5}\\
c(e^{2j})&\deq \mathrm{i}\left( b^j+b^{j\ast}\right)\,,\nonumber
\end{align}
$j=1,\ldots,k$, and, for $p=1$,
\begin{equation}
c(e^n)\deq\pm \mathrm{i}^{k+1}c(e^1)\cdots c(e^{2k})\,,
\label{eq3.6}
\end{equation}
where $b^{1\#},\ldots,b^{k\#}$ act on $S$ and generate $\mathcal{F}_k$, we find that $c(e^1),\ldots,c(e^n)$ define a representation of $\mathrm{Cl}(V)$ on $S$. Eqs.~\eqref{eq3.5},~\eqref{eq3.6} define the \emph{unique}, up to a sign related to space reflection, \emph{irreducible unitary} representation of $\mathrm{Cl}(V)$, which is faithful. Hence
\begin{equation}
\mathrm{Cl}(V)\simeq M\left( 2^k,\C\right)\,.
\end{equation}
A scalar product on $\mathrm{Cl}(V)$ extending the one on $V$ is defined by
\begin{equation}
\scalar{a}{b}\deq 2^{-k} \mathrm{tr}(a\adj b)\,,
\label{eq3.8}
\end{equation}
$a,b\in \mathrm{Cl}(V)$.\\

The \emph{spin group} $\mathrm{Spin}(V)$ is defined by
\begin{equation}
\mathrm{Spin}(V)\deq\big\lbrace a\in \mathrm{Cl}^{\textrm{even}}_{\R}(V)\,\big|\,aa\adj=a\adj a=\umat, a c(V)a\adj \subseteq c(V)\big\rbrace\,,
\label{eq3.9}
\end{equation}
where $\mathrm{Cl}^{\textrm{even}}_{\R}(V)$ denotes the real subalgebra of $\mathrm{Cl}(V)$ generated by products of an even number of elements of the form $c(v),\,v\in V$. We also set  $\mathrm{Spin}(n)=\mathrm{Spin}(\mathbb{E}^n)$. The group $\mathrm{Spin}^{\C}(V)$ is defined by 
\begin{equation}
\mathrm{Spin}^{\C}(V)\deq\big\lbrace e^{\mathrm{i}\alpha}a\,\big|\,\alpha\in\R,\,a\in\mathrm{Spin}(V)\big\rbrace\,.
\label{eq3.10}
\end{equation}
For each $a\in\mathrm{Spin}^{\C}(V)$, we define a linear transformation $\mathrm{Ad}(a)$ of $V$ by
\begin{equation}
c\big(\mathrm{Ad}(a)v\big)\deq ac(v)a\adj\,,\;v\in V\,.
\label{eq3.11}
\end{equation}
Clearly, this linear transformation preserves the scalar product on $V$, and we have the short exact sequence
\begin{equation*}
1\longrightarrow\mathrm{U}(1)\longrightarrow\mathrm{Spin}^{\C}(V)\overset{\mathrm{Ad}}\longrightarrow \mathrm{SO}(V)\longrightarrow 1\,.
\end{equation*}
The Lie algebra $\mathrm{spin}^{\C}(V)$ of $\mathrm{Spin}^{\C}(V)$ is given by
\begin{equation}
\mathrm{spin}^{\C}(V)=\mathrm{spin}(V)\oplus\R\,,
\end{equation}
where
\begin{equation}
\mathrm{spin}(V)=\big\lbrace \xi\in \mathrm{Cl}_{\R}^{\textrm{even}}(V)\big|\,\xi+\xi\adj=0\,,\,\big\lbrack\xi,c(V)\big\rbrack\subseteq c(V)\big\rbrace\,.
\end{equation}
One then finds that
\begin{equation}
\mathrm{spin}(V)=\Big\lbrace \sum_{i,j}\,x_{ij}c(e^i)c(e^j)\,\Big|\,x_{ij}=-x_{ji}\in\R\Big\rbrace\simeq \mathrm{so}(V)\,.
\label{eq3.14}
\end{equation}
Given $V$, let $ \exalg(V\otimes\C)$ denote the exterior algebra over $V\otimes\C$. There is a canonical scalar product on $ \exalg(V\otimes\C)$ extending the one on $V\otimes\C=\bigwedge^{1}(V\otimes\C)$. For $v\in V\otimes\C$, we define operators $a\adj(v)$ and $a(v)$ on $ \exalg(V\otimes\C)$ by setting
\begin{align}
a\adj(v)\,w&\deq v\wedge w\,,\\
a(v)\,w&\deq\imath(G\bar{v})\,w\,,
\end{align}
where $G$ is the metric on $V$ defining the scalar product on $V$, so that $Gv$ is in the dual space of $V$, and $\imath$ denotes interior multiplication. Then $a(v)=(a\adj(v))\adj$, and the operators $a\adj(v),\,a(v),\,v\in V\otimes\C$, are the usual fermionic  creation- and annihilation operators satisfying the CAR, with
\begin{equation}
 \exalg(V\otimes\C)\simeq\textrm{fermionic Fock space over }V\otimes\C\,.
\label{eq3.17}
\end{equation}
The operators
\begin{equation}
\Gamma(v)\deq a\adj(v)-a(v)\,,\hspace{1cm}\overline{\Gamma}(v)\deq \mathrm{i}\big (a\adj(v)+a(v)\big)\,,
\label{eq3.18}
\end{equation}
then define two anti-commuting unitary representations of $\mathrm{Cl}(V)$ on $ \exalg(V\otimes\C)$.\\
	
Let $\mathrm{dim}\,V=2k\;(p=0)$ be even. We set
\begin{equation*}
\gamma=\mathrm{i}^k\Gamma(e^1)\cdots\Gamma(e^n)\,,
\end{equation*}
which \emph{anti-commutes} with all $\Gamma(v)$, and satisfies $\gamma^2=\umat$. Let $S\simeq\C^{2^k}\simeq\overline{S}$. We then have that
\begin{equation*}
 \exalg(V\otimes\C)\simeq S\otimes\overline{S}\,,
\end{equation*}
with
\begin{align}
\Gamma(v)&\simeq c(v)\otimes\umat\,,\\
\overline{\Gamma}(v)&\simeq\gamma\otimes\bar{c}(v)\,,
\end{align}
where $c$ and $\bar c$ denote the irreducible representations of $\mathrm{Cl}(V)$ on $S$ and $\overline{S}$, respectively.\newline

If $\mathrm{dim}\,V=2k+1$ is odd then 
\begin{equation*}
\gamma=\mathrm{i}^{k+1}\Gamma(e^1)\cdots\Gamma(e^n)
\end{equation*}
\emph{commutes} with all $\Gamma(v)$, and satisfies $\gamma^2=\umat$. The operator $\gamma$ has two eigenvalues, $\pm 1$, both with multiplicity $2^{n-1}$. It follows that
\begin{equation*}
 \exalg(V\otimes\C)\simeq S\otimes\C^2\otimes\overline{S}\,,
\end{equation*}
and
\begin{align}
\Gamma(v)&=c(v)\otimes\tau_3\otimes\umat\,,\\
\overline{\Gamma}(v)&=\umat\otimes\tau_1\otimes\bar{c}(v)\,.
\end{align}

\subsection{Pauli's wave equation for an `electron' and for `positronium' in a general differential-geometric formulation -- susy QM}\label{section3.2}

We are ready, now, to formulate \emph{Pauli's wave equation for spinning particles}~\cite{R18} on a space-time $\mathcal{N}=\mathcal M\times\R$, where $\mathcal M$ is a general, $n$-dimensional smooth (compact) $\mathrm{spin}^{\C}$ manifold, e.g., $\mathcal M=\mathbb{E}^n$, $n=2,3$. Let $g=(g_{ij})$ be a \emph{Riemannian metric} on the tangent bundle $\mathrm{T}\mathcal M$ of $\mathcal M$, and let $G=(g^{ij})$ denote the corresponding inverse metric on the cotangent bundle $\mathrm{T}\adj M$. Let $ \exalg\mathcal M$ be the bundle of differential forms on $\mathcal M$, with $ \exalgOmega(\mathcal M)$ the space of \emph{complexified sections} of $ \exalg\mathcal M$. This space is equipped with a natural scalar product $\scalar{\cdot}{\cdot}$, determined by $g$ and by the Riemannian volume form. Let $\mathrm{Cl}(\mathcal{M})$ be the Clifford bundle over $\mathcal M$; its base space is $\mathcal M$ and its fibres are given by $\mathrm{Cl}(\mathrm{T}_x\adj \mathcal M)\simeq \mathrm{Cl}(\mathbb{E}^n)$, with $n=\mathrm{dim}\,\mathcal M$. Let $\mathcal A=C^{\infty}(\mathcal M)$ be the algebra of smooth functions on $\mathcal M$. The space of sections, $\Gamma(E)$, of a vector bundle $E$ over $\mathcal M$ is a finitely  generated, projective module for  $\mathcal A$; $E$ is trivial iff $\Gamma(E)$ is a free $\mathcal A$-module. Our standard  examples for $E$ are
\begin{equation*}
E=\mathrm{T}\mathcal M,\,\mathrm{T}\adj \mathcal M,\, \exalg\mathcal M,\,\mathrm{Cl}(\mathcal M)\,.
\end{equation*}
\noindent The Clifford bundle over $\M$ has two anti-commuting unitary representations, $\Gamma$ and $\overline{\Gamma}$, on the module $\exalgOmega(\M)$, which we define as follows: Given a (complex) 1-form $\omega\in\Omega^{1}(\M)$, we introduce \emph{creation-} and \emph{annihilation operators}, $a\adj(\omega)$ and $a(\omega)$, on $\exalgOmega(\M)$,
\begin{equation}\label{eq3.23}
a\adj(\omega)\sigma\deq\omega\wedge\sigma\,,\hspace{1cm}a(\omega)\sigma\deq\imath(G\omega)\,\sigma\,,
\end{equation}
for $\sigma\in\exalgOmega(\M)$. Then (with $a^{\#}=a$ or $a\adj$)
\begin{equation}
\big\lbrace a^{\#}(\omega_1), a^{\#}(\omega_2)\big\rbrace=0\,,\hspace{1cm}\big\lbrace a(\omega_1), a\adj(\omega_2)\big\rbrace=\scalarr{\omega_1}{\omega_2}\,,
\label{eq3.24}
\end{equation}
for $\omega_1,\,\omega_2\in\Omega^1(\M)$, where $\scalarr{\cdot}{\cdot}$ is the hermitian structure on $\exalg\M$ determined by $G$. We define two anti-commuting representations $\Gamma$ and $\overline{\Gamma}$ of $\mathrm{Cl}(\M)$ on $\exalgOmega(\M)$ by setting
\begin{equation}
\Gamma(\omega)\deq a\adj(\omega)-a(\omega)\,,\hspace{1cm}\overline{\Gamma}(\omega)\deq \mathrm{i}\big (a\adj(\omega)+a(\omega)\big)\,.
\label{eq3.25}
\end{equation}
If the manifold $\M$ is $\mathrm{spin}^\C$ (which we have assumed) then
\begin{equation}
\exalgOmega(\M)=\Gamma(S)\otimes_{{\mathcal A}}\big ( \C^2\otimes\big)\,\Gamma(\overline{S})\,,
\label{eq3.26}
\end{equation}
where $S\equiv S(\M)$ is the \emph{spinor bundle} and $\overline{S}$  the (charge-) conjugate spinor bundle over $\M$. The factor $\C^2$ on the R.S. of~\eqref{eq3.26} only appears if $n=\mathrm{dim}\,\M$ is odd. The modules $\Gamma(S)$ and $\Gamma(\overline{S})$ carry unitary representations $c$ and $\bar{c}$, respectively, of $\mathrm{Cl}(\M)$ with
\begin{align}
\Gamma(\omega)&= c(\omega)  \otimes\big(\tau_3 \otimes\big)\,\umat\,,\label{eq3.27}\\
\overline{\Gamma}(\omega)&= \gamma\otimes\big(\tau_1\otimes\big)\,\bar{c}(\omega)\,,
\label{eq3.28}
\end{align}
with $\gamma=\umat$ if $n$ is odd; see Sect.~\ref{sect3.1}. (Over a coordinate chart of $\M$, eqs~\eqref{eq3.26}~-~\eqref{eq3.28} always make sense, by the results of Sect.~\ref{sect3.1}. But, \emph{globally}, they \emph{only} make sense if $\M$ is $\mathrm{spin}^\C$!)\newline

Let $\nabla$ be the \emph{Levi-Civita connection} on $ \exalg\M$  (unitary with respect to $g$ and torsion-free). A connection $\nabla^S$ on $S$ is called a $\mathrm{spin}^{\C}$ connection iff it satisfies the `\emph{Leibniz rule}'
\begin{equation}
\nabla_X^S\big(c(\xi)\psi\big)=c(\nabla_X\xi)\psi+c(\xi)\nabla_X^S\psi\,,
\end{equation}
where $X$ is a vector field on $\M$, $\xi$ a 1-form and $\psi$ a \emph{spinor} in $\Gamma(S)$, i.e., a section of $S$.\newline
\indent If $\nabla_1^S$ and $\nabla_2^S$ are two hermitian $\mathrm{spin}^{\C}$ connections on $S$ then 
\begin{equation}
\left(\nabla_1^S-\nabla_2^S\right)\psi=\mathrm{i}\alpha\otimes\psi\,,
\end{equation}
for $\psi\in\Gamma(S)$, where $\alpha$ is a real, globally defined 1-form. Physically, $\alpha$ is the difference of two \emph{electromagnetic vector potentials}, $A_1$ and $A_2$, so-called `\emph{virtual $\mathrm{U}(1)$-connections}' on $S$; ($A_i,\,i=1,2$, is `one half times a $\mathrm{U}(1)$-connection' on a line bundle, canonically associated with $S\otimes S$, with magnetic monopoles inside non-contractible 2-spheres in the homology of $\M$).\newline

Given a $\mathrm{spin}^\C$ connection $\nabla^S$ corresponding to a virtual $\mathrm{U}(1)$-connection $A$, the \emph{Pauli (-Dirac) operator} $D_A$ associated with $\nabla^S$ on $S$ is defined by
\begin{equation}
D_A\deq c\circ\nabla^S\,,
\label{eq3.271}
\end{equation}
which is a linear operator on $\Gamma(S)$. Locally, in a coordinate chart of $\M$, with coordinates $x^1,\ldots,x^n$,
\begin{equation}
D_A=\sum_{i=1}^n\,c(\textrm{d}x^i)\,\nabla_i^S\,,
\end{equation}
with
\begin{equation*}
\big\lbrace c(\textrm{d}x^i),c(\textrm{d}x^j)\big\rbrace=g^{ij}(x)\,.
\end{equation*}
To every $\nabla^S$ there corresponds a unique conjugate connection $\overline{\nabla}^S$ on $\overline{S}$, obtained by reversing the electric charge, i.e., $A\rightarrow -A$, and we define
\begin{equation}
\adjOp{D}_{-A}\deq\bar{c}\circ\overline{\nabla}^S\,,
\end{equation}
an operator acting on $\Gamma(\overline{S})$.\newline

The bundles $S$ and $\overline{S}$ are equipped with a  natural hermitian structure. Let $\mathrm{dvol}_g$ denote the Riemannian volume form on $\M$. By $\mathscr{H}_{\textrm{e}}$ we denote the Hilbert-space completion of $\Gamma(S)$ in the scalar product on $\Gamma(S)$ determined by the hermitian structure of $S$ and $\mathrm{dvol}_g$; $\mathscr{H}_{\textrm{p}}$ is defined similarly, with $S$ replaced by $\overline{S}$.\\
\indent We note, in passing, that the closures of $D_A,\,\adjOp{D}_{-A}$ are selfadjoint, elliptic  operators densely defined on $\mathscr{H}_{\textrm{e}},\,\mathscr{H}_{\textrm{p}}$, respectively.\\
\indent Thus, $\M$ equipped with a Riemannian metric $g$, gives rise to what \emph{Alain Connes}~\cite{R22} calls \emph{spectral triples}
\begin{equation}
(\mathcal{A},\,D_A,\,\mathscr{H}_{\textrm{e}})\,,\,(\mathcal{A},\,\adjOp{D}_{-A},\,\mathscr{H}_{\textrm{p}})\,,
\end{equation}
which, in turn, determine $(\M,\,g)$ uniquely. In the special case where $\M=\mathbb{E}^3$, these spectral  triples are familiar to anyone who knows \emph{Pauli's non-relativistic quantum theory of the spinning electron} and its twin, the positron: $\mathcal{A}$ is the algebra of position measurements; $\mathscr{H}_{\textrm{e}}\;\left(\mathscr{H}_{\textrm{p}}\right)$ is the Hilbert space of pure state vectors of a single electron (positron); and $D_A\;\left(\adjOp{D}_{-A}\right)$ is the `square-root' of the \emph{Hamiltonian} generating the unitary time evolution of states of an electron (positron) moving in $\M$ and coupled to an external magnetic field $B=\textrm{d}A$. More precisely, the Hamiltonian is given by
\be
H_A=\frac{{\hbar}^2}{2m}D_{A}^{2}\,,
\label{eq3.31}
\ee
where $m$ is the mass of an electron, $\hbar$ is Planck's constant, and the gyromagnetic factor $g=\gmf=2$. (If $\gmf$ were \emph{different} from 2 then $H_A$ would \emph{not} be the square of $D_A$; there would then appear an additional \emph{Zeeman term} on the R.S. of~\eqref{eq3.31}, as Pauli had introduced it in~\cite{R18}. This term is proportional to $B_{ij}\,c(\textrm{d}x^i)c(\textrm{d}x^j)$, in local coordinates, where $B$ is the field strength corresponding to $A$.) In the presence of an electrostatic potential $\Phi$, the Hamiltonian of Pauli's non-relativistic electron is given by
\be
H_{(\Phi,A)}\deq H_A+\Phi\,,
\ee
and Pauli's version of the time-dependent Schr\"odinger equation reads
\be
\mathrm{i}\hbar\frac{\partial}{\partial t}\psi_t=H_{(\Phi,A)}\,\psi_t\,,
\ee
for  $\psi_t\in\mathscr{H}_{\textrm{e}}$. The corresponding equation for the non-relativistic positron is
\be
\mathrm{i}\hbar\frac{\partial}{\partial t}\chi_t=\Big(\frac{{\hbar}^2}{2m}\adjOp{D}_{-A}^{\;2}-\Phi\Big)\,\chi_t\,,
\ee
for $\chi_t\in\mathscr{H}_{\textrm{p}}$.\newline
	
We observe that when the electrostatic potential $\Phi$ vanishes $H_{(0,A)}=H_A$ is the square of a selfadjoint operator (a `\emph{super charge}')
\begin{equation*}
Q\deq\sqrt{\frac{\hbar^2}{2m}}\,D_A\,.
\end{equation*}
Let the dimension of $\M$ be even, and let $\lbrace\epsilon_1,\ldots,\epsilon_n\rbrace$ be a local, orthonormal basis of $\Omega^1(\M)$; $\big(\lbrace\epsilon_1,\ldots,\epsilon_n\rbrace$ is called an `$n$-bein'$\big)$. We set
\begin{equation*}
\gamma\deq\mathrm{i}^{\frac{n}{2}}c(\epsilon_1)\cdots c(\epsilon_n)\,.
\end{equation*}
Since $\M$ is orientable, $\gamma$ extends to a globally defined involution of $\mathrm{Cl}(\M)$ anti-commuting with $c(\omega),\,\omega\in\Omega^1(\M)$, and hence with $Q$. Then $\left(\gamma,\,Q,\,\mathscr{H}_{\textrm{e}}\right)$ furnishes an example of \emph{supersymmetric quantum mechanics}, with $N=1$ (or (1,0)) supersymmetry. The `super trace'
\be
\mathrm{tr}_{\mathscr{H}_{\textrm{e}}}\left(\gamma\,\mathrm{e}^{-\beta Q^2}\right)\,,\hspace{0.3cm}\beta>0\,,
\ee
is easily seen to be independent of $\beta$ and invariant under small deformations of the metric $g$ and the vector potential $A$. It computes the \emph{index} of the `\emph{Dirac operator}' $D_A$, which is a topological invariant of $\M$.\newline

Next, we study the quantum theory of \emph{positronium}, namely of a bound state of an electron and a positron. We define $\mathscr{H}_{\textrm{e}-\textrm{p}}$ to be the Hilbert space completion of the space $\exalgOmega(\M)$ of differential forms in the scalar product determined by the metric $g$. Then
\begin{equation}
\mathscr{H}_{\textrm{e}-\textrm{p}}\simeq\mathscr{H}_{\textrm{e}}\otimes_{\mathcal{A}}\big(\C^2\otimes\big)\mathscr{H}_{\textrm{p}}\,,
\end{equation}
where the factor $\C^2$ is absent if $\mathrm{dim}\,\M$ is even. We introduce two anti-commuting Pauli (-Dirac) operators $\mathcal{D}$ and $\adjOp{\mathcal D}$ (densely defined and selfadjoint on $\mathscr{H}_{\textrm{e}-\textrm{p}}$):
\be
\mathcal{D}\deq\Gamma\circ\nabla\,,\hspace{1cm}\adjOp{\mathcal D}\deq\overline{\Gamma}\circ\nabla\,,
\ee
where $\nabla$ is the Levi-Civita connection on $\exalgOmega(\M)$, and $\Gamma,\,\overline{\Gamma}$ are the two anti-commuting representations of $\mathrm{Cl}(\M)$ on $\exalgOmega(\M)$ introduced in~\eqref{eq3.23}~-~\eqref{eq3.25}. These operators are easily seen to  satisfy
\be
\big\lbrace\mathcal{D},\adjOp{\mathcal D}\big\rbrace=0\,,\hspace{1cm}{\mathcal{D}\!\!\!\!\!\phantom{\adjOp{\mathcal D}}}^2={\adjOp{\mathcal D}}^2\,.
\ee
Setting
\be
\mathrm{d}\deq\frac{1}{2}\left( \mathcal D-\mathrm{i}\,\adjOp{\mathcal D}\right)\,,\hspace{1cm}\mathrm{d}\adj\deq\frac{1}{2}\left( \mathcal D+\mathrm{i}\,\adjOp{\mathcal D}\right)\,,
\ee
we find that $\mathrm{d}^2={\left(\mathrm{d}\adj\right)}^2=0$. In fact, $\mathrm{d}$ turns out to be the exterior derivative. The Hamiltonian (for the center-of-mass motion of the `groundstates' of a bound electron-positron pair, i.e.,) of \emph{positronium} is given by
\be
H\deq\frac{\hbar^2}{2\mu}\mathcal{D}^2=\frac{\hbar^2}{2\mu}{\adjOp{\mathcal{D}}}^2=\frac{\hbar^2}{2\mu}\left(\mathrm{dd}\adj+\mathrm{d}\adj\mathrm{d}\right)\,,
\label{eq3.35}
\ee
where $\mu=2m$. Note that $\mathcal D,\,\adjOp{\mathcal{D}}$ and $H$ are \emph{independent} of the choice of the vector potential $A$ (and of $\Phi$) which, physically, corresponds to the circumstance that the electric charge of positronium is zero. The data $\left(\mathcal{A},\,\mathcal{D},\,\adjOp{\mathcal D},\,\mathscr{H}_{\textrm{e}-\textrm{p}}\right)$ are thus well defined even if $\M$ does \emph{not} admit a $\mathrm{spin}^\C$ structure. These data, together with~\eqref{eq3.35}, furnish an example of supersymmetric quantum mechanics with $N=(1,1)$ supersymmetry; the supercharges are the operators $\mathcal{D}$ and $\adjOp{\mathcal{D}}$. They completely encode the \emph{de Rham-Hodge theory} and the \emph{Riemannian geometry} of $\M$.\newline

One may wonder how additional geometric structure of $\M$ reveals itself in Pauli's quantum theory of a non-relativistic electron, positron or positronium moving in $\M$. Suppose, e.g., that $\M$ is a symplectic manifold equipped with a symplectic 2-form $\omega$. Let $\Omega$ denote the anti-symmetric bi-vector field associated with $\omega$. We define three operators on $\mathscr{H}_{\textrm{e}-\textrm{p}}$
\be
L_3\deq T-\frac{n}{2}\,,\hspace{0.5cm}L_+\deq\frac{1}{2}\omega\wedge(\,\cdot\,)\,,\hspace{0.5cm}L_-\deq\frac{1}{2}\imath(\Omega)\,,
\ee
where $T\lambda=p\,\lambda$, for any $p$-form $\lambda\in \exalgOmega(\M)$. Then
\be
\big\lbrack L_3,L_{\pm}\big\rbrack=\pm2L_{\pm}\,,\hspace{1cm}\big\lbrack L_+,L_-\big\rbrack=L_3\,,
\ee
i.e. $\big\lbrace L_3,\,L_+,\,L_-\big\rbrace$ define a representation of the Lie algebra $\mathrm{sl}_2$ on $\mathscr{H}_{\textrm{e}-\textrm{p}}$ commuting with the representation of the algebra $\mathcal{A}$ on $\mathscr{H}_{\textrm{e}-\textrm{p}}$. It is actually a unitary representation, because ${L_3}\adj=L_3$ and $(L_{\pm})\adj=L_{\mp}$, in the scalar product of $\mathscr{H}_{\textrm{e}-\textrm{p}}$. Since $\omega$ is closed, we have that $\big\lbrack L_+,\mathrm{d}\big\rbrack=0$, where $\mathrm{d}$ is the exterior derivative. A differential ${\tilde{\mathrm{d}}}\adj$ of degree $-1$ can be defined by
\be
{\tilde{\mathrm{d}}}\adj\deq\big\lbrack L_-,\mathrm{d}\big\rbrack\,.
\ee
One finds that $\big\lbrace{\tilde{\mathrm{d}}}\adj,\mathrm{d}\big\rbrace=0$, $({\tilde{\mathrm{d}}}\adj)^2=0$, and $\big\lbrack L_-,{\tilde{\mathrm{d}}}\adj\big\rbrack=0$. Thus $(\mathrm{d},{\tilde{\mathrm{d}}}\adj)$ transforms as a doublet under the adjoint action of $\mathrm{sl}_2$.\\
\indent One can introduce a second $\mathrm{sl}_2$ doublet, $(\tilde{\mathrm{d}},-{\mathrm{d}}\adj)$, of differentials with the same properties as $(\mathrm{d},{\tilde{\mathrm{d}}}\adj)$. We are \emph{not} claiming that $\lbrace\mathrm{d},{\tilde{\mathrm{d}}}\rbrace=0$; this equation does \emph{not} hold for general symplectic manifolds. It is natural to ask, however, what is special about the geometry of $\M$ if
\be
\big\lbrace\mathrm{d},\tilde{\mathrm{d}}\big\rbrace=0\,.
\ee
It turns out that, in this case, $\M$ is a \emph{K\"ahler manifold}. Defining
\begin{equation*}
\partial\deq\frac{1}{2}\left(\mathrm{d}-\mathrm{i}\,\tilde{\mathrm{d}}\right)\,,\hspace{1cm}\adjpartial\deq\frac{1}{2}\left (\mathrm{d}+\mathrm{i}\,\tilde{\mathrm{d}}\right)\,,
\end{equation*}
one finds that
\begin{equation*}
{\partial\mspace{-10.5mu}\phantom{\adjpartial}}^{2}=\adjpartial^2=0\,,\hspace{0.5cm}\big\lbrace\partial,\adjpartial^{\#}\big\rbrace=0\,,\hspace{0.5cm}\big\lbrace\partial,\partial\adj\big\rbrace=\big\lbrace\adjpartial,\adjpartial^*\big\rbrace\,.
\end{equation*}
The differentials $\partial$ and $\adjpartial$ are the \emph{Dolbeault differentials}. The complex structure $J$ on $\M$ generates a $\mathrm{U}(1)$-symmetry on the differentials:
\begin{equation*}
\big\lbrack\,J,\mathrm{d}\,\big\rbrack=-\mathrm{i}\,\tilde{\mathrm{d}}\,,\hspace{1cm}\big\lbrack\,J,\tilde{\mathrm{d}}\,\big\rbrack=\mathrm{i}\,\mathrm{d}\,.
\end{equation*}
$J$ commutes with the representation of the algebra $\mathcal{A}=C^{\infty}(\M)$ on $\mathscr{H}_{\textrm{e}-\textrm{p}}$.\newline
\indent The data $\left(\mathcal{A},\,\partial,\partial\adj,\,\adjpartial,\,\adjpartial\adj,\,\mathscr{H}_{\textrm{e}-\textrm{p}}\right)$ furnish an example of a supersymmetric quantum theory with $N=(2,2)$ supersymmetry. If the $\mathrm{sl}_2$-symmetry is broken, but the $\mathrm{U}(1)$-symmetry generated by $J$ is preserved then $\M$ may not be symplectic, but it is a \emph{complex-hermitian} manifold.\\

It is possible to reformulate all special geometries of smooth manifolds in terms of the supersymmetric quantum mechanics of a non-relativistic electron or of positronium by analyzing the adjoint action of symmetries  on the Pauli (-Dirac) operators $D_A,\,\adjOp{D}_{-A},\,\mathcal{D}$ and $\adjOp{\mathcal D}$. This mathematical theme is developed in~\cite{R19}. The upshot of that analysis is that the non-relativistic quantum mechanics of the spinning electron and of positronium encodes the differential geometry and topology of Riemannian manifolds $\M$ (`physical space') in a perfect manner. There is a complete dictionary between the \emph{geometry} of $\M$ and the \emph{supersymmetries} of the quantum theory.\\

What about the non-relativistic quantum mechanics of particles with `higher spin'? Let $(\M,g)$ be an $n$-dimensional, oriented, smooth, Riemannian manifold with Riemannian metric $g$ and volume form $\mathrm{dvol}_g$. Let $\rho$ be a finite-dimensional, unitary representation of $\Spinn$ on a Hilbert space $V_{\rho}$. If $\rho$ is a double-valued representation of $\mathrm{SO}(n)$, i.e., $\sigma(\rho)=-1$, then $\M$ must be assumed to be $\mathrm{spin}^\C$; for $\sigma(\rho)=1$, this assumption is not necessary. From the transition functions of the spinor bundle $S$ (or the tangent  bundle $\mathrm{T}\M$, for $\sigma(\rho)=1$) and the representation $\rho$ of $\Spinn$ we can construct a hermitian vector bundle $E_{\rho}$ over $\M$ whose fibres are all isomorphic to $V_{\rho}$. The hermitian structure on $E_{\rho}$ and $\mathrm{dvol}_g$ determine a scalar product $\scalar{\cdot}{\cdot}_{\rho}$ on the space of sections $\Gamma(E_{\rho})$. The completion of $\Gamma(E_{\rho})$ in the norm determined by the scalar product $\scalar{\cdot}{\cdot}_{\rho}$ is a Hilbert space $\mathscr{H}_{\rho}$. A $\mathrm{spin}^{\C}$ connection $\nabla^S$ on $S$ (or the Levi-Civita connection $\nabla$ on $\exalg\M$ if $\sigma(\rho)=1$) determines a connection $\nabla^{\rho}$ on $E_{\rho}$. (As a physicist, I think about these matters in coordinate charts $U$ of $\M$, with $E_{\rho}|_{U}\simeq U\times V_{\rho}$, use a little representation theory of $\Spinn$ and $\mathrm{spin}(n)$, and glue charts together using the transition functions of $S$, or $\mathrm{T}\M$, respectively, in the representation $\rho$). The connection $\nabla^{\rho}$, the hermitian structure on $E_{\rho}$ and $\mathrm{dvol}_g$ determine a \emph{Laplace-Beltrami operator} $-\Delta_{g,A}$ densely defined on $\mathscr{H}_{\rho}$, (e.g., via the Dirichlet form on $\mathscr{H}_{\rho}$ determined by $\nabla^{\rho}$).\\
\indent Pauli's non-relativistic quantum mechanics for a particle moving in physical space $\M$, with an `intrinsic angular momentum' described by the representation $\rho$ of $\Spinn$, is given in terms of the following data: The Hilbert space of pure state-vectors is given by $\mathscr{H}_{\rho}$. A real 2-form $\varphi$ on $\M$ determines a section of the subbundle $\mathrm{spin}(\M)$ of $\mathrm{Cl}(\M)$, whose fibres are all isomorphic to the Lie algebra $\mathrm{spin}(n)\simeq\mathrm{so}(n)$ of $\Spinn$; see~\eqref{eq3.14}. By $\mathrm{d}\rho$ we denote the representation of $\mathrm{spin}(n)$ on $V_{\rho}$.\newline

The \emph{Pauli Hamiltonian} is then given by
\bel{eq3.38}
H_A^{\rho}=-\frac{\hbar^2}{2m}\Delta_{g,A}+\mu_{\rho}\mathrm{d}\rho(B)+\Phi\,,
\ee
where $m$ is the mass of the particle, $\mu_{\rho}$ its `magnetic moment', $B\in\Omega^2(\M)$ the curvature (`magnetic field') of the virtual $\mathrm{U}(1)$-connection $A$ (the electromagnetic vector potential), and $\Phi$ is an external (electrostatic) potential. The second term on the R.S. of~\eqref{eq3.38} is the Zeeman term.\newline\newline

\noindent\emph{Remarks:}
\begin{itemize}
	\item[(1)] Relativistic corrections (spin-orbit interactions) and a variety of further effects can be described in terms of additive contributions to the ($\mathrm{U}(1)$- and) $\mathrm{Spin}(n)$ connection and further Zeeman terms.
	\item[(2)] In relativistic field theory on four-di\-men\-sio\-nal space-time, one encounters \emph{acausality phenomena} in the propagation of fields of spin $>1$ minimally coupled to external electromagnetic fields (`\emph{Velo-Zwanziger} phenomenon')~\cite{R23bis}. This may shed some light on the question why, in Nature, there do not appear to exist any charged elementary particles of spin $>1$. See also section 7.1. It should be noted, however, that the Velo-Zwanziger acausality phenomenon disappears in \emph{locally supersymmetric} field theories~\cite{R23ter}. (I thank N. Straumann for pointing this out to me.)
\end{itemize}
\indent Well, I suppose this is all we might want to know about these general matters, right now.\\
\indent To conclude this general, mathematical section, I want to specialize to the case where $\M=\mathbb{E}^3$, $\mathrm{Spin}(3)=\mathrm{SU}(2)$, which is what we physicists care about most.

\subsection{Back to physics: multi-electron systems, Weyl's theorem, the Dirac equation}

We first specialize the material of section~\ref{section3.2} to the case where $\M=\mathbb{E}^3$. Then $S\equiv S(\M)$ and $\exalg(\M)$ are trivial bundles, and
\be
\label{eq3.39}
\mathscr{H}_{\textrm{e / p}}\simeq L^2\left(\R^3,\,\textrm{d}^3 x\right)\otimes\C^2\,,
\ee
the space of \emph{square-integrable, two-component spinors} on $\R^3$. Choosing Cartesian coordinates $x^1,\,x^2,\,x^3$ on $\mathbb{E}^3$, the Pauli (-Dirac) operator $D_A$ takes the form
\be
\label{eq3.40}
D_A=\sum_{j=1}^3\,\sigma_j\,\left(-\mathrm{i}\frac{\partial}{\partial x^j}+\frac{e}{\hbar c}A_j(x)\right)\,,
\ee
where $\vec{\sigma}=(\sigma_1,\,\sigma_2,\,\sigma_3)$ are the usual Pauli matrices, and $\vecA(x)=(A_1(x),\,A_2(x),\,A_3(x))$ is the electromagnetic vector potential in physical units -- whence the factor $\frac{e}{\hbar c}$ multiplying $A_j(x)$ in~\eqref{eq3.40}, where $-e$ is the charge of an electron and $c$ the speed of light. The Pauli Hamiltonian $H_A$ is given by
\be
\label{eq3.41}
H_A=\frac{\hbar^2}{2m}D_A^2+\Phi\,,
\ee
where $\Phi$ is an external electrostatic potential.\\
\indent We easily find that
\bel{eq3.42}
\frac{\hbar^2}{2m}D_A^2=-\frac{\hbar^2}{2m}\Delta_A+\frac{e}{mc}\vecS\cdot\vecB\,,
\ee
where $\Delta_A$ is the covariant Laplacian, $\vecS=\frac{\hbar}{2}\vec{\sigma}$ is the spin operator of an electron, and $\vecB=\rotA$ is the magnetic field. Thus, for the `supersymmetric' Hamiltonian $H_A$, the gyromagnetic factor $\gmf$ of the electron has the value 2! As long as spin-orbit interactions can be neglected, i.e., in the absence of heavy nuclei, the Hamiltonian $H_A$ in~\eqref{eq3.41} describes the dynamics of a slow electron in an external electromagnetic field with good accuracy. Yet, one may  wonder how the relativistic effects of spin-orbit interactions and the Thomas precession modify the expression~\eqref{eq3.41} for the Pauli Hamiltonian. From~\eqref{eq2.14} and~\eqref{eq2.17} we find that $H_A$ must then be replaced by
\bel{eq3.43}
H_A^{\mathrm{SO}}=-\frac{\hbar^2}{2m}\Delta_A^2+\frac{e}{mc}\vecS\cdot\left(\vecB-\frac{1}{2}\,\frac{\vec{v}}{c}\wedge\vecE\right)+\Phi\,,
\ee
where the (gauge-invariant) \emph{velocity operator} $\vec{v}$ is given by
\bel{eq3.44}
\vec{v}=\frac{\hbar}{m}\left(\,-\mathrm{i}\grad+\frac{e}{\hbar c}\vecA\,\right)\,,
\ee
and $-\frac{\hbar^2}{2m}\Delta_A=\frac{m}{2}\vec{v}^{\,2}$. We introduce a \emph{spin} ($\mathrm{SU}(2)$-) \emph{connection} $w=(w_0,\,\vec{w})$ on $S(\mathbb{E}^3)$ in terms of its components in the `natural orthonormal basis' of sections of $S(\mathbb{E}^3)$:
\begin{align}
w_0(x)&=\mathrm{i}\frac{e}{mc\hbar}\,\vecB(x)\cdot\vecS\,,\label{eq3.45}\\
\vec{w}(x)&=-\mathrm{i}\frac{e}{2mc\hbar}\,\vecE(x)\wedge\vecS\,.\label{eq3.46}
\end{align}
We then define \emph{covariant derivatives},
\bel{eq3.47}
D_0=\frac{1}{c}\,\frac{\partial}{\partial t}+\frac{\mathrm{i}}{\hbar c}\Phi'+w_0\,,
\ee
where
\bel{eq3.48}
\Phi'=\Phi-\frac{\hbar^2}{2m}\,\frac{e^2}{8(mc^2)^2}\vecE^2\,,
\ee
($D_0$ is the \emph{covariant time derivative}), and
\bel{eq3.49}
\vecD=\grad+\mathrm{i}\frac{e}{\hbar c}\vecA+\vec{w}\,.
\ee
Here $(\Phi',e\vecA\,)$ are the components of an \emph{electromagnetic} $\mathrm{U}(1)$-\emph{connection}. Then the Pauli equation,
\begin{equation*}
\mathrm{i}\hbar\frac{\partial}{\partial t}\Psi_t=H_A^{\mathrm{SO}}\,\Psi_t,\hspace{0.3cm}\Psi_t\in\mathscr{H}_{\textrm{e}}\,,
\end{equation*}
can be rewritten in a \emph{manifestly} $\mathrm{U}(1)\times\mathrm{SU}(2)_{\textrm{spin}}$ \emph{gauge-invariant} form
\bel{eq3.50}
\mathrm{i}\hbar c D_0\,\Psi_t=-\frac{\hbar^2}{2m}\,\vecD^2\,\Psi_t\,.
\ee
This observation has been made in~\cite{R24}; (see also the original papers quoted there). When incorporated into the formalism of quantum-mechanical many-body theory the $\mathrm{U}(1)\times \mathrm{SU}(2)_{\textrm{spin}}$ gauge-invariance of Pauli's theory has very beautiful and important applications to problems in condensed-matter physics, which are discussed in much detail in~\cite{R24}. Depending on context, the $\mathrm{U}(1)$- and $\mathrm{SU}(2)$-connections introduced above receive further contributions, e.g., from a \emph{divergence-free velocity field} (quantum mechanics in moving coordinates, with applications, e.g., to superconductivity, super-fluidity, a quantum Hall effect for rotating Bose gases~\cite{R24}, nuclear physics,...), from a non-trivial \emph{spin connection} on $S(\mathbb{E}^3)$ with \emph{curvature} and \emph{torsion} describing disclinations and dislocations in a microscopic crystalline background, and/or from the `\emph{Weiss exchange field}' describing a magnetic background. It is most regrettable that we cannot enter into all these applications, here. But the reader will find a detailed exposition of these topics in~\cite{R24}.\\

Next, we recall the quantum theory of a system of many $(N=1,2,3,\ldots)$ Pauli electrons. The Hilbert space of pure state vectors of such a system is given by
\bel{eq3.51}
\mathscr{H}^{(N)}=\mathscr{H}_{\textrm{e}}\wedge\cdots\wedge\mathscr{H}_{\textrm{e}}\equiv\mathscr{H}_{\textrm{e}}^{\wedge N}\,,
\ee
where $\mathscr{H}_{\textrm{e}}$ is given by~\eqref{eq3.39}, and $\wedge$ denotes an anti-symmetric tensor product. The anti-symmetric tensor product in~\eqref{eq3.51} incorporates the \emph{Pauli exclusion principle}. Let $H^{(1)}$ denote the Pauli Hamiltonian for a single electron, as given in~\eqref{eq3.41} or~\eqref{eq3.43}. In applications to atomic, molecular or condensed matter physics, $\Phi(x)$ is the Coulomb potential  of the electron in the field of $K$ nuclei with charges $eZ_1,\ldots,\,eZ_K$, which we shall usually treat, for simplicity, as \emph{static}, (Born-Oppenheimer approximation); i.e.,
\bel{eq3.52}
\Phi(x)=-\sum_{k=1}^K\frac{e^2 Z_k}{4\pi |x-X_k|}\,,
\ee
where $x$ is the position of the electron, and $X_1,\ldots,\,X_K$ are the positions of the nuclei. Moreover, $\vecB$ is an arbitrary external magnetic field, and $\vecE(x)\simeq -\frac{1}{e}\grad\Phi(x)$ is the electric field created by the nuclei (regularized or cut-off, for $x$ near $X_1,\ldots,\,X_K$).\\
\indent The Hamiltonian for the $N$ electrons is chosen to be
\begin{equation}
\label{eq3.53}
H^{(N)}=\sum_{j=1}^N\,\umat\wedge\cdots\wedge H^{(1)}\wedge\cdots\wedge\umat+V_{\mathrm{C}}\left(x_1,\ldots,\,x_N\right)+V_{\mathrm{C}}^{\mathrm{nuc}}\left(X_1,\ldots,\,X_K\right)\,,
\end{equation}
where, in the $j^{\textrm{th}}$ term of the sum on the R.S. of~\eqref{eq3.53}, $H^{(1)}$ stands in the $j^{\textrm{th}}$ place (factor), with $\umat$'s in other factors, and
\begin{align}
V_{\mathrm{C}}\left(x_1,\ldots,\,x_N\right)\;&=\sum_{1\leq i<j\leq N}\,\frac{e^2}{4\pi |x_i-x_j|}\,,\label{eq3.54}\\
V_{\mathrm{C}}^{\mathrm{nuc}}\left(X_1,\ldots,\,X_K\right)\;&=\sum_{1\leq k<l\leq K}\,\frac{e^2 Z_k Z_l}{4\pi |X_k-X_l|}\,.\label{eq3.55}
\end{align}
Properties of the Hamiltonian $H^{(N)}$ (with $H^{(1)}$ as in~\eqref{eq3.41} and $\Phi$ as in~\eqref{eq3.52}) will be studied in the next section.\newline
\indent We observe that the Hilbert space $\mathscr{H}^{(N)}$ is given by
\bel{eq3.56}
\mathscr{H}^{(N)}=\mathrm{P}_{\mathrm{a}}\,\left(L^2\left(\R^{3N},\textrm{d}^{3N}x\right)\otimes\C^{2^N}\right)\,,
\ee
where $\mathrm{P}_{\mathrm{a}}$ denotes the projection onto the subspace of totally anti-symmetric spinor wave functions. In an obvious sense, $\Hilbert^{(N)}$ carries a tensor product representation of two representations, $V^{\textrm{orbit}}$ and $V^{\textrm{spin}}$, of the permutation group $\mathscr{S}_N$ of $N$ symbols, where 
\begin{align*}
V^{\textrm{orbit}}(\pi)\,&=V_{\omega}(\pi)\otimes\umat\,,\\
V^{\textrm{spin}}(\pi)\,&=\umat\otimes V_{\sigma}(\pi)\,,\hspace{0.3cm}\pi\in\mathscr{S}_N\,,
\end{align*}
in the tensor product decomposition~\eqref{eq3.56}. The projection $\mathrm{P}_{\mathrm{a}}$ selects the alternating representation (multiplication by $\mathrm{sig}(\pi)$, $\pi\in\mathscr{S}_N$) from $V_{\omega}\otimes V_{\sigma}$; only tensor products of subrepresentations, $V_{\omega}^i$ and $V_{\sigma}^j$, of $V_{\omega}$ and $V_{\sigma}$, respectively, are in the range of $\mathrm{P}_{\mathrm{a}}$ for which $V_{\omega}^i(\pi)=\mathrm{sig}(\pi)\,V_{\sigma}^j(\pi)$, (i.e., $V_{\omega}^i$ is `\emph{associated}' to $V_{\sigma}^j$).\newline
\indent The spin space $\C^{2^N}\simeq\left(\C^2\right)^{\otimes N}$ carries the $N$-fold tensor product representation, $\rho$, of the spin $s=\frac{1}{2}$ representation of $\mathrm{SU}(2)$. This representation is a direct sum of irreducible representations with spin $s=s_0,\,s_0+1,\ldots,\,\frac{N}{2}$, where $s_0=0$ if $N$ is even and $s_0=\frac{1}{2}$ if $N$ is odd. It \emph{commutes} with the representation $V_{\sigma}$ of $\mathscr{S}_N$ on $\left(\C^2\right)^{\otimes N}$.\newline

\emph{Hermann Weyl} has proven the following 
\begin{theorem}
\bel{eq3.57}
\left(\C^2\right)^{\otimes N}\simeq\,\bigoplus_{(\Delta,\,s)}\,\Hilbert_{\Delta}\otimes\Hilbert_s\,,
\ee
with
\begin{align}
V_{\sigma}\,&=\bigoplus_{(\Delta,\,s)}\,\Delta\big|_{\Hilbert_{\Delta}}\otimes\umat\big|_{\Hilbert_s}\label{eq3.58}\\
\rho\,&=\bigoplus_{(\Delta,\,s)}\,\umat\big|_{\Hilbert_{\Delta}}\otimes\rho_s\big|_{\Hilbert_s}\label{eq3.59}\,,
\end{align}
where the $\Delta$'s are irreducible representations of the group $\mathscr{S}_N$ labeled by Young diagrams with \emph{one} or \emph{two} rows and a total of $N$ boxes, and $\rho_s$ is the irreducible representation of $\mathrm{SU}(2)$ with spin $s\in\lbrace s_0,\,s_0+1,\ldots,\,\frac{N}{2}\rbrace$. Moreover, in~\eqref{eq3.57}, every $\Delta$ and every $s$ occur only \emph{once}, i.e., a $\Delta$ on the R.S. of~\eqref{eq3.57}~-~\eqref{eq3.59} paired with a spin $s$ is \emph{uniquely determined} by $s$, $\Delta=\Delta(s)$, and conversely. (The spin $s=s(\Delta)$ corresponding to a representation $\Delta$ is given by half the number of columns in the Young diagram of $\Delta$ that consist of a single box.)\newline
\end{theorem}
\indent Weyl's theorem is a special case of a general theory of `\emph{dual pairs}' of groups; see~\cite{R25}. Weyl has shown that the groups $\mathscr{S}_N$ and $\mathrm{SU}(n)$, $N=1,\,2,\,3,\ldots$, $n=2,\,3\,\ldots$ are `dual pairs'. From our previous discussion we understand that a subrepresentation $\Delta$ of $V_{\sigma}$ can only be paired with a subrepresentation $\adjop{\Delta}$ of $V_{\omega}$ given by
\begin{equation*}
\adjop{\Delta}(\pi)=\mathrm{sig}(\pi)\Delta(\pi),\hspace{0.3cm}\pi\in\mathscr{S}_N\,,
\end{equation*}
in order for the tensor product representation $\adjop{\Delta}\otimes\Delta$ to `survive' the projection $\mathrm{P}_{\mathrm{a}}$. This, together with Weyl's theorem, implies that the spin $s$ of an $N$-electron wave function completely determines its symmetry properties under exchange of electron positions or momenta (the `race' of the orbital wave function) and under exchange of electron spins (the `race' of the spin wave function). This explains why in the classification of atomic spectra the permutation groups do not appear; (see section 2). In a system of many electrons moving in a shell of an atom or in a crystalline background, one might expect that, by a conspiracy of electron motion (kinetic energy) and Coulomb repulsion between electrons (potential energy) the energies of those states are particularly low that correspond to totally anti-symmetric orbital wave functions, i.e., $\adjop{\Delta}(\pi)=\mathrm{sig}(\pi),\,\pi\in\mathscr{S}_N$. Then the spin wave functions must be totally symmetric, i.e., $\Delta$ must be the trivial representation of $\mathscr{S}_N$. This implies that the spin $s$ of such a state is \emph{maximal}, i.e., $s=\frac{N}{2}$ (for $N$ electrons). The expectation described here is at the core of explanations of \emph{Hund's first rule} and of \emph{ferromagnetism}. While, in many situations, this expectation is quite plausible it is still poorly understood, mathematically.\newline

What is missing? Well, maybe, a few comments on \emph{Dirac's relativistic electron equation}. But I will cut this short, since everybody is familiar with it! A nice approach to the Dirac equation can be extracted from the theory of projective, unitary, irreducible representations of the \emph{Poincar\'e group} $\mathcal{P}_{+}^{\uparrow}$, which is the semi-direct product of the group of proper, orthochronous Lorentz transformations of Minkowski space $\mathbb{M}^4$ and the group of space-time translations. The Poincar\'e group has \emph{two Casimir operators},
\begin{itemize}
	\item[(i)] \begin{equation}M^2=P_0^2-\vec{P}^2\label{eq3.60}\,,\end{equation} where $P_0\equiv H$ (the Hamiltonian) is the generator of time-translations, and $\vec{P}$ (the momentum operator) is the generator of space-translations; and
	\item[(ii)] \begin{equation}W_0^2-\vecW^2\label{eq3.61}\,,\end{equation} where $(W_0,\,\vecW)$ is the \emph{Pauli-Lubanski pseudo vector}; see, e.g.,~\cite{R26}.
\end{itemize}
For purposes of quantum physics, we are only interested in projective, unitary representations of $\mathcal{P}_{+}^{\uparrow}$ for which $M^2\geq0$ and $W_0^2-\vecW^2$ is finite. In an \emph{irreducible}, projective unitary representation of $\mathcal{P}_{+}^{\uparrow}$,
\begin{align*}
M^2\,&=m^2\,\umat\,,\\
W_0^2-\vecW^2\,&=-m^2s(s+1)\,\umat\,,
\end{align*}
where $m\geq0$ is the mass of the representation and (for $m>0$) $s$ is the \emph{spin} of the representation of the subgroup of space rotations. All projective, unitary, irreducible representations of $\mathcal{P}_{+}^{\uparrow}$ corresponding to a given mass $m\geq0$ and a finite $s$ can be constructed by the method of \emph{induced representations} developed by Wigner and generalized by \emph{George Mackey}. We consider an energy-momentum vector $p=(p_0,\,\vec{p})$ with $p^2=p_0^2-{\vec{p}}^{\hspace{0.04cm}2}=m^2$. By $\mathrm{H}_{\textrm{p}}$ we denote the subgroup of all those Lorentz transformations that leave $p$ fixed. For $m>0$,
\begin{equation*}
\mathrm{H}_{\textrm{p}}\simeq\mathrm{SO}(3)\,,
\end{equation*}
while, for  $m=0$,
\begin{equation*}
\mathrm{H}_{\textrm{p}}\simeq\mathrm{E}(2)\,,
\end{equation*}
the group of Euclidian motions of the plane. Representations of $\mathrm{SO}(3)$ and $\mathrm{E}(2)$ then determine representations theory $\mathcal{P}_{+}^{\uparrow}$. The Hilbert space of pure state vectors of a free, relativistic particle of mass $m\geq0$ is the representation space of an irreducible unitary representation of the quantum-mechanical Poincar\'e group with mass $m\geq0$ and a finite eigenvalue for $W_0^2-\vecW^2$. For an electron or positron, $m$ is positive, and hence $W_0^2-\vecW^2=-m^2s(s+1)\,\umat$, where $s$ is the spin of the representation of the little group $\mathrm{H}_{\textrm{p}}\simeq\mathrm{SO}(3)$. For the electron or positron, $s=\frac{1}{2}$! If we insist that space reflections should be a symmetry of the theory, we must glue together two unitary, irreducible representations of the quantum-mechanical Poincar\'e group with $m>0$ and $s=\frac{1}{2}$. Considering that $p_0$ can be $\geq m$ or $\leq -m$, we find the \emph{Dirac equation} for the relativistic electron hiding in the representation theory of  $\mathcal{P}_{+}^{\uparrow}$  with mass $m>0$ and spin $s=\frac{1}{2}$. The second-quantized Dirac theory for free electrons and positrons is obtained by considering anti-symmetric tensor products of the positive-energy representation of $\mathcal{P}_{+}^{\uparrow}$ for single electrons and positrons in a rather standard fashion; see, e.g.,~\cite{R26}. All this is so exceedingly well-known that I do not want to enter into details. One might note, perhaps, that, for massless particles ($m=0$), the helicity is not `quantized' in general, but can be an arbitrary real number. However, helicities that are not integers or half-integers do \emph{not} appear in quantum field theories formulated in terms of field operators localizable in space-time points; (see section 7).\newline

\indent The results and methods just alluded to, above, can be generalized to Minkowski space-times of arbitrary dimension $d=n+1\geq2$. \emph{Formally}, a \emph{local quantum field theory} of electrons and positrons moving in quite general \emph{Lorentzian} space-time manifolds and coupled to external electromagnetic fields can be written down without difficulty. However, in contrast to the theory of Pauli electrons and positrons moving in a general physical space, the number of electrons and positrons is no longer conserved (electron-positron pair creation processes happen), and one encounters serious analytical problems when one attempts to develop Dirac theory on general Lorentzian space-times and coupled to general electromagnetic fields. These problems are only partially solved, and I do not wish to enter into this matter.\\
\indent Pauli's non-relativistic theory of the spinning electron, along with a systematic treatment of relativistic corrections, can be recovered by studying the limit of Dirac's theory, as the speed of light $c$ tends to $\infty$. Relativistic corrections can be found by perturbation theory in $c^{-1}$. A mathematically careful treatment of such matters can be found in~\cite{hunziker}.


\section[Stability of Non-Relativistic Matter in Arbitrary External Magnetic Fields]{Stability of Non-Relativistic Matter in Arbitrary External Magnetic Fields}
In order to get a first idea of the importance of \emph{electron spin} and the \emph{Pauli principle} in the physics of systems of many electrons moving in the Coulomb field of static (light) nuclei and coupled to an arbitrary external magnetic field, I review some fairly recent results on the \emph{stability} of such systems. \emph{The reference} for such results is~\cite{R27}.\newline

Let us consider a system of $N$ electrons and $K$ static nuclei with nuclear charges $eZ_1,\,\ldots,\,eZ_k$. with $\sum_{k=1}^K Z_k\sim N$. The Hilbert space of the system is the space $\Hilbert^{(N)}$ introduced in~\eqref{eq3.51}, the Hamiltonian is the operator $H^{(N)}$ defined in~\eqref{eq3.53}, where the one-electron operator $H^{(1)}$ is the Pauli operator of eq.~\eqref{eq3.41}, with $D_A$ as in~\eqref{eq3.40} and $\Phi$ as in~\eqref{eq3.52}.\\
\indent \emph{Units:} The energy unit is $\mathrm{Ry}=2mc^2\alpha^2$, where $\alpha=\frac{e^2}{\hbar c}\sim \frac{1}{137}$ is Sommerfeld's fine structure constant. The unit of length is half the Bohr radius, i.e., $l=\frac{\hbar^2}{2me^2}$. The magnetic field $\vecB=\rotA$ is in units of $\frac{e}{l^2\alpha}$; the magnetic field energy is given by $\epsilon\int\vecB^2\,\textrm{d}^3x$, with $\epsilon=\frac{1}{2\alpha^2}$.\\
\indent The Pauli operator $D_A$ is given, in our units, by
\bel{eq4.1}
D_A=\vec{\sigma}\cdot\left(\,-\mathrm{i}\grad+\vecA\,\right)\,.
\ee
It is convenient to work in the \emph{Coulomb gauge}, 
\bel{eq4.2}
\grad\cdot\vecA=0\,.
\ee
For a vector field $\vecX$ on $\R^3$ or a spinor $\psi\in L^2(\R^3,\textrm{d}^3x)\otimes\C^2$, we say that $\vecX\in L^p\;(\psi\in L^p)$ iff
\begin{align*}
{\big({\vecX\cdot\vecX}\big)}^{1/2}\,&\in L^p(\R^3,\textrm{d}^3x)\,,\\
\scalarr{\psi}{\psi}^{1/2}\,&\in L^p(\R^3,\textrm{d}^3x)\,.
\end{align*}
It is shown in~\cite{R28} that if $\vecB$ has \emph{finite field energy}, i.e., $\vecB\in L^2$, then there exists a \emph{unique} $\vecA$ such that
\begin{equation*}
\rotA=\vecB\,,\hspace{0.3cm}\grad\cdot\vecA=0\,,\hspace{0.3cm}\vecA\in L^6\,.
\end{equation*}

\subsection{Zero-modes of the Pauli operator}
Loss and Yau~\cite{R29} have proven, by a fairly explicit construction, the following important result:
\begin{theorem}\label{theorem4.1}
There exists a single-electron two-component spinor wave function $\psi\in H^1(\R^3)$ (the usual Sobolev space) and a vector potential $\vecA\in L^6$, with $\grad\cdot\vecA=0$ and $\vecB=\rotA\in L^2$ such that
\bel{eq4.4}
D_A\,\psi=0\,,
\ee
i.e, $\psi$ is a zero-mode of the Pauli operator $D_A$.
\end{theorem}
\noindent An explicit choice of a magnetic field leading to a zero-mode, in the sense of eq.~\eqref{eq4.4} is 
\begin{equation*}
\vecB(x)=\frac{12}{(1+x^2)^3}\Big\lbrack(1-x^2)n+2\left(n\cdot x\right)\,x+2n\wedge x\Big\rbrack\,,
\end{equation*}
where $n$ is a unit vector.\\
\indent This result, whose proof we omit, has some rather remarkable consequences that we will discuss next. (The proof relies on a three-dimensional analogue of the celebrated Seiberg-Witten equations.)

\subsection{Stability and instability of atoms and one-electron molecules}

We consider the Pauli Hamiltonian for a one-electron ion in a general external magnetic field $\vecB$ of finite field energy:
\bel{eq4.4*}
H_A=D_A^2-\frac{Z}{4\pi\,|x|}\,.
\ee
Let $E_0\,(\vecB,\,Z)$ denote the \emph{infimum of the spectrum of} $H_A$. If $\vecB$ is a constant external magnetic field, $\vecB=(0,\,0,\,B)$, then it is known from work of \emph{Avron, Herbst} and \emph{Simon} quoted in~\cite{R28} that
\begin{equation*}
E_0\,(\vecB,\,Z)\sim -\mathrm{const}\,(\ln\,B)^2\,.
\end{equation*}
This implies that $E_0\,(\vecB_n,\,Z)\longrightarrow -\infty$ even for a sequence of suitably chosen magnetic fields $\vecB_n$ of finite, but ever larger field energy. It is then natural to ask whether
\bel{eq4.5}
E_0\,(\vecB,\,Z)+\epsilon\int\textrm{d}^3x\,|{{\vecB}(x)}|^2\,
\ee
is bounded below, uniformly in $\vecB$, and for what range of values of the nuclear charge.\\

\indent The answer is worked out in~\cite{R28}. We define a convenient space, $\mathcal{C}$, of configurations $(\psi,\,\vecA)$,
\begin{multline}\label{eq4.6}
\mathcal{C}\deq\Bigl\lbrace\left(\psi,\,\vecA\,\right)\,\Big|\,\psi\in H^1(\R^3)\,,\,\|\psi\|^2_2=1\,,\,\vecA\in L^6\,,
\divA=0\,,\,\rotA\in L^2\,\Bigr\rbrace
\end{multline}
and a space $\mathcal{N}$ of `zero modes',
\bel{eq4.7}
\mathcal{N}\deq\Big\lbrace\,\left(\psi,\,\vecA\right)\,\big|\,\left(\psi,\,\vecA\right)\in\mathcal{C}\,,\,D_{A}\psi=0\,\Big\rbrace\,.
\ee
We then define a critical nuclear charge $Z_c$ by
\bel{eq4.8}
\normalsize Z_c\deq\inf_{(\psi,\,\vecA)\in\mathcal{N}}\,\Big\lbrace\,\epsilon\,\|\vecB\|_2^2 \;\Big/\,\small{\scalarB{\psi}{\frac{1}{4\pi|x|}\psi}}\normalsize\Big\rbrace\,.
\ee
(Note that, by scaling, the analogue of $Z_c$ vanishes in more than three dimensions.)\\
\indent The following result has been shown in~\cite{R28}.
\begin{theorem}\label{theorem4.2}
$Z_c$ is positive and finite.\newline

\noindent For $Z>Z_c$,
\begin{equation*}
\inf_{\vecB\in L^2}\,\Big\lbrace E_{0}\left(\vecB,\,Z\right)+\epsilon\,\|\vecB\|^2_2\Big\rbrace=-\infty\,.
\end{equation*}
For $Z<Z_c$,
\begin{equation*}
\inf_{\vecB\in L^2}\,\Big\lbrace E_{0}\left(\vecB,\,Z\right)+\epsilon\,\|\vecB\|^2_2\Big\rbrace>-\infty\,,
\end{equation*}
and the infimum is a minimum reached for some pair $\left(\psi\,,\vecA\right)\in\mathcal{C}$.\\
\noindent Furthermore, the infimum on the R.S. of~\eqref{eq4.8} is reached on a pair $\left(\psi\,,\vecA\right)\in\mathcal{N}$.
\end{theorem}

In~\cite{R28}, $Z_c$ is estimated for the physical value of the fine structure constant and comes out to be $Z_c\sim17'900$. Thus, a single-electron ion coupled to an arbitrary magnetic field $\vecB$ of finite field energy is \emph{stable} (the total energy is bounded from below) if the nuclear charge $Z$ is smaller than $Z_c$, while it is \emph{unstable} if $Z>Z_c$. This result crucially depends on the fact that electrons have \emph{spin} and a \emph{magnetic moment} with a \emph{gyromagnetic factor} $\gmf=2$, (as long as radiative (QED) corrections are neglected). If $\gmf<2$ then
\begin{equation*}
\inf_{\vecB\in L^2} E_0\,\left(\vecB,\,Z\right)>-\mathrm{const}\,Z^2>-\infty\,,
\end{equation*}
for \emph{all} values of $Z$, by Kato's `\emph{diamagnetic inequality}', while for $\gmf>2$, ions would \emph{always} be unstable\newline
\indent In~\cite{R30}, the results summarized in Theorem~\ref{theorem4.2} are extended to many-electron atoms and to a system consisting of a single electron moving in the Coulomb field of arbitrarily many static nuclei, (one-electron molecule in the Born-Oppenheimer approximation). For this purpose, one considers the energy functional
\bel{eq4.8*}
\large{\mathcal{E}}\normalsize\big(\,\Psi,\,\vecB,\,\niceunderline{X},\,\niceunderline{Z}\,\big)\deq\scalarB{\Psi}{H_{\vecA}^{(N)}\,\Psi}+\epsilon\|\vecB\|^2_2\,,
\ee
where $\Psi\in\Hilbert^{(N)}$, see~\eqref{eq3.51}, is an $N$-electron wave function with $\scalar{\Psi}{\Psi}=1$, and $H_{\vecA}^{(N)}\equiv H^{(N)}$ is the $N$-electron Hamiltonian introduced in~\eqref{eq3.53}~-~\eqref{eq3.55}, with $H^{(1)}$ as in~\eqref{eq3.41} and~\eqref{eq3.52}, (see also~\eqref{eq4.4*}, with $\frac{Z}{4\pi|x|}$ replaced by the Coulomb potential~\eqref{eq3.52} of many nuclei). There is an obvious extension of the definition~\eqref{eq4.6} of the space $\mathcal{C}$ to an $N$-electron system. We are interested in studying the lowest possible energy
\bel{eq4.9}
\normalsize E_0\deq\large{\inf_{\scriptsize{\begin{array}{c} (\Psi,\,\vecA)\in\mathcal{C} \\ \niceunderline{X}\in\R^{3K} \end{array}}}}\,\large{\mathcal{E}}\normalsize\big(\,\Psi,\,\vecB,\,\niceunderline{X},\,\niceunderline{Z}\,\big)\,.
\ee
It is shown in~\cite{R30} that, for $K=1$ (\emph{one} nucleus) and $N$ arbitrary (arbitrarily many electrons), or for $K$ arbitrary and $N=1$, 
\begin{equation*}
E_0>-\infty\,,
\end{equation*}
provided $Z_j<\tilde{Z}_c<\infty$, for all $j=1,\ldots,\,K$, \emph{and} provided
\begin{equation}\label{eq4.10*}
\alpha<\alpha_c\,,
\end{equation}
with $0.32<\alpha_c<6.7$, i.e., \emph{provided the fine structure constant} $\alpha$ \emph{is sufficiently small}. The bound~\eqref{eq4.10*} comes from studying 1-electron molecules and is `\emph{real}': If $\alpha>\alpha_c$ there are configurations of $K$ identical nuclei with arbitrary $Z<\tilde{Z}_c=\mathcal{O}(\alpha^{-2})$ such that, for some choice of $K$, $E_0=-\infty$, for a 1-electron molecule. Again, \emph{the} crucial role in the proofs of these results is played by the electron spin and the fact that $\gmf=2$!\\
\indent The punchline in this analysis of stability of non-relativistic matter was reached, a little more than ten years ago, in works of \emph{Charles Fefferman}~\cite{R31} and of \emph{Elliott H. Lieb}, \emph{Michael Loss} and \emph{Jan Philip Solovej}~\cite{R32} (whose treatment is considerably simpler than Fefferman's, but came a little later)\footnote{All this work came after ground-breaking work of \emph{Dyson} and \emph{Lenard} in the 1960's, and of Lieb and \emph{Thirring}; see~\cite{R27} and references given there.}. It is summarized in the next subsection.

\subsection{Stability of matter in magnetic fields}

Consider the energy functional $\mathcal{E}\,(\,\Psi,\,\vecB,\,\niceunderline{X},\,\niceunderline{Z}\,)$ introduced in~\eqref{eq4.8*} -- with $N$ electrons moving in the Coulomb field of $K$ static nuclei at positions ${X}_1,\ldots,\,{X}_K$, with nuclear charges $Z_1,\ldots,\,Z_K$, and coupled to an arbitrary external magnetic field $\vecB$ of finite field energy $\epsilon\,\|\vecB\|^2_2$. Let
\bel{eq4.10}
\normalsize E_0\equiv E_0(\alpha,\,\niceunderline{Z})\deq\large{\inf_{\scriptsize{\begin{array}{c} (\Psi,\,\vecA)\in\mathcal{C} \\ \niceunderline{X}\in\R^{3K} \end{array}}}}\,\large{\mathcal{E}}\normalsize\big(\,\Psi,\,\vecB,\,\niceunderline{X},\,\niceunderline{Z}\,\big)\,.
\ee
The following result is proven in~\cite{R32}.
\begin{theorem}\label{theorem4.3}
Suppose that $Z_k\leq Z<\infty$, for all $k=1,\dots,\,K$, and that
\bel{eq4.11}
Z\alpha^2<0.041\hspace{0.2cm}\textrm{and}\hspace{0.2cm}\alpha<0.06\,.
\ee
Then
\bel{eq4.12}
E_0(\alpha,\,\niceunderline{Z})\geq-C\,(N+K)\,,
\ee
for some finite constant $C$ depending on $Z$ and $\alpha$, but \emph{independent} of $N$ and $K$.
\end{theorem}
\indent \emph{Remarks:} The bound~\eqref{eq4.12} expresses \emph{stability of matter} in the sense that the \emph{energy per particle} (electrons and nuclei) has a lower bound ($\geq-\mathrm{const}\,Z^2 \mathrm{Ry}$) \emph{independent} of the number of electrons and nuclei in the system. This is an expression of \emph{thermodynamic stability} of such systems, which is a pillar on which all of condensed-matter physics rests; (`independence' of condensed-matter physics of nuclear form factors and cut-offs imposed on the magnetic field).\\
\indent For stability of matter, i.e., for the validity of~\eqref{eq4.12}, it is crucial that electrons are \emph{fermions}, i.e., that they satisfy Pauli's exclusion principle. In Lieb-Thirring type proofs of stability of matter, the Pauli principle enters in the form of generalized Sobolev inequalities (bounding the electron kinetic energy from below by the Thomas-Fermi kinetic energy) \emph{only valid for fermions}; see~\cite{R27}.\\
\indent We know from the results in the last two subsections that $E_0(\alpha,\,\niceunderline{Z})=-\infty$, i.e., the system becomes \emph{unstable}, if either $Z\gg\alpha^{-2}$ or if $\alpha$ is `large' ($\alpha>6.7$). It is somewhat tantalizing that \emph{electron spin and the fact that} $\gmf=2$ \emph{would render systems of many electrons and nuclei} -- as they are studied in atomic, molecular and condensed-matter physics -- \emph{unstable if} $\alpha>6.7$ \emph{and/or if} $Z\alpha^2$ is \emph{very `large'}. This is reminiscent of the possibility that the \emph{Landau pole} in relativistic QED will descend to the non-relativistic regime if $\alpha$ is large enough.\\
\indent Let us see what the source of the potential instability is! It is actually a short-distance or \emph{ultraviolet instability}: If in the definition of $H_{\vecA}^{(N)}$, the electromagnetic vector potential $\vecA$ in the Coulomb gauge is replaced by a mollified potential
\begin{equation*}
\vecA_{\kappa}(x)\deq\int\,\textrm{d}^3y\,\kappa(x-y)\vecA(y)\,,
\end{equation*}
where $\kappa$ is an arbitrary positive, smooth function, with $\int\,\kappa=1$, (i.e., a smooth approximate $\delta$-function) then the bound
\begin{equation*}
E_0(\alpha,\,\niceunderline{Z})\geq-C\,(N+K)\,
\end{equation*}
is true for arbitrary $\alpha$ and $Z$, but the constant $C$ now depends on $\kappa$, and if $\alpha>6.7$ and/or $Z\alpha^2$ is large enough, then $C=C_{\kappa}\longrightarrow\infty$, as $\kappa$ approaches a $\delta$-function. In order to arrive at a deeper understanding of these matters, we should \emph{quantize the electromagnetic field}, too. 


\section{Electrons Interacting with the Quantized Electromagnetic Field; Radiative Corrections to the Gyromagnetic Factor}

It is important to ask what becomes of the results in the last section if the electromagnetic field is treated quantum mechanically. One of my strong scientific interests, during the past fifteen years, has been to find mathematically precise answers to this question; see~\cite{R33,R34,R35,R36,R37,R38,R39,R40,R401,R41,R42,R43,R44}, and~\cite{R45} for a review of some of these and other results.\\
\indent We return to the Hamiltonian~\eqref{eq3.53}, i.e.
\begin{multline}\label{eq5.1}
H^{(N)}=\sum_{j=1}^N\,\biggl\lbrace{\Big\lbrack\vec{\sigma}_j\cdot\left(-\mathrm{i}\grad_j+\vecA(x_j)\right)\Big\rbrack}^2-\sum_{k=1}^{K}\,\frac{Z_k}{4\pi|x_j-X_k|}\biggr\rbrace\\+\sum_{1\leq i<j\leq N}\frac{1}{4\pi|x_i-x_j|}+\sum_{1\leq k<l\leq K}\,\frac{Z_kZ_l}{4\pi|X_k-X_l|}\,,
\end{multline}
acting on the $N$-electron Hilbert space
\bel{eq5.2}
\Hilbert^{(N)}=\Big( L^2\left(\R^3,\,\textrm{d}^3x\right)\otimes\C^2\Big)^{\wedge N}\,.
\ee
We are interested in studying the dynamics of such systems when the electromagnetic field is \emph{quantized}, i.e., electrons can emit and absorb photons. We quantize the electromagnetic field in the Coulomb gauge, i.e.,
\bel{eq5.3}
\divA=0\,.
\ee
Then
\begin{equation}\label{eq5.4}
\qquad\vecA(x)=\frac{1}{(2\pi)^{3/2}}\,\sum_{\lambda=\pm1}\,\int\,\frac{\textrm{d}^3k}{\sqrt{2|k|}}\Big\lbrack\,\vec{\epsilon}_{\lambda}(k)a\adj_{\lambda}(k)\mathrm{e}^{-\mathrm{i}k\cdot x}+\overline{\vec{\epsilon}_{\lambda}(k)}a_{\lambda}(k)\mathrm{e}^{\mathrm{i}k\cdot x}\,\Big\rbrack\,,
\end{equation}
where $a\adj_{\lambda}(k)$, $a_{\lambda}(k)$ are the usual creation and annihilation operators for a photon with wave vector $k\in\R^3$ and helicity $\lambda=\pm$, satisfying the canonical commutation relations (CCR),
\bel{eq5.5}
\lbrack a^{\#}_{\lambda}(k),a^{\#}_{\mu}(l)\rbrack=0\,,\hspace{0.3cm}\lbrack a_{\mu}(k),a\adj_{\lambda}(l)\rbrack=\delta_{\mu\lambda}\delta^{(3)}(k-l)\,,
\ee
and $\vec{\epsilon}_{\lambda}(k)\perp k$, $\lambda=\pm$, are two orthonormal polarization vectors. We consider the Fock representation of the commutation relations~\eqref{eq5.5} uniquely characterized by the existence of a vacuum state $\Omega$ in which none of the field modes is excited, so that
\bel{eq5.6}
a_{\lambda}(k)\,\Omega=0\,,\:\textrm{for all }\lambda\textrm{ and }k\,,
\ee
and $\scalar{\Omega}{\Omega}=1$. Fock space $\mathscr{F}$ is the Hilbert space completion of the linear space obtained by applying arbitrary polynomials in creation operators smeared out with square-integrable functions to the vacuum $\Omega$. The Hamiltonian of the free electromagnetic field generating the time evolution of vectors in $\mathscr{F}$ is given, in our units, by the operator
\begin{align}\label{eq5.7}
H_{\textrm{f}}\deq&\frac{1}{2\alpha^2}\int\,\textrm{d}^3x\,\Big\lbrace :{\vecE(x)}^2:+:{\vecB(x)}^2:\Big\rbrace\nonumber\\
=&\alpha^{-2}\,\sum_{\lambda=\pm}\,\int\,\textrm{d}^3k\;a\adj_{\lambda}(k)|k|a_{\lambda}(k)\,,
\end{align}
where
\begin{equation*}
\vecE(x)=\frac{1}{(2\pi)^{3/2}}\,\sum_{\lambda=\pm1}\,\int\,\textrm{d}^3k\sqrt{\frac{|k|}{2}}\Big\lbrack\,\mathrm{i}\vec{\epsilon}_{\lambda}(k)a\adj_{\lambda}(k)\mathrm{e}^{-\mathrm{i}k\cdot x}-\mathrm{i}\overline{\vec{\epsilon}_{\lambda}(k)}a_{\lambda}(k)\mathrm{e}^{\mathrm{i}k\cdot x}\,\Big\rbrack\,,
\end{equation*}
are the transverse components of the electric field, $\vecB=\rotA$ is the magnetic field, the double colons indicate standard Wick ordering, and $\alpha^{-2}|k|$ is the energy of a photon with wave vector $k$ (in our units).\\
\indent The total Hilbert space of electrons and photons is given by
\bel{eq5.8}
\Hilbert\deq\Hilbert^{(N)}\otimes\mathscr{F}\,,
\ee
and the Hamiltonian is given by
\bel{eq5.9}
H\deq H^{(N)}+\umat\otimes H_{\textrm{f}}\,.
\ee
Alas, this operator is \emph{ill-defined}. To arrive at a mathematically well defined expression for the Hamiltonian (selfadjoint on $\Hilbert$ and bounded from below), we must replace the vector potentials $\vecA(x_j)$ on the R.S. of~\eqref{eq5.1} by ultraviolet regularized potentials $\vecA_{\Lambda}(x_j)$, $j=1,\ldots,\,N$, where
\begin{equation*}
\vecA_{\Lambda}(x)=\int\,\textrm{d}^3y\,\kappa_{\Lambda}(x-y)\vecA(y)\,,
\end{equation*}
and $\kappa_{\Lambda}$ is the Fourier transform of, e.g., a normalized Gaussian
\begin{equation*}
\frac{1}{{(2\pi\Lambda^2)}^{3/2}}\,\mathrm{e}^{-\left(|k|^2/{2\Lambda^2}\right)}\,,
\end{equation*}
where $\Lambda$ is an ultraviolet cutoff energy that one may choose to be of the order of the rest energy of an electron. Of course one will ultimately be interested in studying the limit, as $\Lambda\longrightarrow\infty$. This limit is only meaningful if the mass and the chemical potential of an electron are \emph{renormalized}. To study the renormalization theory of the model of quantum electrodynamics (QED) considered in this section, we must replace the Pauli Hamiltonians, ${\big\lbrack\vec{\sigma}_j\cdot\big(-\mathrm{i}\grad_j+\vecA(x_j)\big)\big\rbrack}^2$ on the R.S. of~\eqref{eq5.1} by operators
\bel{eq5.10}
\frac{1}{M_{\Lambda}}{\Big\lbrack\vec{\sigma}_j\cdot\left(-\mathrm{i}\grad_j+\vecA_{\Lambda}(x_j)\right)\Big\rbrack}^2+\mu_{\Lambda}\,,
\ee
for $j=1,\ldots,\,N$, where $M_{\Lambda}$ is the ratio between the `\emph{bare mass}' of an electron and its observed (physical) mass, and $\mu_{\Lambda}$ is the bare self-energy (or chemical potential) of an electron. The Hamiltonians obtained after the replacement~\eqref{eq5.10} are denoted by $H^{(N)}_{\Lambda}\equiv H^{(N)}_{\Lambda}(M_{\Lambda},\,\mu_{\Lambda})$, see~\eqref{eq5.1}, and $H_{\Lambda}\equiv H_{\Lambda}(M_{\Lambda},\,\mu_{\Lambda})$, see~\eqref{eq5.9}, respectively. A fundamental question in renormalization theory is whether $M_{\Lambda}>0$ and $\mu_{\Lambda}$ can be chosen to depend on the cutoff energy $\Lambda$ in such a way that the limiting Hamiltonian
\bel{eq5.11}
H_{\textrm{ren}}=\textrm{``}{\lim_{\Lambda\rightarrow\infty}\,H_{\Lambda}}\textrm{''}\,
\ee
exists as a selfadjoint operator on $\Hilbert$.\\
\indent A mathematically rigorous answer to this question remains to be found. (I rather bet it might be `no'.) However, there are indications of various kinds as to how to choose $M_{\Lambda}$ and $\mu_{\Lambda}$ and plenty of perturbative calculations (perturbation theory in $\alpha$), which we briefly summarize next.

\begin{itemize}
\item[(1)] Since, in our model of QED, the number of electrons and nuclei is conserved -- electron-positron pair creation processes are suppressed -- there is no vacuum polarization, and hence the fine structure constant $\alpha$ is independent of $\Lambda$.
	
\item[(2)] (Non-rigorous) perturbative renormalization group calculations suggest that
	\bel{eq5.12}
	M_{\Lambda}\sim\Lambda^{-(8\alpha/3\pi)+\mathcal{O}(\alpha^2)}\,,
	\ee
	i.e., the bare mass of an electron must approach  $0$ like a small inverse power of $\Lambda$, as $\Lambda\longrightarrow\infty$; or, in other words, the physical mass of an electron consists entirely of radiative corrections\footnotemark[12].\footnotetext[12]{In these calculations, the Zeeman terms in $H^{(N)}_{\Lambda}$ are \emph{neglected}.}
\item[(3)] There are some rather crude bounds on the self-energy $\mu_{\Lambda}$:
		\begin{equation*}
			c_1\Lambda^{3/2}\leq\mu_{\Lambda}\leq c_2\Lambda^{12/7}\,,
		\end{equation*}
		for constants $c_1$ and $c_2$ (but derived under the assumption that $M_{\Lambda}=1$); see~\cite{R45} and references given there.
	\item[(4)] Perturbatively, a finite Lamb shift is found, as $\Lambda\longrightarrow\infty$, which is in rough agreement with experimental data\footnotemark[12]; (an improved version of Bethe's calculation of 1947).
	\item[(5)] The \emph{gyromagnetic factor} $\gmf$ \emph{of the electron} is affected by radiative corrections. In low-order perturbation theory in $\alpha$, it remains \emph{finite}, as $\Lambda\longrightarrow\infty$, and is given by
	\bel{eq5.13}
	\gmf=2\Big\lbrack1+\frac{8}{3}\,\frac{\alpha}{2\pi}+\mathcal{O}(\alpha^2)\Big\rbrack\,;
	\ee
	see~\cite{R45,R46}. This result should be compared to the value for $\gmf$ predicted by perturbative \emph{fully relativistic} QED,
	\bel{eq5.14}
	\gmf=2\Big\lbrack1+\frac{\alpha}{2\pi}+\mathcal{O}(\alpha^2)\Big\rbrack\,,
	\ee
	where the lowest-order correction, $\frac{\alpha}{2\pi}$, was first calculated by \emph{Julian Schwinger}. Experiment favours Schwinger's result! This can be viewed -- if one likes -- as a high-precision confirmation of, among other things, the \emph{special theory of relativity}.\\
\end{itemize}

\indent No matter whether electrons are treated non-relativistically or relativistically, we find that $\gmf>2$! For a single, freely moving electron with Hamiltonian $H_A$ given by~\eqref{eq3.41} (with $\Phi=0$), this results in a breaking of the `\emph{supersymmetry}' (see section 3.2) of the quantum theory, and the effects of `supersymmetry breaking' offer a handle on \emph{precision measurements} of $\gmf-2$; (see section 6).\\
\indent The fact that $\gmf>2$ and the results in section 4 apparently imply that QED with non-relativistic matter ultimately only yields a mathematically meaningful description of physical systems if a (large, but finite) \emph{ultraviolet cutoff} is imposed on the interactions between electrons and photons, no matter how small $\alpha$ is chosen. For large values of $\alpha$ $(\alpha>6.7)$, this theory is expected to exhibit cutoff dependence already at atomic and molecular energies.\\
\indent The need for an ultraviolet cutoff in QED with non-relativistic matter is reminiscent of the problem of the \emph{Landau pole} in \emph{relativistic} QED.\\
\indent The following results are non-perturbative and mathematically rigorous:
\begin{itemize}
	\item[(6)] \emph{Stability of Matter:} For an arbitrary number $N$ of electrons and $K$ static nuclei with nuclear charges $Z_k\leq Z<\infty$, for all $k=1,\ldots,\,K$ and arbitrary $K<\infty$,
	\bel{eq5.15}
	H_{\Lambda}\geq-C_{\alpha,\,Z}\,K\Lambda\,,
	\ee
	for a finite constant $C_{\alpha,\,Z}$ independent of $\Lambda$ and $K$. While~\eqref{eq5.15} proves stability of matter if an ultraviolet cutoff $\Lambda$ is imposed on the theory, the linear dependence on $\Lambda$ on the R.S. of~\eqref{eq5.15} is disastrous, physically speaking. It is not understood, at present, whether a lower bound on $H_{\Lambda}\left(M_{\Lambda}\,,\mu_{\Lambda}\right)$ can be found that is \emph{uniform} in $\Lambda$, provided $M_{\Lambda}$ and $\mu_{\Lambda}$ are chosen appropriately!
\end{itemize}
Present mathematically rigorous efforts towards understanding QED with non-relativistic matter are therefore aimed at an analysis of $H_{\Lambda}^{(N)}$, for a fixed ultraviolet cutoff $\Lambda$ ($\sim$ rest energy of an electron), and at tackling the so-called \emph{infrared problem} that is caused by the masslessness of the photons. Here there has been tremendous progress, during the past fifteen years; see e.g.~\cite{R33,R34,R35,R36,R37,R38,R39,R40,R41,R42,R43,R44,R45}.

\indent The most remarkable results that have been found, during the last ten years, are, perhaps, the following ones:\\

\indent We choose an arbitrary, but fixed ultraviolet cutoff $\Lambda$.
\begin{itemize}
	\item[(7)] Atoms have stable ground states;~\cite{R39,R40,R401}.
	\item[(8)] Excited states of atoms are turned into resonances (meta-stable states) whose energies and widths (inverse life times) can be calculated to arbitrary precision by a constructive and convergent algorithm. These energies and life times agree, to leading order in $\alpha$, with those first calculated by Bethe in order to explain the Lamb shift,~\cite{R39,R40}.
	\item[(9)] Scattering amplitudes, $S_{fi}$, for Rayleigh scattering of photons at atoms (below the ionization threshold)  have asymptotic expansions of the form
	\begin{equation*}
	S_{fi}=\sum_{n=0}^{N}\upsigma_{fi,\,n}(\alpha)\,\alpha^n+o(\alpha^N)\,,
	\end{equation*}
	where
	\begin{equation*}
	\lim_{\alpha\rightarrow 0}\alpha^{\delta}\upsigma_{fi,\,n}(\alpha)=0\,,
	\end{equation*}
	for an arbitrarily small $\delta>0$. It is expected (and can be verified in examples) that
	\begin{equation*}
	\upsigma_{fi,\,n}(\alpha)=\sum_{k=0}^n\upsigma_{fi,\,n,\,k}\left(\ln{\frac{1}{\alpha}}\right)^k\,.
	\end{equation*}
	The powers of $\ln\frac{1}{\alpha}$ come from infrared singularities that render ordinary perturbation theory \emph{infrared-divergent} in large, but finite orders in $\alpha$; see~\cite{R43}. Our results yield, among many other insights, a mathematically rigorous justification of Bohr's frequency condition for radiative transitions.
	\item[(10)] Infrared-finite, constructive, convergent algorithms have been developed to calculate the amplitudes for ionization of atoms by Laser pulses (unpublished work of the author and \emph{Schlein} based on earlier work by \emph{Fring, Kostrykin} and \emph{Schader}) and for \emph{Compton scattering} of photons at a freely moving electron; see \emph{Pizzo} et al.~\cite{R44}.
\end{itemize}
Most proofs of the results reviewed in this section rely on complex spectral deformation methods, multi-scale perturbation theory and/or operator-theoretic renormalization group methods; see~\cite{R39,R40,R43} and references given there.

I now leave this thorny territory and sketch how the gyromagnetic factor of the electron can be measured experimentally.\\


\section{Three Methods to Measure $\gmf$}

We have already seen in section 2 that atomic spectroscopy in a magnetic field (Zeeman splittings) offers a possibility to measure the gyromagnetic factor $\gmf$ of the electron.\\
\indent Another possibility originating in condensed-matter physics is to exploit the {\emph{Einstein--de Haas effect}}.

\subsection[The Einstein--de Haas effect]{The Einstein--de Haas effect; (see, e.g.,~\cite{R24})}
Consider a cylinder of iron magnetized in the direction of its axis and suspended in such a way that it can freely rotate around its axis. Should this cylinder rotate, then it is advisable to treat the quantum theory of the electrons (and nuclei) in the iron in a rotating frame.\\
\indent Let $\vecV(\vec{y},\,t)$ be a (divergence-free) vector field on physical space that generates an incompressible flow $\phi_t\,:\,\mathbb{E}^3\rightarrow\mathbb{E}^3$ with the property that $\vec{y}=(y^1,\,y^2,\,y^3)$, given by
\bel{eq6.1}
\vec{y}=\phi^{-1}_t(\vec{x})\,,
\ee
are coordinates in the \emph{moving frame} at time $t$, with $\vec{x}=(x^1,\,x^2,\,x^3)$ the Cartesian laboratory coordinates. If $\vecV$ generates space rotations around a point $\vec{x}_0$ in space with a fixed angular velocity $\vec{\omega}$ then
\bel{eq6.2}
\vecV(\yv,\,t)=\vec{\omega}\wedge(\vec{y}-\vec{x_0})\,.
\ee
\indent The quantum theory of electrons in the moving frame is described by a (in general time-dependent) Hamiltonian
\begin{multline}\label{eq6.3}
H_{\vecV}^{(N)}=\,\sum_{j=1}^N\,\Big\lbrace\,\frac{m}{2}\left(\vec{\sigma}_j\cdot\vec{v}_j\right)^2+(\gmf-2)\frac{e}{2mc}\,\frac{\hbar}{2}\,\vec{\sigma}_j\cdot\vecB(\vec{y}_j,\,t)\\-\frac{e}{c}\vecA(\vec{y}_j,\,t)\cdot\vecV(\vec{y}_j,\,t)-\frac{m}{2}{\vecV(\vec{y}_j,\,t)}^2\Big\rbrace\\
+U_{\textrm{Coulomb}}\left(\phi_t(\vec{y}_1),\ldots,\,\phi_t(\vec{y}_N),\,\vecX_1,\ldots,\,\vecX_K\right)\,,
\end{multline}
where the velocity operators $\vec{v}_j$ are given by
\bel{eq6.4}
\vec{v}_j=\frac{\hbar}{m}\left(-\mathrm{i}\grad_j+\frac{e}{\hbar c}\vecA(\vec{y}_j,\,t)+\frac{m}{\hbar}\vecV(\vec{y}_j,\,t)\right)\,,
\ee
and $U_{\textrm{Coulomb}}$ is the total Coulomb potential of electrons and nuclei, expressed in laboratory coordinates. The term $-\frac{m}{2}{\vecV(\vec{y}_j,\,t)}^2$ appearing in~\eqref{eq6.3} is the potential of the \emph{centrifugal force} at the position $\vec{y}_j$ of the $j^{\textrm{th}}$ electron in the moving frame. We observe that in~\eqref{eq6.3} and~\eqref{eq6.4}
\bel{eq6.5}
\frac{e}{c}\vecA\hspace{0.3cm}\textrm{and}\hspace{0.3cm} m\vecV\,
\ee
play perfectly analogous r\^oles, at least if $\gmf=2$. As one will easily guess, $m\vecV$ is the vector potential generating the \emph{Coriolis force}, which can be obtained from the \emph{Lorentz force} by replacing $\frac{e}{c}\vecB=\frac{e}{c}\rotA$ by $m\rotV$. Note that
\begin{multline}\label{eq6.6}
\frac{m}{2}\left(\vec{\sigma}\cdot\vec{v}\right)^2+(\gmf-2)\frac{e}{2mc}\,\vecS\cdot\vecB\\
=\frac{1}{2m}\left(-\mathrm{i}\hbar\grad+\frac{e}{c}\vecA+m\vecV\right)^2+\frac{\gmf e}{2mc}\,\vecS\cdot\vecB+\vecS\cdot\vecOmega\,,
\end{multline}
where $\vecS=\frac{\hbar}{2}\vec{\sigma}$ is the spin operator of an electron and $\vecOmega=\rotV$ is twice the vorticity of $\vecV$.\\
\indent What we are describing here is the \emph{quantum-mechanical Larmor theorem}: (see, e.g.,~\cite{R24} for details).\\
\indent Let us now imagine that a magnetized iron cylinder is initially at rest in the laboratory frame. An experimentalist then turns on a constant external magnetic field $\vecB$ in the direction opposite to that of the spontaneous magnetization of the cylinder (parallel to its axis), so as to \emph{demagnetize} the cylinder. This causes an \emph{increase} in the free energy of the cylinder, which can be released in the form of mechanical energy. What is this mechanical energy? Well, the cylinder starts to rotate about its axis with an angular velocity $\vec{\omega}$ it chooses so as to cancel the effect of the external magnetic field $\vecB$ as best it can. By formula~\eqref{eq6.6}, the total Zeeman term in the electron Hamiltonian in the rotating frame, \emph{vanishes} if
\bel{eq6.7}
2\vec{\omega}=\vecOmega=-\frac{\gmf e}{2mc}\,\vecB\,
\ee
and the total vector potential affecting orbital motion of the electrons is then given by $\frac{e}{c}\vecA+m\vecV=\mathcal{O}(\gmf-2)\simeq 0$. The total Coulomb potential $U_{\textrm{Coulomb}}$ is invariant under the transformation $\vec{x}_j\rightarrow\vec{y}_j$, $\vecX_j\rightarrow\vec{Y}_j$. Thus, in the moving frame, the free energy of the electrons in a cylinder rotating with an angular velocity $\vec{\omega}$ given by~\eqref{eq6.7} is approximately the same as the free energy in the laboratory frame before the field $\vecB$ was turned on and $\vec{\omega}=0$. This explains the Einstein--de Haas effect. \\
\indent \emph{By measuring} $\vecB$ \emph{and} $\vec{\omega}$, \emph{one can determine} $\gmf$!\\
\indent The \emph{Barnett effect} describes the phenomenon that an iron cylinder can be magnetized by setting it into rapid rotation; (see~\eqref{eq6.6}).\\
\indent Other effects based on the same ideas are encountered in cyclotron physics, two-di\-men\-sio\-nal electron gases exhibiting the quantum Hall effect, molecular and nuclear physics; see~\cite{R24} and references given there.

\subsection{Accelerator measurement of $\gmf$}

Consider an electron circulating in an accelerator ring of radius $R$. It is kept in the ring by a constant external magnetic field $\vecB$ perpendicular to the plane of the ring. Its angular velocity $\vec{\omega}_C\parallel\vecB$ is found by balancing the centrifugal with the Lorentz force. Thus, its angular velocity is obtained by solving the equation
\bel{eq6.8}
|\vec{\omega}_C|=\frac{e}{\gamma mc}\,|\vecB|\,,
\ee
where $\gamma=\left(1-\frac{|\vec{v}|^2}{c^2}\right)^{-1/2}$, $|\vec{v}|=R\,|\vec{\omega}_C|$.\\
\indent This means that the velocity $\vec{v}$ of the electron precesses around the direction of $\vecB$ with an angular frequency $|\vec{\omega}_C|$ given by~\eqref{eq6.8}. What does its \emph{spin} $\vecS$ do? The precession of $\vecS$ around $\vecB$ is described by the so-called \emph{Bargmann--Michel--Telegdi} (BMT) \emph{equation}. In the special situation considered here, this equation simplifies to 
\bel{eq6.9}
\frac{\textrm{d}\vecS}{\textrm{d}t}=\frac{e}{mc}\,\vecS\wedge\left(\frac{\gmf-2}{2}+\frac{1}{\gamma}\right)\vecB\,,
\ee
see, e.g.,~\cite{R9}. Thus, the precession frequency of the spin is found to be
\bel{eq6.10}
\vec{\omega}_S=\frac{e}{\gamma mc }\,\vecB+\frac{e}{2mc}\,(\gmf-2)\,\vecB\,.
\ee
We find that, for $\gmf=2$, $\vec{\omega}_S=\vec{\omega}_C$; but if $\gmf\neq2$ the spin- and velocity precession frequencies differ by an amount
\bel{eq6.11}
\frac{e}{2mc}\,(\gmf-2)\,|\vecB|\,.
\ee
(If $\gmf>2$ then the spin precesses faster than the velocity.) By measuring the spin polarization of a bunch of electrons, with the property that, \emph{initially}, their spins were parallel to their velocities, after many circulations around the accelerator ring, one can determine $\gmf-2$ with very high accuracy.\\
\indent Of course, the formula for the Thomas precession encountered in section 2 can be found as an application of the general BMT equation. How watertight the derivation of the BMT equation is, mathematically, is still a matter of debate~\cite{R47}.\\

\subsection{Single-electron synchrotron measurement of $\gmf$}

Consider a single electron in a constant external magnetic field $\vecB=(0,\,0,\,B)$ in the $z$-direction whose motion in the $z$-direction is quantized by a confining (electrostatic) potential $\Phi(z)$. The time-independent Schr\"odinger equation for this particle is
\bel{eq6.12}
H^{(1)}\psi=E\psi\,,
\ee
where $\HO{1}$ is given by
\bel{eq6.13}
\HO{1}=\frac{\hbar^2}{2m}\left(-\mathrm{i}\grad+\frac{e}{\hbar c}\vecA(\vec{x})\right)^2+\frac{\gmf e}{2mc}\,S^{(3)}B+\Phi(z)\,,
\ee
where $\Av(\xv)=\frac{1}{2}(-yB,\,xB,\,0)$, $\xv=(x,\,y,\,z)$. Eq.~\eqref{eq6.12} can be solved by separating variables:
\begin{equation*}
\psi(x,\,y,\,z)=\chi(x,\,y)\,h(z)\,,
\end{equation*}
where $\chi$ is a two-component spinor wave function only depending on $x$ and $y$, and $h(z)$ is a scalar wave function satisfying
\bel{eq6.14}
\left(-\frac{\hbar^2}{2m}\,\frac{\textrm{d}^2}{\textrm{d}z^2}+\Phi(z)\right)\,h(z)=\mathcal{E}\,h(z)\,,
\ee
with $h\in L^2(\R,\,\textrm{d}z)$. Let $\mathcal{E}_0<\mathcal{E}_1<\mathcal{E}_2<\ldots$ be the energy eigenvalues of the eigenvalue problem~\eqref{eq6.14}. As shown by \emph{Lev Landau}, the energy spectrum of the operator $\HO{1}$ is then given by the energies
\bel{eq6.15}
E_{n,\,s,\,k}=\hbar\omega_C\left(n+\frac{1}{2}\right)+\frac{\gmf}{2}\hbar\omega_C\,s+\mathcal{E}_k\,,
\ee
where $\omega_C=\frac{|eB|}{mc}$, $n=0,\,1,\,2\,\ldots$, $s=\pm\frac{1}{2}$, $k=0,\,1,\,2\,\ldots$, and $\mathcal{E}_k$ as in~\eqref{eq6.14}. All these eigenvalues are infinitely degenerate. Their eigenfunctions corresponding to a degenerate energy level $E_{n,\,s,\,k}$ can be labeled by the eigenvalues of the $z$-component, $L_z$, of the orbital angular momentum operator, which are given by
\begin{equation*}
\hbar l,\textrm{ with }l=-n,\,-n+1,\ldots,\,0,\,1,\,2\ldots\,.
\end{equation*}
We observe that if $\gmf$ were exactly equal to 2 then
\bel{eq6.16}
E_{n,\,-\frac{1}{2},\,k}=E_{n-1,\,+\frac{1}{2},\,k}\,,
\ee
and
\begin{equation*}
E_{0,\,-\frac{1}{2},\,k}=\mathcal{E}_k\,.
\end{equation*}
These equations are an expression of the `supersymmetry' of Pauli's non-relativistic quantum theory of an electron with $\gmf=2$; (see section 3). If $\gmf\neq2$ this supersymmetry is broken, and we have that
\bel{eq6.17}
E_{m-1,\,\frac{1}{2},\,k}-E_{n,\,-\frac{1}{2},\,k}=\hbar\omega_C(m-n)+\frac{\gmf-2}{2}\,\hbar\omega_C\,.
\ee
By measuring such energy differences with great precision in very slow radiative transitions, one can determine $\gmf$ with astounding accuracy. The life times of the excited states can be made long, and hence the energy uncertainties tiny, by using cavities obeying non-resonance conditions. Very beautiful high-precision measurements of $\gmf$ based on these ideas have recently been performed by \emph{Gerald Gabrielse} and collaborators; see~\cite{R48}.


\section{KMS, Spin and Statistics, CPT}
In this last section, we study the general connection between the spin of particles and their quantum statistics~--~particles with half-integer spin are fermions, particles with integer spin are bosons~--~and the related connection between the spin of fields and their commutation relations within the framework of local relativistic quantum field theory. Our approach to this subject yields, as a byproduct, a proof of the celebrated CPT theorem, namely of the statement that the product of the discrete operations of charge conjugation $(C)$, space reflection $(P)$ and time reversal $(T)$ is an \emph{anti-unitary symmetry} of any local quantum field theory on an even-dimensional space-time. This symmetry maps \emph{states of matter} onto corresponding \emph{states of anti-matter}. Thus the prediction of the existence of the positron by Dirac and Weyl, on the basis of Dirac's hole theory, can be viewed, in hindsight, as a corollary of the locality of quantized Dirac theory and of the general CPT theorem\\
\indent I should like to mention that in a three-dimensional space-time, e.g., in the physics of two-dimensional electron gases exhibiting the quantum Hall effect, or of films, one may encounter (quasi-) particles with \emph{fractional spin} $\not\in\frac{1}{2}\Z$ and a type of `\emph{fractional}'\emph{quantum statistics} described by representations of the \emph{braid groups}, or braid groupoids (originally introduced in mathematics by \emph{Emil Artin}). Moreover, in two- and three-dimensional local quantum field theories, there are fields of fractional spin whose commutation relations give rise to representations of the braid groups or groupoids. It is conceivable that this exotic type of quantum statistics is relevant in the context of the fractional quantum Hall effect, and there are people who hope to exploit it for the purpose of (topological) \emph{quantum computing}.\footnote{An idea probably first suggested by myself.}\\
\indent It may be appropriate to make some sketchy remarks on the history of the discoveries of the connection between spin and statistics, of the CPT theorem and of braid statistics.\\

\indent The general connection between spin and statistics for free fields was discovered, on the basis of earlier work by Heisenberg and Pauli and by Pauli and Weisskopf, by \emph{Markus Fierz} in 1939,~\cite{R49}. His result was later rederived more elegantly by Pauli. In axiomatic field theory, a general result was found by L\"uders and Zumino; see~\cite{R50,R51}. A much more general analysis of the statistics of superselection sectors, based on the algebraic formulation of local quantum field theory, was carried out by \emph{Doplicher}, \emph{Haag} and \emph{Roberts}; see~\cite{R52,R53}. They showed that general para-Bose or para-Fermi statistics can always be converted into ordinary Bose or Fermi statistics by introducing `internal degrees of freedom' on which a compact topological group of internal symmetries acts, and they rederived the general connection between spin and statistics. All these results only hold in space-times of dimension $\ge4$.\\
\indent The CPT theorem, i.e., the statement that the product of $C$, $P$ and $T$ is an anti-unitary symmetry of any local, relativistic quantum field theory, was first derived in~\cite{R54} and then, in its general form, by \emph{Res Jost} in~\cite{R55}; see also~\cite{R50,R51}. Based on Jost's analysis and on the KMS condition~\cite{R56} characterizing thermal equilibrium states, it was rederived in a general setting by \emph{Bisognano} and \emph{Wichmann}~\cite{R57}, who established a connection with Tomita-Takesaki theory~\cite{R58}.\\
\indent We will see that the general connection between spin and statistics and the CPT theorem are consequences of the fact that the vacuum state of a local relativistic quantum field theory is a \emph{KMS} (equilibrium) \emph{state} for all one-parameter subgroups of the Poincar\'e group consisting of \emph{Lorentz boosts} in a two-dimensional plane containing a time-like direction. This observation has been made in~\cite{R57}. Incidentally, it is at the core of the theory of the \emph{Unruh effect}.\\
\indent Exotic commutation relations between fields carrying `fractional charges' in local relativistic quantum field theories with soliton sectors in \emph{two} space-time dimensions first appeared in work of \emph{R. Streater} and \emph{I. Wilde}~\cite{R59} and of the author~\cite{R60}, in the early seventies. (They gave rise to certain abelian representations of the braid groups.) In 1977, \emph{M. Leinaas} and \emph{J. Myrheim}~\cite{R61} discovered the first example of a system of quantum particles moving in the plane and exhibiting braid (or `fractional') statistics: Charged point particles carrying magnetic vorticity. The braid statistics of such particles is a consequence of the Aharonov-Bohm effect. Their analysis was generalized in~\cite{R62} and~\cite{R63}. Within the context of abelian gauge (Higgs) theories in three dimensions, particles with fractional spin and braid statistics were analyzed in~\cite{R64}. The general theory of (abelian and \emph{non-abelian}) braid statistics was initiated by the author in~\cite{R65} and completed in~\cite{R66,R67}, and references given there. A general connection between fractional spin and braid statistics was established in~\cite{R67}, and it was shown that, in local theories in three-dimensional space-time, ordinary Bose or Fermi statistics implies that all spins are integer of half-integer, and that braid statistics implies the breaking of parity $(P)$ and time reversal $(T)$.

\subsection{SSC, KMS and CPT}

I will now first recall the \emph{connection between spin and statistics} (SSC) in the general framework of \emph{local relativistic quantum field theory} (RQFT), as formalized in the so-called \emph{(G\aa r\-ding-) Wightman axioms}~\cite{R50,R51}; (see also~\cite{R511}). As a corollary, I will then show that the vacuum state of an arbitrary local RQFT is a \emph{KMS} (equilibrium) \emph{state}~\cite{R56} for any one-parameter group of Lorentz boosts at inverse temperature $\beta=2\pi$,~\cite{R57}. The \emph{CPT theorem} and \emph{SSC} turn out to be consequences of the KMS condition.\\
\indent I will follow methods first introduced in~\cite{R55,R57}, and my presentation is similar to that in~\cite{R68}, where various mathematical details can be found.\\

\indent We consider a local RQFT on Minkowski space $\mathbb{M}^d,\,(d=n+1)$, at zero temperature satisfying the Wightman axioms~\cite{R50,R51}. Let $\Hilbert$ denote the Hilbert space of pure state vectors of the theory and $\Omega\in\Hilbert$ the \emph{vacuum vector}. The space $\Hilbert$ carries a projective, unitary representation, $U$, of $\mathcal{P}_{+}^{\uparrow}$. We first consider RQFT's with fields localizable in points and transforming covariantly under the adjoint action of $U$; a more general framework is considered in the next subsection, (see~\cite{R69} for a general analysis of the localization properties of fields). Let $\Psi_1,\ldots,\,\Psi_N$ be the fields of the theory. Smearing out these fields with test functions in the Schwartz space over $\mathbb{M}^d$, one obtains operators densely defined on $\Hilbert$. In fact, $\Hilbert$ turns out to be the norm-closure of the linear space obtained by applying arbitrary polynomials in $\Psi_1,\ldots,\,\Psi_N$ (smeared out with Schwartz space test functions) to the vacuum $\Omega$. Let $\Pi\subset\mathbb{M}^d$ be a two-dimensional plane containing a time-like direction. Without loss of generality, we can choose coordinates $x^0,\,x^1,\ldots,\,x^{d-1}$ in $\mathbb{M}^d$ such that $\Pi$ is the $(x^0,\,x^1)$-coordinate plane. We consider the one-parameter subgroup of Lorentz boosts given by
\begin{align}\label{eq7.1}
x_{\theta}^0\,=&\cosh(\theta)x^0+\sinh(\theta)x^1\,,\nonumber\\
x_{\theta}^1\,=&\sinh(\theta)x^0+\cosh(\theta)x^1\,,\\
x_{\theta}^j\,=&\,x^j\,,\,\textrm{ for }j\ge2\,,\nonumber
\end{align}
with $\theta\in\R$ the rapidity of the boost. Let $M_{\Pi}=M_{\Pi}\adj$ denote the generator of the boosts~\eqref{eq7.1} in the projective, unitary representation $U$ of $\mathcal{P}_{+}^{\uparrow}$ on $\Hilbert$. To each field $\Psi_j$ of the theory, there is associated a finite-dimensional, irreducible projective representation $S_j$ of the group $\mathrm{L}_+^{\uparrow}$ of proper, orthochronous Lorentz transformations of $\mathbb{M}^d$ such that
\bel{eq7.2}
\mathrm{e}^{\mathrm{i}\theta M_{\Pi}}\,\Psi_j(x^0,\,x^1,\,\xv)\,\mathrm{e}^{-\mathrm{i}\theta M_{\Pi}}=S_j^{-1}(\theta)\,\Psi_j(x^0_{\theta},\,x^1_{\theta},\,\xv)\,,
\ee
with $\xv=(x^2,\ldots,\,x^{d-1})$, or, in components,
\bel{eq7.3}
\mathrm{e}^{\mathrm{i}\theta M_{\Pi}}\,\Psi_j^{A}(x^0,\,x^1,\,\xv)\,\mathrm{e}^{-\mathrm{i}\theta M_{\Pi}}=\sum_{B}\,S_j^{-1}(\theta)^{A}_{\phantom{A}B}\,\Psi_j^{B}(x^0_{\theta},\,x^1_{\theta},\,\xv)\,,
\ee
where $\Psi_j^{A}$ is the $A^{\textrm{th}}$ component of $\Psi_j$.\\
\indent A theorem due to \emph{Bargmann, Hall} and \emph{Wightman}~\cite{R50,R51} guarantees that, for an RQFT satisfying the Wightman axioms, the \emph{Wick rotation} from real times to purely \emph{imaginary times} $ct=\mathrm{i}\tau$, $\tau\in\R$, is \emph{always possible}. The vacuum vector $\Omega$ turns out to be in the domain of all the operators $\prod_{k=1}^n\,\op{\Psi}_{j_k}(x_k)$, where  $x_k=(\tau_k,\,x^1_k,\,\xv)\in\mathbb{E}^d$ ($d$-dim. Euclidean space),
\bel{eq7.4}
\op{\Psi}_j(\tau,\,x^1,\,\xv)\deq\Psi_j(\mathrm{i}\tau,\,x^1,\,\xv)=\mathrm{e}^{-\tau H}\,\Psi_j(0,\,x^1,\,\xv)\,\mathrm{e}^{\tau H}\,,
\ee
with $H\ge0$ the \emph{Hamiltonian} of the theory, provided that
\bel{eq7.5}
0<\tau_1<\tau_2<\ldots<\tau_n\,;
\ee
see~\cite{R50,R70}. The Euclidian Green- or \emph{Schwinger functions} are then defined by
\bel{eq7.6}
S^{(n)}(j_1,\,x_1,\ldots,\,j_n,\,x_n)\deq\scalarb{\Omega}{\op{\Psi}_{j_1}(x_1)\cdots\op{\Psi}_{j_n}(x_n)\,\Omega}\,.
\ee
By Bargmann-Hall-Wightman, the Schwinger functions $S^{(n)}$ are defined on all of
\bel{eq7.7}
\mathbb{E}^{dn}_{\not =}\deq\Big\lbrace (x_1,\ldots,\,x_n)\,\Big|\,x_j\in\E^d,\;j=1,\ldots,\,n,\;x_i\not= x_j,\textrm{ for }i\not = j\Big\rbrace\,.
\ee
It is convenient to introduce polar coordinates, $(\alpha,\,r,\,\xv)$, with $r>0$, $\alpha\in\lbrack 0,2\pi)$, in the $(\tau,\,x^1)$-plane by setting
\bel{eq7.8}
\tau=r\sin\alpha,\;x^1=r\cos\alpha,\;\xv=(x^2,\ldots,\,x^{d-1})\,;
\ee
(the angle $\alpha$ is an imaginary rapidity).\\
\indent
Let $\mathscr{S}_+$ denote the Schwartz space of test functions $f(r,\,\xv)$ with support in $\R_+\times\R^{d-2}$. We define functions $\Phi^{(n)}$ of $n$ angles as follows:
\begin{multline}\label{eq7.9}
\Phi^{(n)}(j_1,\,f_1,\,\alpha_1,\ldots,\,j_n,\,f_n,\,\alpha_n)\deq\\
\int\,S^{(n)}(j_1,\,\alpha_1,\,r_1,\,\xv_1,\ldots,\,j_n,\,\alpha_n,\,r_n,\,\xv_n)\prod_{k=1}^n\,f_k(r_k,\,\xv_k)\,\textrm{d}r_k\,\textrm{d}^{d-2}x_k\,.
\end{multline}
As shown in~\cite{R68} (see  also~\cite{R70}), using Bargmann-Hall-Wightman (see~\eqref{eq7.6},~\eqref{eq7.7}) -- among other things -- these functions are given by
\begin{equation}\label{eq7.10}
\Phi^{(n)}(j_1,\,f_1,\,\alpha_1,\ldots,\,j_n,\,f_n,\,\alpha_n)
=\scalarb{\Omega}{\op{\Psi}_{j_1}(f_j,\,\alpha_1)\cdots\op{\Psi}_{j_n}(f_n,\,\alpha_n)\,\Omega}\,,
\end{equation}
provided $\alpha_1<\alpha_2<\ldots<\alpha_n$, with $\alpha_n-\alpha_1<2\pi$. On the R.S. of~\eqref{eq7.10},
\bel{eq7.11}
\op{\Psi}_j(f,\,\alpha+\beta)=\mathrm{e}^{-\alpha M_{\Pi}}\,R_j(\alpha)\,\op{\Psi}_j(f,\,\beta)\,\mathrm{e}^{\alpha M_{\Pi}}\,,
\ee
for arbitrary angles $\alpha>0$, $\beta\ge0$ with $\alpha+\beta<\pi$, where
\bel{eq7.12}
R_j(\alpha)\deq S_j(\mathrm{i}\alpha)
\ee
is the finite-dimensional, irreducible representation of $\mathrm{Spin}(d)$ obtained from $S_j$ by analytic continuation in the rapidity. Formally, ~\eqref{eq7.10} and~\eqref{eq7.11} follow from~\eqref{eq7.2},~\eqref{eq7.3} and~\eqref{eq7.6}; (the details required by mathematical rigor are a little complicated; but see~\cite{R57,R68}). We note that the vacuum $\Omega$ is invariant under Poincar\'e transformations; in particular
\bel{eq7.13}
\mathrm{e}^{\mathrm{i}\theta M_{\Pi}}\,\Omega=\Omega\,,\;\textrm{ for all }\theta\in\C\,.
\ee
We also note that two points $(\alpha_1,\,r_1,\,\xv_1)$ and $(\alpha_2,\,r_2,\,\xv_2)$ in $\E^d$ are \emph{space-like separated} whenever $\alpha_1\not =\alpha_2$. Thus, the local commutation relations of fields at space-like separated points~\cite{R50,R51,R70} imply that, for $\alpha_k\neq\alpha_{k+1}$,
\begin{multline}\label{eq7.14}
\Phi^{(n)}(\ldots,\,j_k,\,f_k,\,\alpha_k,\,j_{k+1},\,f_{k+1},\alpha_{k+1},\ldots)\\
=\exp(\mathrm{i} 2\pi\theta_{j_k\,j_{k+1}})\,\Phi^{(n)}(\ldots,\,j_{k+1},\,f_{k+1},\,\alpha_{k+1},\,j_{k},\,f_{k},\alpha_{k},\ldots)\,,
\end{multline}
for arbitrary $1\leq k<n$, where, for $d\geq 4$,
\begin{align}
\theta_{j\,j'}&=0\textrm{ mod }\Z\textrm{ if }\Psi_j\textrm{ or }\Psi_{j'}\textrm{ is a \emph{Bose field}}\,,\label{eq7.15}\\
\theta_{j\,j'}&=\frac{1}{2}\textrm{ mod }\Z\textrm{ if }\Psi_j\textrm{ \emph{and} }\Psi_{j'}\textrm{ are \emph{Fermi fields}}\,.\label{eq7.16}
\end{align}
For details see~\cite{R70} and~\cite{R71}. In two space-time dimensions, the statistics of fields localizable in points can be more complicated; see subsection 7.2, and~\cite{R65,R66,R67}. In particular, the phases $\theta_{j\,j'}$ can be arbitrary real numbers, and this is related to the fact that $\mathrm{Spin}(2)=\widetildeSO=\R$, which implies that the \emph{spin} (parity) $s_j$ of a field $\Psi_j$ can be an arbitrary real number. The \emph{spin} (parity) $s_j$ of a field $\Psi_j$ is defined as follows: Since $R_j$ is a finite-dimensional, irreducible representation of $\mathrm{Spin}(d)$,
\bel{eq7.17}
R_j(2\pi)=\mathrm{e}^{\mathrm{i}2\pi s_j}\,\umat\,,
\ee
where $s_j=0,\,\frac{1}{2}$ mod $\Z$, for $d\geq3$, while $s_j\in\lbrack 0,\,1)$ mod $\Z$, for $d=2$.\\
\indent Given a field index $j$, we define the `adjoint' index $\jadj$ through the equation
\bel{eq7.18}
{\big(\Psi_j^B(g)\big)}\adj=\Psi_{\jadj}^B(\overline{g}),\;g\in\mathscr{S}(\mathbb{M}^d)\,,
\ee
where $A\adj$ is the adjoint of the operator $A$ on $\Hilbert$ in the scalar product of $\Hilbert$.\\

\indent We are now prepared to prove the general \emph{spin-statistics-connection} (SSC) for fields of a local RQFT localizable in space-time points. We first note that, by~\eqref{eq7.11} and~\eqref{eq7.18},
\begin{align*}
\op{\Psi}_j(f,\,\alpha)\adj&={\left(\mathrm{e}^{-\alpha M_{\Pi}}R_j(\alpha)\,\op{\Psi}_j(f,\,0)\,{\mathrm{e}}^{\alpha M_{\Pi}}\right)}\adj\\
&={\mathrm{e}}^{\alpha M_{\Pi}}{R_j(\alpha)}\adj\,\op{\Psi}_{\jadj}(\barfunction{f},\,0)\,{\mathrm{e}}^{-\alpha M_{\Pi}}\\
&={R_j(\alpha)}\adj R_{\jadj}^{-1}(-\alpha)\,\op{\Psi}_{\jadj}(\barfunction{f},\,-\alpha)\\
&\stackrel{!}{=}\op{\Psi}_{\jadj}(\barfunction{f},\,-\alpha)\,,
\end{align*}
by~\eqref{eq7.2},~\eqref{eq7.3} and~\eqref{eq7.18}. Thus
\bel{eq7.19}
{R_j(\alpha)}\adj=R_{\jadj}(-\alpha)\,.
\ee
Furthermore, by~\eqref{eq7.3},~\eqref{eq7.11} and~\eqref{eq7.13},
\begin{eqnarray}\label{eq7.20}
\lefteqn{\Phi^{(n)}(j_1,\,f_1,\,\alpha_1+\alpha,\ldots,\,j_n,\,f_n,\,\alpha_n+\alpha)\nonumber}\hspace{1cm}\\
& &=\scalarB{\Omega}{\prod_{k=1}^n\,\op{\Psi}_{j_k}(f_k,\,\alpha_k+\alpha)\,\Omega}\nonumber\\
& &=\scalarB{\Omega}{\prod_{k=1}^n\,\left({\mathrm{e}}^{-\alpha M_{\Pi}}\,R_{j_k}(\alpha)\op{\Psi}_{j_k}(f_k,\,\alpha_k)\,{\mathrm{e}}^{\alpha M_{\Pi}}\right)\,\Omega}\nonumber\\
& &=R_{j_1}(\alpha)\otimes\cdots\otimes R_{j_n}(\alpha)\,\Phi^{(n)}(j_1,\,f_1,\,\alpha_1,\ldots,\,j_n,\,f_n,\,\alpha_n)\,,
\end{eqnarray}
which expresses the rotation covariance of the functions $\Phi^{(n)}$, (a consequence of the \emph{Poincar\'e covariance} of the fields $\Psi_j$ and the \emph{Poincar\'e invariance} of the vacuum $\Omega$). Thus, using the positivity of the scalar product $\scalar{\cdot}{\cdot}$ on $\Hilbert$, we find that, for $0<\alpha<\pi$,
\begin{eqnarray}
\lefteqn{0<\scalarb{{\mathrm{e}}^{-\alpha M_{\Pi}}\,\op{\Psi}_j(f,\,0)\,\Omega}{{\mathrm{e}}^{-\alpha M_{\Pi}}\,\op{\Psi}_j(f,\,0)\,\Omega}\nonumber }\\
&\overset{\eqref{eq7.10},\eqref{eq7.11}}=&R_{\jadj}^{-1}(-\alpha)\otimes R_{j}^{-1}(\alpha)\,\Phi^{(2)}(\jadj,\,\barfunction{f},\,-\alpha,\,j,\,f,\,\alpha)\nonumber\\
&\overset{\eqref{eq7.14},\eqref{eq7.19}}=&R_j^{-1}(\alpha)\otimes R_{\jadj}^{-1}(-\alpha)\,{\mathrm{e}}^{\mathrm{i}2\pi\theta_{\jadj\,j}}\,\Phi^{(2)}(j,\,f,\,\alpha,\,\jadj,\,\barfunction{f},\,-\alpha)\nonumber\\
&=&R_j^{-1}(\alpha)\otimes R_{\jadj}^{-1}(-\alpha)\,{\mathrm{e}}^{\mathrm{i}2\pi\theta_{\jadj\,j}}\,\Phi^{(2)}(j,\,f,\,\alpha,\,\jadj,\,\barfunction{f},\,2\pi-\alpha)\nonumber\\
&\overset{\eqref{eq7.20}}=&R_j^{-1}(\alpha-\pi)\otimes R_{\jadj}^{-1}(-\alpha-\pi)\,{\mathrm{e}}^{\mathrm{i}2\pi\theta_{\jadj,j}}\,\Phi^{(2)}(j,\,f,\,\alpha-\pi,\,\overline{j},\,\barfunction{f},\,\pi-\alpha)\nonumber\\
&\overset{\eqref{eq7.17}}=&{\mathrm{e}}^{\mathrm{i}2\pi\theta_{\jadj\,j}}\,{\mathrm{e}}^{\mathrm{i}2\pi s_{\jadj}}\,R_j^{-1}(\alpha-\pi)\otimes R_{\jadj}^{-1}(\pi-\alpha)\,\Phi^{(2)}(j,\,f,\,\alpha-\pi,\,\jadj,\,\barfunction{f},\,\pi-\alpha)\nonumber\\
&\overset{\eqref{eq7.11}}=&{\mathrm{e}}^{\mathrm{i}2\pi\theta_{\jadj\,j}}\,{\mathrm{e}}^{\mathrm{i}2\pi s_{\jadj}}\,\scalarb{{\mathrm{e}}^{(\alpha-\pi)M_{\Pi}}\,\op{\Psi}_{\jadj}(\barfunction{ f},\,0)\,\Omega}{{\mathrm{e}}^{(\alpha-\pi)M_{\Pi}}\,\op{\Psi}_{\jadj}(\barfunction{f},\,0)\,\Omega}\label{eq7.21}\,.
\end{eqnarray}
Note that the L.S. and the scalar product ($3^{\textrm{rd}}$ factor) on the very R.S. of~\eqref{eq7.21} are well defined and \emph{strictly positive}, for $0<\alpha<\pi$. It then follows that 
\bel{eq7.22}
s_j=-s_{\jadj}=\theta_{\jadj\,j}\textrm{ mod }\Z\,,
\ee
which is the usual connection between spin and statistics:
\begin{align}
s_j\textrm{ \emph{half-integer} }&\longleftrightarrow\,\Psi_j\textrm{ a \emph{Fermi field}}\,,\nonumber\\
s_j\textrm{ \emph{integer} }&\longleftrightarrow\,\Psi_j\textrm{ a \emph{Bose field}}\,,\label{eq7.23}
\end{align}
and, for $d=2$,
\begin{equation*}
s_j\textrm{ \emph{fractional} }\longleftrightarrow\,\Psi_j\textrm{ a field with \emph{fractional} (braid) \emph{statistics}.}\\
\end{equation*}

\indent Next, we show that our results imply that the vacuum $\Omega$ is a \emph{KMS state} at inverse temperature $\beta=2\pi$ for the one-parameter group of Lorentz boosts in the plane $\Pi$.\\
\indent We consider the Schwinger function
\bel{eq7.24}
\Phi^{(n)}(j_1,\,f_1,\,\alpha_1,\ldots,\,j_n,\,f_n,\,\alpha_n)=\scalarB{\Omega}{\prod_{k=1}^n\,\op{\Psi}_{j_k}(f_k,\,\alpha_k)\,\Omega}\,,
\ee
for $\alpha_1<\cdots<\alpha_n$, with $\alpha_n-\alpha_1<2\pi$. For simplicity, we assume that $d\geq3$, so that all spins are half-integer or integer and, by~\eqref{eq7.22}, only Fermi- or Bose statistics is possible. Then $\Phi^{(n)}(j_1,\,f_1,\,\alpha_1,\ldots,\,j_n,\,f_n,\,\alpha_n)$ vanishes, unless an \emph{even} number of the fields $\Psi_{j_1},\ldots,\,\Psi_{j_n}$ are Fermi fields. For every $1\leq m<n$, we define the phase
\bel{eq7.25}
\varphi_m=\sum_{\scriptsize{\begin{array}{c} k=1,\ldots,\,m \\ 
l=m+1,\ldots,\,n \end{array}}}\theta_{j_k\,j_l}\,,
\ee
with $\theta_{j_k\,j_l}$ as in~\eqref{eq7.14}.\\
\indent Using eqs.~\eqref{eq7.15} and~\eqref{eq7.16} and the fact that the total number of Fermi fields among $\Psi_{j_1},\ldots,\Psi_{j_n}$ is even, one easily deduces from the spin statistics connection~\eqref{eq7.23} that
\bel{eq7.26}
\varphi_m=\sum_{k=1}^m\,s_{j_k}\textrm{ mod }\Z\,.
\ee
\indent Next, by repeated use of~\eqref{eq7.14}, we find that 
\begin{eqnarray}
\lefteqn{\Phi^{(n)}(j_1,\,f_1,\,\alpha_1,\ldots,\,j_n,\,f_n,\,\alpha_n)\nonumber}\\
&=&{\mathrm{e}}^{\mathrm{i}2\pi\varphi_m}\,\Phi^{(n)}(j_{m+1},\,f_{m+1},\,\alpha_{m+1},\ldots,\,j_n,\,f_n,\,\alpha_n,\,j_1,\,f_1,\,\alpha_1,\ldots,\,j_m,\,f_m,\,\alpha_m)\nonumber\\
&\overset{\eqref{eq7.26}}=&\exp{\Big(\mathrm{i}2\pi\sum_{k=1}^m\,s_{j_k}\Big)}\,\Phi^{(n)}(j_{m+1},\,f_{m+1},\,\alpha_{m+1},\ldots,\,j_1,\,f_1,\,\alpha_1,\ldots)\nonumber\\
&\overset{\eqref{eq7.17}}=&\umat\otimes\cdots\otimes\umat\otimes R_{j_1}(2\pi)\otimes\cdots\otimes R_{j_m}(2\pi)\nonumber\\
&&\hspace{0.3cm}\cdot\,\Phi^{(n)}(j_{m+1},\,f_{m+1},\,\alpha_{m+1},\ldots,\,j_1,\,f_1,\,\alpha_1+2\pi,\ldots,\,j_m,\,f_m,\,\alpha_m+2\pi)\,.\nonumber\\
\label{eq7.27}
\end{eqnarray}
Note that $\alpha_{m+1}<\ldots<\alpha_n<\alpha_1+2\pi<\ldots<\alpha_{2m}+2\pi$, with $\alpha_m+2\pi-\alpha_{m+1}<2\pi)\;(\Leftrightarrow\alpha_m<\alpha_{m+1})$. Thus, by~\eqref{eq7.24} (applied to the L.S. and the R.S. of~\eqref{eq7.27}), we arrive at the identity
\begin{multline}\label{eq7.28}
\scalarB{\Omega}{\prod_{k=1}^m\,\op{\Psi}_{j_k}(f_k,\,\alpha_k)\,\prod_{l=m+1}^n\,\op{\Psi}_{j_l}(f_l,\,\alpha_l)\,\Omega}\\
=\scalarB{\Omega}{\prod_{l=m+1}^n\,\op{\Psi}_{j_l}(f_l,\,\alpha_l)\prod_{k=1}^m\,\left({\mathrm{e}}^{-2\pi M_{\Pi}}\,\op{\Psi}_{j_k}(f_k,\,\alpha_k)\,{\mathrm{e}}^{2\pi M_{\Pi}}\right)\Omega}\,,
\end{multline}
which is the celebrated \emph{KMS condition}.\\
\indent Defining
\bel{eq7.29}
\omega(A)\deq\scalar{\Omega}{A\,\Omega}\,,
\ee
and
\bel{eq7.30}
\tau_{\theta}(A)\deq {\mathrm{e}}^{\mathrm{i}\theta M_{\Pi}}\,A\,{\mathrm{e}}^{-\mathrm{i}\theta M_{\Pi}}\,,
\ee
with $(\tau_{\theta}(A))\adj=\tau_{\theta}(A\adj)$ and $\tau_{\theta}(A_1\cdot A_2)=\tau_{\theta}(A_1)\tau_{\theta}(A_2)$, where $A,\,A_1,\,A_2$ are operators on $\Hilbert$, we find, setting
\begin{equation*}
\prod_{k=1}^m\,\op{\Psi}_{j_k}(f_k,\,\alpha_k)\eqd B\,,
\end{equation*}
and
\begin{equation*}
\prod_{l=m+1}^n\,\op{\Psi}_{j_l}(f_l,\,\alpha_l)\eqd C\,,
\end{equation*}
that
\begin{align}
\omega(B\cdot C)=&\,\omega(C\tau_{2\pi \mathrm{i}}(B))\nonumber\\
=&\,\omega(\tau_{-2\pi \mathrm{i}}(C)B)\,,\label{eq7.31}
\end{align}
a more familiar form of the KMS condition for $(\omega,\,\tau_{\theta})$ at inverse temperature $\beta=2\pi$; see~\cite{R56}.\\
\indent It deserves to be noticed that the KMS condition~\eqref{eq7.28},~\eqref{eq7.31} implies the spin-statistics connection. We calculate formally: For $0<\epsilon<\pi$,
\begin{eqnarray}
\lefteqn{\omega\big(\op{\Psi}_{j_1}(f_1,\,0)\op{\Psi}_{j_2}(f_2,\,\epsilon)\big)\nonumber}\hspace{0.5cm}\\
&\overset{\textrm{KMS},\eqref{eq7.11}}=&{\mathrm{e}}^{-\mathrm{i}2\pi s_{j_2}}\,\omega\big(\op{\Psi}_{j_2}(f_2,\,2\pi+\epsilon)\op{\Psi}_{j_1}(f_1,\,0)\big)\nonumber\\
&\overset{\eqref{eq7.14}}=&{\mathrm{e}}^{-\mathrm{i}2\pi s_{j_2}}\,{\mathrm{e}}^{\mathrm{i}2\pi\theta_{j_1\,j_2}}\,\omega\big(\op{\Psi}_{j_1}(f_1,\,0)\op{\Psi}_{j_2}(f_2,\,2\pi+\epsilon)\big)\nonumber\\
&\overset{\eqref{eq7.11}}=&{\mathrm{e}}^{-\mathrm{i}2\pi s_{j_2}}\,{\mathrm{e}}^{\mathrm{i}2\pi\theta_{j_1\,j_2}}\,\omega\big(\op{\Psi}_{j_1}(f_1,\,0)\op{\Psi}_{j_2}(f_2,\,\epsilon)\big)\label{eq7.32}\,.
\end{eqnarray}
Thus,
\bel{eq7.33}
s_{j_2}=\theta_{j_1\,j_2}\textrm{ mod }\Z\,,
\ee
unless $\omega(\op{\Psi}_{j_1}(f_1,\,0)\op{\Psi}_{j_2}(f_2,\,\epsilon))\equiv 0$. If this quantity does \emph{not} vanish (and in $d\geq 3$) then either $\Psi_{j_1}$ and $\Psi_{j_2}$ are both Fermi fields ($\theta_{j_1\,j_2}=\frac{1}{2}\,\textrm{mod}\,\Z$) or they are both Bose fields ($\theta_{j_1\,j_2}=0\,\textrm{mod}\,\Z$). Thus,~\eqref{eq7.33} proves (a special case of ) SSC!\\

\indent It turns out that the \emph{CPT theorem} (for $d$ \emph{even}) is a direct consequence of the KMS condition~\eqref{eq7.31}. This claim can be viewed as a corollary of the general Tomita-Takesaki theory~\cite{R58}. But, in our concrete context, it is easy to directly define an \emph{anti-unitary involution} $J$ acting on $\Hilbert$, which, thanks to the KMS condition~\eqref{eq7.31}, turns out to be a symmetry of the theory: We define
\bel{eq7.34}
B\deq\op{\Psi}_{j_1}(f_1,\,\alpha_1)\cdots\op{\Psi}_{j_n}(f_n,\,\alpha_n)\,,
\ee
with $0<\alpha_1<\ldots<\alpha_n<\pi$, and
\bel{eq7.35}
C\deq\op{\Psi}_{l_1}(g_1,\,\beta_1)\cdots\op{\Psi}_{l_m}(g_m,\,\beta_m)\,,
\ee
with $0<\beta_1<\ldots<\beta_n<\pi$. We define
\bel{eq7.36}
JB\,\Omega\deq {\mathrm{e}}^{-\pi M_{\Pi}}B\adj\,\Omega\,,
\ee
or
\begin{eqnarray}
\lefteqn{J\op{\Psi}_{j_1}(f_1,\,\alpha_1)\cdots\op{\Psi}_{j_n}(f_n,\,\alpha_n)\,\Omega\nonumber}\hspace{1cm}\\
&=&{\mathrm{e}}^{-\pi M_{\Pi}}\,\op{\Psi}_{\jadj_n}(\barfunction{f}_n,\,-\alpha_n)\cdots\op{\Psi}_{\jadj_1}(\barfunction{f}_1,\,-\alpha_1)\,\Omega\nonumber\\
&=&R_{\jadj_n}^{-1}(\pi)\,\op{\Psi}_{\jadj_n}(\barfunction{f}_n,\,\pi-\alpha_n)\cdots R_{\jadj_1}^{-1}(\pi)\,\op{\Psi}_{\jadj_1}(\barfunction{f}_1,\,\pi-\alpha_1)\,\Omega\label{eq7.37}\,,
\end{eqnarray}
with $0<\pi-\alpha_n<\pi-\alpha_{n-1}<\ldots<\pi-\alpha_1<\pi$. By analytic continuation of~\eqref{eq7.37} in the angles $\alpha_1,\ldots,\,\alpha_n$ to the imaginary axis, see~\cite{R68}, we see that $J$ has the interpretation of the product $CP_1T$, where $P_1$ is the space reflection $x^1\mapsto -x^1,\,\xv\mapsto\xv,\,\xv=(x^2,\,\ldots,\,x^{d-1})$; (geometrically, the action of $J$ only involves a reflection in the plane $\Pi$). Using~\eqref{eq7.36}, we find that
\begin{eqnarray}
\scalarb{JC\,\Omega}{JB\,\Omega}&\overset{\eqref{eq7.36}}=&\scalarb{{\mathrm{e}}^{-\pi M_{\Pi}}\,C\adj\Omega}{{\mathrm{e}}^{-\pi M_{\Pi}}\,B\adj\Omega}\nonumber\\
&=&\scalarb{{\mathrm{e}}^{-2\pi M_{\Pi}}C\adj\Omega}{B\adj\Omega}\nonumber\\
&=&\omega(\tau_{-2\pi \mathrm{i}}(C)B\adj)\nonumber\\
&\overset{\eqref{eq7.31}}=&\omega(B\adj C)\nonumber\\
&=&\scalarb{B\,\Omega}{C\,\Omega}\,,\nonumber
\end{eqnarray}
which tells us that $J$ is \emph{anti-unitary}. Moreover, 
\begin{align}
J(JB\,\Omega)&=J\left({\mathrm{e}}^{-\pi M_{\Pi}}\,B\adj {\mathrm{e}}^{\pi M_{\Pi}}\right)\Omega\nonumber\\
&={\mathrm{e}}^{-\pi M_{\Pi}}\left({\mathrm{e}}^{\pi M_{\Pi}}\,B {\mathrm{e}}^{-\pi M_{\Pi}}\right)\Omega\nonumber\\
&=B\,\Omega\nonumber\,,
\end{align}
i.e., $J$ is an \emph{involution}.\\
\indent In \emph{even} space-time dimension, the product $P_1P$, where $P$ is space reflection, has determinant $=1$ and can be represented as a space rotation. Hence $P_1P$ is a \emph{symmetry} of the theory. It follows that the CPT operator $\Theta$ defined by
\bel{eq7.38}
\Theta\deq JP_1P
\ee
is an \emph{anti-unitary symmetry} of the theory. This is the celebrated \emph{CPT theorem}~\cite{R55}. In a space-time of \emph{odd} dimension, the operators $J_j=CP_jT,\,j=1,\ldots,\,d-1$ are always anti-unitary symmetries, but, in general, $\Theta$ is \emph{not} a symmetry.\\
\indent For an analysis of SSC and CPT for local RQFT's on a class of curved space-time manifolds with `large' groups of Killing symmetries (Schwarzschild, de Sitter, AdS), see, e.g.,~\cite{R68}.\\

\indent I conclude my discussion with a result due to \emph{Steven Weinberg} and \emph{Edward Witten},~\cite{R80}: In a four-di\-men\-sio\-nal local RQFT \emph{without gravity}, but with well defined current- and charge density operators, there are no massless charged (asymptotic) particles of spin $>\frac{1}{2}$; and there are no massless (asymptotic) particles of spin $>1$ if the theory admits a well defined energy-momentum tensor.

\subsection[Braid statistics in two and three space-time dimensions and SSC]{Braid statistics in two and three space-time dimensions and SSC\footnote{Sources for this section are~\cite{R52,R53,R65,R66,R67,R69,R72,R73}.}}

Two-dimensional electron gases in a transversal external magnetic field exhibiting the fractional quantum Hall effect appear to be examples of quantum-mechanical systems with fractionally charged quasi-particles having fractional spin $s\not\in\frac{1}{2}\Z$ and fractional or braid statistics; see, e.g.,~\cite{R74,R75}, and references given there. The analysis of these particles is important in order to calculate, e.g., the value of the Hall conductivity $\upsigma_H$ (a rational multiple of $\frac{e^2}{h}$). Certain systems exhibiting the fractional quantum Hall effect (e.g., the ones with $\upsigma_H=\frac{5}{2}\,\frac{e^2}{h}$) are believed to be of interest for purposes of quantum computation. All this is quite fascinating and has been among my more serious scientific interests in the 1990's. Thus, it would have been tempting to give a rather detailed account of the theory of planar systems exhibiting fractional electric charges, fractional spin and fractional or braid statistics.\\
\indent However, after much agonizing,  I have come to the conclusion that it is impossible to give an account of fractional spin and braid statistics that is \emph{accurate} (mathematically precise), \emph{comprehensible}, and \emph{short}. I therefore decided, with considerable regrets, to limit my account of these matters to some very sketchy remarks. \\

\indent The pure physical states of a quantum-mechanical system with infinitely many degrees of freedom at \emph{zero temperature}, described, e.g., by a local RQFT, fall into different irreducible (`simple') \emph{superselection sectors}. These sectors are invariant under the action of operators corresponding to local observable quantities (`measurements') of the theory. (The action of the algebra of all `local observables' on every superselection sector of the theory is usually irreducible.) Superpositions of states from different superselection sectors are therefore \emph{incoherent:} Their relative phases are not observable, and interference terms vanish (`decoherence').\\
\indent Let $I=\lbrace e,\,2,\,3,\ldots,\,N\rbrace$, $N\leq\infty$, be a set of indices labeling the different irreducible superselection sectors of such a system, with $e$ labeling the sector containing the \emph{ground state} (or vacuum) $\Omega$ of the system. Let $U_j$, $j\in I$, denote the unitary representation of the quantum-mechanical rotation group $\mathrm{Spin}(d-1)$ on (the Hilbert space $\Hilbert_j$ of pure states corresponding to) the superselection sector $j$. Since the algebra of local observables is assumed to act irreducibly on $\Hilbert_j$, and because observables commute with rotations through an angle $2\pi$, one can show that $U_j(R(2\pi))$, where $R(2\pi)$ is a space rotation through an angle $2\pi$, is a multiple of the identity, i.e.,
\bel{eq7.39}
U_j\big(R(2\pi)\big)=\mathrm{e}^{\mathrm{i}2\pi s_j}\,\umat_j\,,
\ee
where $s_j$ is called the `\emph{spin (parity) of sector $j$}'. For $d\ge4$, $s_j\in\frac{1}{2}\Z$, but, for $d=3$,
\bel{eq7.40}
\mathrm{Spin}(2)\simeq\R\,,
\ee
so that $s_j$ can, in principle, be an \emph{arbitrary real number} (mod $\Z$).\\
\indent If the theory describing the system has a \emph{local structure} (see~\cite{R52,R66,R67,R69}) and the vacuum sector $e$ has appropriate properties (`\emph{Haag duality}', see~\cite{R52}) then one can show that sectors can be \emph{composed}, i.e., with two sectors, $i$ and $j$, one can associate their composition, $i\otimes j$, (a kind of \emph{tensor product}), and the sector $i\otimes j$ can be decomposed into a direct sum of irreducible sectors with multiplicities, according to
\bel{eq7.41}
i\otimes j=\bigoplus_{k\in I}\,N_{ij}^k\cdot k\equiv\bigoplus_{k\in I}\,\left(\bigoplus_{\alpha=1}^{N_{ij}^k}\,k^{(\alpha)}\right)\,,
\ee
where $N_{ij}^k=0,\,1,\,2,\ldots$ is the \emph{multiplicity} of the irreducible sector $k$ in the tensor product sector $i\otimes j$, and $k^{(\alpha)}\simeq k$. The integers $N_{ij}^k$ are called `\emph{fusion rules}'. If the theory describing the systems has a local structure one can show that:
\begin{itemize}
	\item[\small{\textbullet}] $N_{ij}^k=N_{ji}^k$ and $i\otimes j\simeq j\otimes i$;
	\item[\small{\textbullet}] to every irreducible sector $j\in I$ one can uniquely associate a (charge-) \emph{conjugate sector} $\jadj$ such that $\jadj\otimes j\simeq j\otimes\jadj$ contains the vacuum (groundstate) sector $e$, \emph{exactly once}, i.e.,
	\begin{equation}\label{eq7.44}
	j\otimes\jadj=e\oplus\Big(\bigoplus_{\scriptsize{\begin{array}{c} k\in I \\ 
	k\neq e \end{array}}}N_{ij}^k\cdot k\Big)\,;
	\end{equation}
\end{itemize}
and
\begin{itemize}
	\item[\small{\textbullet}] $e\otimes j\simeq j\otimes e\simeq j$, for all $j\in I$.
\end{itemize}
\indent Since $i\otimes j\simeq j\otimes i$, there must exist an \emph{intertwiner} (morphism) $\epsilon_{ij}$ intertwining $i\otimes j$ with $j\otimes i$:
\bel{eq7.45}
\epsilon_{ij}\,:\,i\otimes j\overset{\simeq}\longrightarrow j\otimes i\,.
\ee
Focusing on systems in two or three space-time dimensions -- which we will do in the following -- we find, after some serious reflection, that there are usually \emph{two} distinguished intertwiners $\epsilon_{ij}^{+}$ and $\epsilon_{ij}^{-}$ satisfying~\eqref{eq7.45}. (In two space-time dimensions, this can be understood to be a consequence of the fact that the complement of a light cone has two disjoint components; in three space-time dimensions, it is related to the circumstance that two points in the plane can be exchanged either clockwise or anti-clockwise.) It turns out that, thanks to the associativity of the composition of sectors (the tensor product $\otimes$), the operators $\epsilon_{ij}^{\pm}$ obey the \emph{Yang-Baxter equations} (as first observed in~\cite{R65}), and
\bel{eq7.46}
\epsilon_{ij}^{+}\,\epsilon_{ji}^{-}=\textrm{identity}\,.
\ee
It follows from these properties that the intertwiners $\lbrace\epsilon_{ij}^{\pm}\,|\,i,j\in I\rbrace$ determine a unitary representation of the groupoid of colored braids on $n$ strands (the colors are the labels of the irreducible sectors, i.e., the elements of $I$), for arbitrary $n=2,\,3,\ldots$. These representations describe the \emph{quantum statistics} of the system. If
\bel{eq7.47}
\epsilon_{ij}^{+}=\epsilon_{ij}^{-}\textrm{ for all }i,j\in I\,,
\ee
then the representations of the braid groupoids are actually representations of the \emph{permutation groups}, and the quantum statistics ultimately reduces to ordinary \emph{Bose / Fermi statistics}. In $d\ge 4$ space-time dimensions, eq.~\eqref{eq7.47} always holds.\\
\indent 
Let $\mathbb{N}_i$ denote the $|I|\times|I|$ matrix with positive integer matrix elements
\bel{eq7.48}
\left(\mathbb{N}_i\right)_j^k=N_{ij}^k\,.
\ee
The matrices $\mathbb{N}_i$, $i\in I$, all commute and have a common Perron-Frobenius eigenvector $\Delta$, with components $\Delta_i\ge 0$, $i\in I$. It is quite easy to show, using~\eqref{eq7.41}~-~\eqref{eq7.45}, that
\bel{eq7.49}
\mathbb{N}_i\,\Delta=\Delta_i\,\Delta\,,
\ee
i.e., $\Delta_i$ is the largest eigenvalue of the matrix $\N_i$; $\Delta_i$, is called the \emph{statistical} (or \emph{quantum}) \emph{dimension} of the sector $i$. Clearly $\N_e=\umat$ and hence $\Delta_e=1$. If \emph{all} statistical dimensions $\Delta_i$, $i\in I$, are \emph{positive integers} then the quantum statistics is ordinary Bose / Fermi statistics or \emph{abelian} braid statistics. Thus \emph{non-abelian} braid statistics is only encountered in theories with some fractional quantum dimensions.\\
\indent Next, we introduce the `\emph{monodromy operators}'
\bel{eq7.50}
\mu_{ij}\deq\epsilon_{ij}^{+}\,\epsilon_{ji}^{+}\,.
\ee
One aspect of the general \emph{connection between spin and statistics} is that the spectrum of the monodromy operator $\mu_{ij}$ consists of the eigenvalues
\bel{eq7.51}
\exp{\lbrack \mathrm{i}2\pi(s_i+s_j-s_k)\rbrack}\,,\;k\in I\,,
\ee
and the multiplicity of the eigenvalue $\exp{\lbrack \mathrm{i}2\pi(s_i+s_j-s_k)\rbrack}$ is given by $N_{ij}^k$; see~\cite{R67,R72}. Let $v_{ij}^k$ be an intertwiner (`Clebsch-Gordan operator') intertwining the sector $i\otimes j$ with a subsector $k$; see~\eqref{eq7.45}. There are precisely $N_{ij}^k$ \emph{linearly independent} such intertwiners.\\
\indent Then
\bel{eq7.52}
\mu_{ij}\,v_{ij}^k=\exp{\lbrack \mathrm{i}2\pi(s_i+s_j-s_k)\rbrack}\,v_{ij}^k\,.
\ee
In particular, for $i=\jadj$, $k=e$, we have that
\bel{eq7.53}
\mu_{\jadj j}\,v_{\jadj j}^e=\exp{\lbrack \mathrm{i}2\pi(s_j+s_{\jadj})\rbrack}\,v_{\jadj j}^e\,,
\ee
because $s_e=0$ mod $\Z$. One can show that
\bel{eq7.54}
s_j=-s_{\jadj}\textrm{ mod }\Z\,,
\ee
or, equivalently,
\begin{equation*}
\mu_{\jadj j}\,v_{\jadj j}^e=v_{\jadj j}^e\,.
\end{equation*}
This is a weaker form of eq.~\eqref{eq7.22}, subsection 7.1. We conclude this brief survey with the following result valid (for local RQFT in ) \emph{three} space-time dimensions and established in~\cite{R67}; (see also references given there).
\begin{theorem}
\noindent
\begin{itemize}
	\item[(1)] If $I$ is a finite set then $s_j$ is a \emph{rational number}, for all $j\in I$.
	\item[(2)] If either \emph{space reflection in a line} or \emph{time reversal} is a \emph{symmetry} of the theory on \emph{all} its superselection sectors $j\in I$ then the quantum statistics of the theory is ordinary \emph{permutation-group} (Bose / Fermi) \emph{statistics}, and
	\bel{eq7.55}
	s_j\in\frac{1}{2}\Z\,,\textrm{ for all }j\in I\,.
	\ee
	\item[(3)] The following two statements are \emph{equivalent}:
		\begin{itemize}
			\item[(i)] The quantum statistics of the theory is ordinary permutation-group (Bose / Fermi) statistics.
			\item[(ii)] $\exp{\lbrack\mathrm{i}2\pi(s_i+s_j-s_k)\rbrack}=1$, for all $i,\,j,\,k$ in $I$ with $N_{ij}^k\ge 1$.
		\end{itemize}
	Moreover, both statements imply that
	\begin{equation*}
	s_j\in\frac{1}{2}\Z\,,\textrm{ for all }j\in I\,.
	\end{equation*}\\
\end{itemize}
\end{theorem}
\newpage
\noindent\emph{Remarks:}
\indent
\begin{itemize}
	\item[(1)] The \emph{rationality} of the Hall conductivity, i.e., $\upsigma_H=r\frac{e^2}{h}$, $r\in\Q$, in two-dimensional, incompressible electron gases exhibiting the fractional quantum Hall effect is intimately connected to part {\it (1)} of the theorem; see~\cite{R75}.
	\item[(2)] Space reflections in a line and time reversal are \emph{not} symmetries of a two-dimensional electron gas in a transversal, external magnetic field. In view of part {\it (2)} of the theorem, this explains why such systems may exhibit quasi-particles with braid statistics.
	\item[(3)] The precise hypotheses under which the theorem is proven (e.g., local RQFT satisfying `Haag duality') can be found in~\cite{R67}.
\end{itemize}

\indent It is not entirely easy to translate the contents of this theorem into purely field theoretic jargon, at least if one desires to be precise, mathematically. The remark may help the reader that `physical' examples of sectors with fractional spin and braid statistics can be found in the realm of abelian and non-abelian Chern-Simons theories; see, e.g.,~\cite{R64,R76}. In these theories, sectors with fractional spin and statistics can be constructed by applying field operators with Mandelstam flux strings to the vacuum sector. In the theory of the quantum Hall effect topological versions of these theories play a fundamental r\^ole; see~\cite{R75}. They also appear in the theoretical description of graphene.\\

\indent Well, I guess it is time to claim victory!

\newpage
\addcontentsline{toc}{section}{References}

\label{section:references}


\begin{thebibliography}{99}

\small

\bibitem{R1}
	A. Pais, {\it Inward Bound}, Oxford University Press, New York, 1986
\bibitem{R2}
	 H. Kragh, {\it Quantum Generations}, Princeton University Press, Princeton, 1999
\bibitem{R3}
	N. Straumann, {\it \"Uber Paulis wichtigste Beitr\"age zur Physik}, preprint,  arXiv:physics/001003
\bibitem{R4}
	D. Giulini, {\it Electron Spin or `Classically Non-Describable Two-Valuedness'}, preprint, arXiv:hist-ph/0710.3128
\bibitem{R5}
	J. Fr\"ohlich, {\it R\'eflexions sur Wolfgang Pauli}, proceedings of the ``Colloque 2000: Pens\'ee et Science'' of the Fondation F. Gonseth, Eric Emery (ed.), Rev. Synt. \textbf{126}:443-450, 2005
\bibitem{R6}
	N. Straumann, {\it Quantenmechanik}, Springer-Verlag, Berlin, Heidelberg , 2002
\bibitem{R7}
	W. Pauli, {\it Z. Physik }\textbf{16}:155-164, 1923
\bibitem{R8}
	W. Pauli, {\it Z. Physik }\textbf{31}:373-385, 1925
	
\bibitem{R8bis} 
	S. Ferrara, M. Porrati, V.L. Telegdi, {\it Phys. Rev. D }\textbf{46}:3529-3537, 1992

\bibitem{R9}
	J.D. Jackson, {\it Classical Eletromagnetism}, John Wiley \& Sons, New York, 1975
\bibitem{R10}
	W. Pauli, {\it Z. Physik }\textbf{31}:765-783, 1925
\bibitem{R11}
	E.C. Stoner, {\it Phil. Magazine} \textbf{48}:719-736, 1924
\bibitem{R12}
	W. Heisenberg, {\it Zeitschrift f\"ur Physik }\textbf{33}:879-893, 1925
\bibitem{R13}
	M. Born, P. Jordan, {\it Zeitschrift f\"ur Physik }\textbf{34}:858-888, 1925
\bibitem{R14}
	M. Born, W. Heisenberg, P. Jordan, {\it Zeitschrift f\"ur Physik }\textbf{35}:557-615, 1926
\bibitem{R15}
	P.A.M. Dirac, {\it Proc. Royal Soc. } (London) A \textbf{109}:642-653, 1925
\bibitem{R16}
	E. Schr\"odinger, {\it Annalen der Physik }\textbf{79}:361-376, 1926; {\it Annalen der Physik} \textbf{76}:146-147, 1926; {\it Annalen der Physik} \textbf{80}:437-490, 1926 
\bibitem{R17}
	P.A.M. Dirac, {\it Proc. Royal Soc.} A \textbf{117}:610, 1928; A \textbf{118}:351, 1928
\bibitem{R18} W. Pauli, {\it Z. Physik} \textbf{43}:601-623, 1927

\bibitem{R19} J. Fr\"ohlich, O. Grandjean, A. Recknagel, {\it Comm. Math. Phys.} \textbf{193}:527-594, 1998

\bibitem{R20} J. Fr\"ohlich, {\it The Electron is Inexhaustible}, Amer. Math. Soc. Publ., Providence RI, 1999

\bibitem{R21} D. Salamon, {\it Spin Geometry and Seiberg-Witten Invariants}, preprint, 1995

\bibitem{R22} A. Connes, {\it Noncommutative Geometry}, Academic Press, New York, 1994

\bibitem{R23} J. Fuchs, Chr. Schweigert, {\it Symmetries, Lie Algebras and Representations}, Cambridge University Press, Cambridge, New York, 1997

\bibitem{R23bis} G. Velo, D. Zwanziger, {\it Phys. Rev. }\textbf{186}: 1337-1341, 1969; {\it Phys. Rev. }\textbf{188}:2218-2222, 1969\newline\newline
A.Z. Capri, R.L. Kobes, {\it Phys. Rev. D }\textbf{22}:1967-1978, 1980

\bibitem{R23ter} S. Deser, B. Zumino, {\it Phys. Lett. }\textbf{62}B: 335, 1976\newline\newline
K. Vonlanthen, {\it Supergravitation und Velo-Zwanziger Ph\"anomene}, ETH diploma thesis 1978 (N. Straumann, advisor)

\bibitem{R24} J. Fr\"ohlich, U.M. Studer, E. Thiran, {\it Quantum Theory of Large Systems of Non-Relativistic Matter}, in: {\it Fluctuating Geometries in Statistical Mechanics and Field Theory}, Les Houches, Session LXII (1994), F. David, P Ginsparg, J. Zinn-Justin (eds.), Elsevier, New York, 1996

\bibitem{R25} R. Howe, {\it Lect. Appl. Math.} \textbf{21}:179, 1985

\bibitem{R26} S. Weinberg, {\it The Quantum Theory of Fields}, Vol. 1, Cambridge University Press, Cambridge, New York, 1995\newline\newline
	J. Fr\"ohlich, {\it Einf\"uhrung in die Quantenfeldtheorie}, ETH Lecture Notes, 1986

\bibitem{hunziker} W. Hunziker, {\it Commun. Math. Phys.} \textbf{40}:215-222, 1975

\bibitem{R27} E.H. Lieb, {\it The Stability of Matter: From Atoms to Stars}, 4$^{\textrm{th}}$ edition, Springer-Verlag, Berlin, Heidelberg, New York, 2005

\bibitem{R28} J. Fr\"ohlich, E.H. Lieb, M. Loss, {\it Commun. Math. Phys.} \textbf{104}:251-270, 1986

\bibitem{R29} M. Loss, H.-T. Yau, {\it Commun. Math. Phys.} \textbf{104}:283-290, 1986

\bibitem{R30} E.H. Lieb, M. Loss, {\it Commun. Math. Phys.} \textbf{104}:271-282, 1986

\bibitem{R31} C. Fefferman, {\it Proc. Natl. Acad. Science USA} \textbf{92}:5006-5007, 1995: and Lecture Notes

\bibitem{R32} E.H. Lieb, M. Loss, J.-Ph. Solovej, {\it Phys. Rev. Letters} \textbf{75}:985-989, 1995

\bibitem{R33} J. Fr\"ohlich, {\it Ann. Inst. H. Poincar\'e} \textbf{19}:1-103, 1974; {\it Fortschritte der Physik} \textbf{22}:159-198, 1974

\bibitem{R34} L. Bugliaro Goggia, J. Fr\"ohlich, G.M. Graf, {\it Phys. Rev. Letters} \textbf{77}:3494-3497, 1996

\bibitem{R35} C. Fefferman, J. Fr\"ohlich, G.M. Graf, {\it Proc. Natl. Acad. Sci.} \textbf{93}:15009-15011, 1996

\bibitem{R36} C. Fefferman, J. Fr\"ohlich, G.M. Graf, {\it Commun. Math. Phys.} \textbf{190}:309-330, 1999

\bibitem{R37} L. Bugliaro Goggia, C. Fefferman, J. Fr\"ohlich, G.M. Graf, J. Stubbe, {\it Commun. Math. Phys.} \textbf{187}:567-582, 1997

\bibitem{R38} L. Bugliaro Goggia, C. Fefferman, G.M. Graf, {\it Revista Matematica Iberoamericana} \textbf{15}:593-619, 1999

\bibitem{R39} V. Bach, J. Fr\"ohlich, I.M. Sigal, {\it Adv. Math.} \textbf{137}:205-298, 1998; \textbf{137}:299-395, 1998

\bibitem{R40} V. Bach, J. Fr\"ohlich, I.M. Sigal, {\it Commun. Math. Phys.} \textbf{207}:249-290, 1999

\bibitem{R401} M. Griesemer, M. Loss, E.H. Lieb, {\it Inventiones Math.} \textbf{145}:557-587, 1999

\bibitem{R41} J. Fr\"ohlich, M. Griesemer, B. Schlein, {\it Adv. Math.} \textbf{164}:349-398, 2001

\bibitem{R42} J. Fr\"ohlich, M. Griesemer, B. Schlein, {\it Ann. Henri Poincar\'e} \textbf{3}, No. 1:107-170, 2002

\bibitem{R43} V. Bach, J. Fr\"ohlich, A. Pizzo, {\it Comm. Math. Phys.} \textbf{264}:145--165,2006; {\it Comm. Math. Phys.} \textbf{274}:457-486, 2007; {\it Adv. Math.} (to appear)

\bibitem{R44} T. Chen, J. Fr\"ohlich, A. Pizzo, {\it Infraparticle Scattering States in Non-Relativistic QED: I \& II}, preprints 2007

\bibitem{R45} H. Spohn, {\it Dynamics of Charged Particles and Their Radiation Field}, Cambridge University Press, Cambridge, New York, 2004

\bibitem{R46} T. Chen, {\it ETH Diploma Thesis}, 1994

\bibitem{R47} R. Stora, {\it private communication}

\bibitem{R48} G. Gabrielse et al.; see Gabrielse's contribution to these proceedings

\bibitem{R49} M. Fierz, {\it Helv. Phys. Acta} \textbf{12}:3, 1939

\bibitem{R50} R. Jost, {\it The General Theory of Quantized Fields}, AMS Publ., Providence RI, 1965

\bibitem{R51} R.F. Streater, A.S. Wightman, {\it PCT, Spin and Statistics and All That}, Benjamin, New York, 1964

\bibitem{R511} J. Glimm, A. Jaffe, {\it Quantum Physics: A functional Integral Point of View}, Springer-Verlag, Berlin, Heidelberg, New York, 1987

\bibitem{R52} S. Doplicher, R. Haag, J.E. Roberts, {\it Commun. Math. Phys.} \textbf{33}:199, 1971; {\it Commun. Math. Phys.} \textbf{35}:49, 1974

\bibitem{R53} S. Doplicher, J.E. Roberts, {\it Commun. Math. Phys.} \textbf{131}:51, 1990

\bibitem{R54} G. L\"uders, {\it Kong. Dansk. Vid. Selskab, Mat.-Fys. Medd.} \textbf{28}:5, 1954; {\it Ann. Phys.} \textbf{2}:1, 1957\newline\newline
	W. Pauli, {\it Nuovo Cimento} \textbf{6}:204, 1957

\bibitem{R55} R. Jost, {\it Helv. Phys. Acta} \textbf{30}:409, 1957

\bibitem{R56} R. Kubo, {\it J. Phys. Soc. Japan} \textbf{12}:570, 1957\newline\newline
		P.C. Martin, J. Schwinger, {\it Phys. Rev.} \textbf{115}:1342, 1959\newline\newline
		R. Haag, N. Hugenholtz, M. Winnink, {\it Commun. Math. Phys.} \textbf{5}:215, 1967

\bibitem{R57} J.J. Bisognano, E.H. Wichmann, {\it J. Math. Phys.} \textbf{16}:985-1007, 1975

\bibitem{R58} M. Takesaki, {\it Tomita's Theory of Modular Hilbert Algebras and its Applications}, Lecture Notes in Mathematics \textbf{128}, Springer-Verlag, Berlin, Heidelberg, New York, 1970\newline\newline
	O. Bratteli, D.W. Robinson, {\it Operator Algebras and Quantum Statistical Mechanics}, Springer-Verlag, Berlin, Heidelberg, New York, 1979, 1981

\bibitem{R59} R.F. Streater, I.F. Wilde, {\it Nucl. Phys. B} \textbf{24}:561, 1970

\bibitem{R60} J. Fr\"ohlich, {\it Commun. Math. Phys.} \textbf{47}:269-310, 1976

\bibitem{R61} M. Leinaas, J. Myrheim, {\it Il Nuovo Cimento} \textbf{37 B}:1, 1977

\bibitem{R62} G.A. Goldin, R. Menikoff, D.H. Sharp, {\it J. Math. Phys.} \textbf{22}:1664, 1981

\bibitem{R63} F. Wilczeck, {\it Phys. Rev. Letters} \textbf{48}:1144, 1982; \textbf{49}:957, 1982

\bibitem{R64} J. Fr\"ohlich, P.A. Marchetti, {\it Lett. Math. Phys.} \textbf{16}:347, 1988; {\it Commun. Math. Phys.} \textbf{121}:177, 1988

\bibitem{R65} J. Fr\"ohlich, {\it Statistics of Fields, the Yang-Baxter Equation and the Theory of Knots and Links}, in: {\it Non-Perturbative Quantum Field Theory}, Carg\`ese 1987, G. 't Hooft et al. (eds.), Plenum Press, New York, 1988\newline\newline
	J. Fr\"ohlich, {\it Statistics and Monodromy in Two- and Three-Dimensional Quantum Field Theory}, in: {\it Differential Geometrical Methods in Theoretical Physics}, K. Bleuler, M. Werner (eds.), Kluwer Academic Publ., Dordrecht, 1988

\bibitem{R66} K. Fredenhagen, K.H. Rehren, B. Schroer, {\it Commun. Math. Phys.} \textbf{125}:201, 1989

\bibitem{R67} J. Fr\"ohlich, F. Gabbiani, {\it Rev. Math. Phys.} \textbf{2}:251, 1990\newline\newline
	J. Fr\"ohlich, P.A. Marchetti, {\it Nucl. Phys. B} \textbf{356}:533, 1991

\bibitem{R68} L. Birke, J. Fr\"ohlich, {\it Rev. Math. Phys.} \textbf{14}:829, 2002

\bibitem{R80} S. Weinberg, E. Witten, {\it Phys. Letters B} \textbf{96}:59, 1980

\bibitem{R69} D. Buchholz, K. Fredenhagen, {\it Commun. Math. Phys.
 } \textbf{84}:1, 1982
\bibitem{R70} K. Osterwalder, R. Schrader, {\it Commun. Math. Phys.
 } \textbf{42}:281, 1975; see also\newline\newline
 V. Glaser, {\it Commun. Math. Phys.} \textbf{37}:257, 1974

\bibitem{R71} H. Araki, {\it J. Math. Phys. } \textbf{2}:267, 1961\newline\newline
		W. Schneider, {\it Helv. Phys. Acta} \textbf{42}:201, 1969

\bibitem{R72} J. Fr\"ohlich, T. Kerler, {\it Quantum Groups, Quantum Categories and Quantum Field Theory}, Lecture Notes in Mathematics, Vol. 1542, Springer-Verlag, Berlin, Heidelberg, New York, 1993


\bibitem{R73} J. Fuchs, I. Runkel, Chr. Schweigert, {\it Nucl. Phys. B} \textbf{624}:452, 2002; {\it Nucl. Phys. B} \textbf{646}:353, 2002 

\bibitem{R74} {\it The Quantum Hall Effect}, R.E. Prange, S.M. Girvin (eds.), Graduate Texts in Contemporary Physics, Springer-Verlag, Berlin, Heidelberg, New York, 1990\newline\newline
		{\it Quantum Hall Effect}, M. Stone (ed.), World Scientific Publ., Singapore, London, Hong Kong, 1992

\bibitem{R75} J. Fr\"ohlich, {\it The Fractional Quantum Hall Effect, Chern-Simons Theory, and Integral Lattices}, in: {\it Proc. of ICM '94}, S.D. Chatterji (ed.), Birkh\"auser Verlag, Basel, Boston, Berlin, 1995\newline\newline
	J. Fr\"ohlich, B. Pedrini, Chr. Schweigert, J. Walcher, {\it J. Stat. Phys.} \textbf{103}:527, 2001\newline\newline
	J. Fr\"ohlich, B. Pedrini, in: {\it Statistical Field Theory}, Como 2001, A. Cappelli, G. Mussardo (eds.), Kluwer, New York, Amsterdam, 2002

\bibitem{R76} R. Jackiw, S. Templeton, {\it Phys. Rev. D} \textbf{23}:2291, 1981\newline\newline
		S. Deser, R. Jackiw, S. Templeton, {\it Phys. Rev. Letters
 } \textbf{48}:975, 1982\newline\newline
R. Pisarski, S. Rao, {\it Phys. Rev. D} \textbf{32}:2081, 1985

\end{thebibliography}
\end{document}